\PassOptionsToPackage{table,xcdraw}{xcolor}
\documentclass[sigconf]{acmart}
\usepackage{float}
\usepackage{graphicx}
\usepackage{subcaption}
\usepackage{booktabs} 
\usepackage{comment}
\usepackage[utf8]{inputenc}
\usepackage{fontenc}
\DeclareUnicodeCharacter{202F}{ }
\AtBeginDocument{%
  }

\copyrightyear{2026}
\acmYear{2026}
\setcopyright{cc}
\setcctype{by}
\acmConference[CUI '26]{ACM Conversational User Interfaces 2026}{July 21--24, 2026}{Bremen, Germany}
\acmBooktitle{ACM Conversational User Interfaces 2026 (CUI '26), July 21--24, 2026, Bremen, Germany}
\acmDOI{10.1145/3816046.3816223}
\acmISBN{979-8-4007-2741-2/2026/07}

\begin{document}

\title{Toward Metaphor-Fluid Conversation Design for Voice User Interfaces}

\author{Smit Desai}
\authornote{Corresponding author}
\orcid{0000-0001-6983-8838}
\affiliation{%
  \institution{Northeastern University}
  \city{Boston}
  \state{Massachusets}
  \country{USA}
}
\email{sm.desai@northeastern.edu}

\author{Jessie Chin}
\affiliation{%
  \institution{University of Illinois, Urbana-Champaign}
  \city{Urbana-Champaign}
  \state{Illinois}
  \country{USA}}
\email{chin5@illinois.edu}

\author{Dakuo Wang}
\affiliation{%
  \institution{Northeastern University}
  \city{Boston}
  \state{Massachusets}
  \country{USA}}
\email{d.wang@northeastern.edu}

\author{Benjamin Cowan}
\affiliation{%
  \institution{University College Dublin}
  \city{Dublin}
  \country{Ireland}}
\email{benjain.cowan@ucd.ie}

\author{Michael Twidale}
\affiliation{%
  \institution{University of Illinois, Urbana-Champaign}
  \city{Urbana-Champaign}
  \country{USA}}
\email{twidale@illinois.edu}

\renewcommand{\shortauthors}{Desai et al.}

\begin{abstract}

Metaphors play a critical role in shaping user experiences with Voice User Interfaces (VUIs), yet existing designs rely on static, human-centric metaphors that fail to adapt to diverse contexts and user needs. We introduce Metaphor-Fluid Design, a novel approach that adjusts metaphorical representations based on conversational use-contexts. We compare this to a Default VUI, which models commercial VUIs commonly designed around the assistant persona, offering uniform interaction across contexts. In Study 1 (N=130), metaphors were mapped to four key use-contexts—commands, information seeking, sociality, and error recovery—along dimensions of formality and hierarchy, revealing distinct preferences for task-specific metaphorical designs. Study 2 (N=91) evaluates a Metaphor-Fluid VUI against a Default VUI, showing that Metaphor-Fluid Design enhances perceived intention to adopt, enjoyment, and likability. However, individual differences in metaphor preferences highlight the need for personalization. These findings challenge the one-size-fits-all paradigm and demonstrate Metaphor-Fluid Design's potential for more adaptive human-AI interactions.

\end{abstract}

%%
%% The code below is generated by the tool at http://dl.acm.org/ccs.cfm.
%% Please copy and paste the code instead of the example below.
%%

%%
%% Keywords. The author(s) should pick words that accurately describe
%% the work being presented. Separate the keywords with commas.

\begin{teaserfigure}
   \centering
    \includegraphics[width=\textwidth]{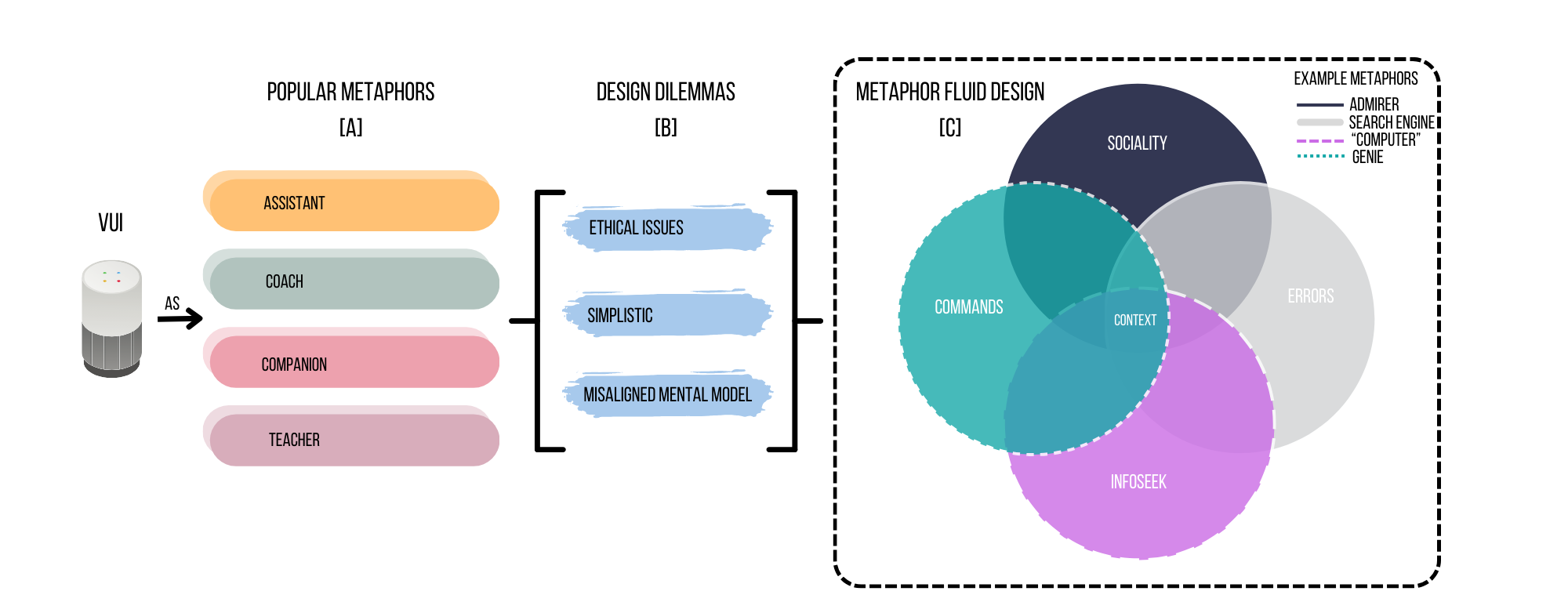}
    \Description[Schematic of Metaphor-Fluid Design evolution]{The schematic depicts three stages from left to right. Stage A shows a Voice User Interface described using popular metaphors including Assistant, Coach, Companion, and Teacher. Stage B identifies three design issues: ethical issues, simplistic design, and misaligned mental models. Stage C illustrates Metaphor-Fluid Design as a central diagram with four conversational use-contexts arranged around it: Sociality, Commands, Errors, and Infoseek, connected by a Context node. Example metaphors are listed: Admirer, Search Engine, Computer, and Genie, each represented by distinct line styles.}
   \caption{The schematic depicts the proposed evolution toward Metaphor-Fluid Design, starting from the description of Voice User Interfaces (VUIs) in different social roles using popular metaphors [A], leading to the identification of design issues [B], and ultimately inspiring Metaphor-Fluid design [C].}
   \label{fig:mfd}
\end{teaserfigure}

%%
%% This command processes the author and affiliation and title
%% information and builds the first part of the formatted document.
\maketitle

\section{Introduction}
Voice User Interfaces (VUIs) are becoming an integral part of daily interactions, embedded in smart speakers, mobile devices, automobiles, and wearable technologies. As these technologies proliferate, the ways in which users engage with and conceptualize them evolve dynamically. However, contemporary VUI design remains largely constrained by a single dominant metaphor—humanness \cite{Doyle_Edwards_Dumbleton_Clark_Cowan_2019}. Most commercial VUIs are designed to mimic human conversational abilities, using speech, turn-taking, and natural language processing to create an interaction model that feels intuitive \cite{Edlund_2019, Pradhan_Lazar_2021}. Within this overarching metaphor, VUIs are often framed as social entities that take on various roles, including teachers, therapists, coaches, advisors, and companions \cite{Desai_Chin_2023, Jung_Kim_So_Kim_Oh_2019, Desai_Hu_Lundy_Chin_2023, Wang_Yang_Shao_Abdullah_Sundar_2020, Desai_Lundy_Chin_2023, Lee_Frank_IJsselsteijn_2021, Motalebi_Cho_Sundar_Abdullah_2019, Trajkova_Martin-Hammond_2020, Purington_Taft_Sannon_Bazarova_Taylor_2017}. Among these, the assistant persona has emerged as the most prevalent, dominating commercial implementations and positioning the VUI as a helpful, subservient entity designed to execute user commands \cite{Desai_Twidale_2023, Doyle_Clark_Cowan_2021}. While this metaphor provides a familiar interaction paradigm, it also imposes significant limitations. It assumes a fixed, one-size-fits-all identity for the VUI, failing to account for the fluid or context-dependent nature of human-VUI interactions across different tasks and contexts \cite{Pradhan_Lazar_2021, desai2024cui, Desai_Chin_2023, Desai_Twidale_2022}. This rigidity leads to usability challenges \cite{Cowan_Pantidi_Coyle_Morrissey_Clarke_Al-Shehri_Earley_Bandeira_2017}, mismatches between user expectations and system behavior \cite{Doyle_Clark_Cowan_2021}, and broader concerns related to identity surrounding anthropomorphic design choices \cite{McMillan_Jaber_2021, Desai_Twidale_2022, Simpson_Crone_2022, Pradhan_Lazar_2021, Luger_Sellen_2016, Turk_2016}.

These challenges reflect a holistic issue in interface design: the role of metaphors in shaping user interaction. Metaphors have played a foundational role in shaping Human-Computer Interaction (HCI) \cite{Neale_Carroll_1997, Laurel_1997, Laurel_Mountford_1990, Carroll_Thomas_1982, Carroll_Mack_Kellogg_1988, Don_Brennan_Laurel_Shneiderman_1992}. In Graphical User Interfaces (GUIs), they are structured into primary and secondary layers: the desktop metaphor, for instance, serves as the primary organizing concept, with secondary metaphors—such as files, folders, and recycling bins—providing finer interaction affordances \cite{Blackwell_2006}. These layered metaphors help users form coherent mental models of system functionality. In contrast, VUIs lack a stable visual representation, making their metaphorical framing less explicit. Instead, this framing is operationalized through persona design, where a VUI’s voice, linguistic style, and response behaviors collectively establish its identity \cite{Pradhan_Lazar_2021}. Unlike GUIs, where metaphors map interactions onto familiar objects, VUIs inherently imply humanness, reinforcing anthropomorphic interpretations. The presence of human-like speech cues, names, and conversational norms encourages users to assign social roles (e.g., therapist or companion) to VUIs, often expecting them to behave with intelligence, adaptability, and even emotional awareness \cite{Luger_Sellen_2016, Sutton_Foulkes_Kirk_Lawson_2019}.

Despite this anthropomorphic framing, users do not engage with VUIs as static entities \cite{Pradhan_Findlater_Lazar_2019}. Instead, they shift their metaphorical framings depending on use-contexts—the functional settings in which interactions take place \cite{Desai_Twidale_2023}. Prior research has identified several recurring use-contexts in VUI interactions, including commands, where users expect efficiency and compliance (e.g., "Turn off the lights"); information seeking, where they seek structured knowledge retrieval (e.g., "What is the capital of Japan?"); sociality, where they engage in casual conversation (e.g., "Tell me a joke"); and error recovery, where they attempt to correct or refine system responses (e.g., "That’s not what I meant—try again") \cite{Sciuto_Saini_Forlizzi_Hong_2018, Desai_Chin_2023, ammari2019music, Bentley_Luvogt_Silverman_Wirasinghe_White_Lottridge_2018}. In each of these use-contexts, users adopt distinct metaphorical descriptions: they conceptualize a VUI as a butler or secretary when issuing commands, as an expert librarian when retrieving information, and as a companion in social exchanges \cite{Desai_Twidale_2023}. However, commercial VUIs impose a singular, unchanging persona across all contexts, failing to adapt to the shifting expectations users bring to different types of interactions. This rigidity results in fundamental misalignments—when users anticipate authoritative expertise but receive generic, non-committal responses, or when they expect a social exchange but encounter the same neutral, task-oriented demeanor, the interaction feels unnatural and disengaging \cite{Desai_Hu_Lundy_Chin_2023}. These breakdowns not only present usability challenges but also raise deeper ethical concerns.

The assistant metaphor, in particular, has been widely critiqued for reinforcing problematic socio-cultural assumptions \cite{McMillan_Jaber_2021, Brahnam_De_Angeli_2008, Brahnam_Karanikas_Weaver_2011}. Acting as the Default VUI in commercial systems such as Amazon Alexa, Google Assistant, and Apple Siri, this metaphor frames VUIs as subservient entities designed to efficiently execute user commands \cite{Turk_2016}. These systems are marketed as digital assistants, emphasizing compliance, convenience, and deference—qualities that are often reinforced through feminized voices and polite, deferential speech patterns \cite{Kuzminykh_Sun_Govindaraju_Avery_Lank_2020}. This design choice perpetuates outdated power dynamics that position technology as a passive, compliant agent rather than an adaptive, context-aware tool. Prior studies have highlighted the risks of over-anthropomorphizing VUIs, where users may develop unrealistic expectations of intelligence or emotional understanding, leading to frustration when the system fails to respond appropriately \cite{Cowan_Pantidi_Coyle_Morrissey_Clarke_Al-Shehri_Earley_Bandeira_2017, Trajkova_Martin-Hammond_2020}. At the same time, users may dehumanize VUIs, treating them with excessive aggression or dismissiveness—behaviors that some researchers argue could spill over into human social interactions \cite{Laurel_1997, Brahnam_De_Angeli_2008}. Furthermore, the universal application of a single persona disregards the diverse ways in which users naturally interpret and interact with these systems. 

Building on prior work, we introduce and propose a shift toward Metaphor-Fluid Design (as illustrated in Figure \ref{fig:mfd}), a novel design approach that moves beyond static persona models by enabling VUIs to dynamically shift their metaphorical framing based on conversational context. Rather than adhering to a singular metaphor—such as the assistant—this approach allows the system to adopt multiple personas, aligning its interaction style with the expectations of the task at hand. Crucially, Metaphor-Fluid Design also incorporates non-human and fictional metaphors, which have been shown to be both relevant and frequently used by users \cite{desai2024cui, Desai_Twidale_2023, Jung_Qiu_Bozzon_Gadiraju_2022}. While commercial VUIs default to anthropomorphic personas, users often describe them in non-human terms, likening them to machines, calculators, encyclopedias, or even mythical entities like genies or ghosts \cite{Desai_Twidale_2023}. By broadening the metaphorical landscape beyond human roles, this approach creates greater flexibility in interaction design, allowing VUIs to respond more intuitively to diverse user expectations. This work is guided by two primary research questions:

\begin{itemize}
\item RQ1: What metaphors commonly align with users' interactions in use-contexts such as commands, information seeking, sociality, and error recovery during conversations with VUIs?
\item RQ2: How does a Metaphor-Fluid VUI compare to a Default VUI in terms of perceived enjoyment, intention to adopt, trust, likability, and intelligence, and in the characteristics of the metaphorical descriptions it generates? 

\end{itemize}

These questions are explored through two studies. The first study (N=130) maps user-generated metaphors to different conversational contexts, revealing systematic variations in metaphorical preferences across tasks. Findings show that users do not conceptualize VUIs through a singular metaphor but instead shift their framings based on the use-context. %%Specifically, users associate command-based interactions and error recovery with Guides (e.g., Genie, Search Engine), information-seeking tasks with Aides (e.g., Encyclopedia, "Computer" from Star Trek), and social exchanges with Companions (e.g., Friend, Flmatmate). These patterns demonstrate that users instinctively align VUI metaphors with the demands of the task at hand, challenging the assumption that a single persona is sufficient for all interactions. 
The second study (N=91) empirically evaluates a Metaphor-Fluid VUI, which adapts its metaphorical framing dynamically across use-contexts, against a Default VUI that retains a single static assistant metaphor. Results indicate that Metaphor-Fluid design leads to significantly higher perceived likability, enjoyment, and intention to adopt, showing that aligning a VUI’s persona with context-specific expectations positively impacts user perceptions. However, findings also highlight individual differences in metaphor preferences.

Findings from this work challenge the dominant paradigm of static persona design in VUIs. By formalizing Metaphor-Fluid Design as a structured approach to VUI development, this paper contributes to a growing discourse on how metaphor selection influences user interaction. These insights have broader implications for the design of conversational agents, underscoring the importance of flexibility, contextual awareness, and ethical considerations in shaping future human-AI conversational interactions.

\section{Related Work}

\subsection{Metaphors—A Brief Primer}

Metaphors are more than just literary flourishes; they are fundamental tools for understanding and interpreting the world \cite{Indurkhya_2013}. In linguistics and cognitive science, metaphors have been shown to shape not only how we communicate but also how we think \cite{Gentner_Hoyos_2017}. According to Lakoff and Johnson's theory of conceptual metaphors, our understanding of abstract concepts is often structured by metaphors derived from concrete experiences \cite{Lakoff_Johnson_1980}. For example, when we say "time is money," we are using a metaphor to conceptualize time as a valuable resource, which affects how we perceive and manage it \cite{Lakoff_Johnson_1980}.

Metaphors are powerful because they influence cognition and behavior, often in subtle ways. They help bridge the gap between the familiar and the unfamiliar, making complex ideas more accessible \cite{Gentner_Hoyos_2017}. By providing a conceptual structure, metaphors allow individuals to grasp abstract concepts by relating them to more concrete experiences. In cognitive psychology, metaphors have been found to shape reasoning, decision-making, and problem-solving processes, thereby playing an integral role in how knowledge is constructed and applied \cite{Moser_2000, Cameron_Maslen_2010}. Through metaphors, we can connect domains that are seemingly unrelated, fostering creativity and enabling us to generate novel ideas \cite{Lockton_Singh_Sabnis_Chou_Foley_Pantoja_2019}. In science and technology, metaphors often serve as foundational frameworks that influence both how researchers theorize and how laypeople understand complex phenomena \cite{Hofstadter_1995}. Alan Turing, for example, used a metaphor of a human computer following a set of instructions to describe his model for computation, which became the basis for modern computer science \cite{Piccinini_2003}. Similarly, metaphors like "genetic code" or "cellular machinery" have shaped how scientists and the public conceptualize biological processes, providing familiar reference points that help make sense of otherwise abstract concepts \cite{Keller_2003}. Metaphors thus provide a cognitive shortcut that allows individuals to comprehend and manipulate ideas beyond the immediate grasp of direct experience \cite{Black_1962}.

Moreover, metaphors are deeply embedded in everyday language, often so much so that we are unaware of their influence. Phrases such as "grasping an idea" or "feeling down" illustrate how metaphorical thinking is woven into our conceptual system \cite{Lakoff_Johnson_1980}.

\subsection{Metaphors in HCI}

Metaphors in HCI have profoundly shaped how users conceptualize and interact with technology. By leveraging familiar concepts (e.g., files, folders), metaphors bridge the gap between users' prior knowledge and new digital environments, simplifying interactions and reducing cognitive load \cite{Neale_Carroll_1997, Carroll_Thomas_1982}. The desktop metaphor stands out as a pivotal example in GUIs, drawing on the physical analogy of an office desk to provide users with a coherent mental model for navigating digital spaces \cite{Johnson-Laird_1983}. This foundational desktop metaphor is supplemented with secondary metaphors like folders, files, bins, scrolling, and icons and translates abstract computational processes into familiar actions, such as "dragging a file to the trash" to discard a document \cite{Colburn_Shute_2008}.

The impact of the desktop metaphor extends far beyond mere visual representation. It has played a crucial role in making computers more approachable and familiar, significantly contributing to their widespread adoption \cite{Carroll_Mack_Kellogg_1988}. By aligning user expectations with system capabilities, the desktop metaphor has fostered increased usability and user satisfaction. Furthermore, it has provided a common language for developers and users to communicate about new technologies, facilitating the ongoing evolution of human-computer interfaces. The desktop metaphor's influence on early computing systems was so profound that it shaped users' mental models for decades. For a detailed exploration of its history and significance, Alan Blackwell's work offers an in-depth analysis that underscores the metaphor's lasting impact on HCI design principles \cite{Blackwell_2006}.

%The application of metaphors in interface design extends beyond mere aesthetics, serving dual roles of familiarization and transportation, as outlined by Paul Heckel in his work on friendly software design~\cite{Heckel_1984}. Familiarization employs metaphors to introduce recognizable concepts, facilitating user comprehension and mental model construction. Transportation, on the other hand, utilizes these mental models to create immersive experiences. In the context of VUIs, the concept of humanness frequently serves as the primary metaphor, acting as a conduit for familiarization and transportation effects.

\subsection{Metaphors in VUIs}

Although metaphors have a long lineage in HCI, their systematic study in VUIs is comparatively recent. Contemporary VUI design is largely anchored in a humanness metaphor, where systems are positioned in recognizable social roles such as assistants, educators, coaches, storytellers, and even therapists \cite{Desai_Chin_2023, Jung_Kim_So_Kim_Oh_2019, Desai_Hu_Lundy_Chin_2023, Wang_Yang_Shao_Abdullah_Sundar_2020, Desai_Lundy_Chin_2023, Lee_Frank_IJsselsteijn_2021, Motalebi_Cho_Sundar_Abdullah_2019}. Commercial systems such as Amazon Alexa and Apple Siri operationalize this framing through the digital assistant metaphor, a designer-intended and manufacturer-marketed positioning that emphasizes helpfulness, approachability, and social familiarity \cite{Sciuto_Saini_Forlizzi_Hong_2018, Turk_2016, McMillan_Jaber_2021, Desai_Twidale_2023}. Anthropomorphism—the attribution of human traits to non-human entities—thus becomes both a design strategy and an interpretive lens through which users make sense of these systems \cite{Epley_Waytz_Cacioppo_2007}.

In practice, metaphors are instantiated through system personas and remain central to Conversation Design workflows \cite{Sadek_Calvo_Mougenot_2023}. Industry guidelines, such as Google's Conversation Design framework\footnote{https://developers.google.com/assistant/conversation-design/create-a-persona}, explicitly encourage designers to model agents after familiar real-world roles (e.g., bank tellers, exercise coaches) to ground interaction in established social scripts \cite{McMillan_Jaber_2021}. However, metaphor selection is not neutral. It shapes user expectations, influences trust calibration, and mediates long-term engagement \cite{Desai_Twidale_2023, Khadpe_Krishna_Fei-Fei_Hancock_Bernstein_2020, Pradhan_Findlater_Lazar_2019}. For instance, competence-signaling metaphors may initially attract users, whereas lower-competence framings can sustain long-term preference \cite{Khadpe_Krishna_Fei-Fei_Hancock_Bernstein_2020}. Conversational formality alters metaphor interpretation \cite{Chin2024Like}, and users often prefer systems whose projected personalities align with their own traits \cite{Braun_Mainz_Chadowitz_Pfleging_Alt_2019}. Although "persona" has become the more common term in VUI design through these industry frameworks, we use "metaphor" throughout this work because it connects to the broader HCI lineage and, as we argue in Section 3, avoids the implicit assumption of personhood that "persona" carries, which constrains the design space to human-centric framings.

At the same time, users do not passively adopt designer-imposed metaphors. They develop folk theories—often metaphorical in nature—to explain VUI behavior and rationalize breakdowns \cite{DeVito_Birnholtz_Hancock_French_Liu_2018, Kim_Choudhury_2021, Kuzminykh_Sun_Govindaraju_Avery_Lank_2020, Desai_Twidale_2022}. When these user-generated interpretations diverge from intended personas, the resulting misalignment can widen Norman’s “gulf of execution” \cite{Norman_2013, Luger_Sellen_2016}, affecting usability, trust, and continued adoption.

Despite these tensions, there are practical reasons why static metaphors persist. A consistent metaphorical framing provides a stable anchor for user expectations, supports long-term relationship building, and reduces the risk of disorienting users through unexpected behavioral shifts \cite{Doyle_Clark_Cowan_2021, Clark_Doyle_Garaialde_Gilmartin_Schlögl_Edlund_Aylett_Cabral_Munteanu_Edwards_et_al._2019, Doyle_Edwards_Dumbleton_Clark_Cowan_2019}. These benefits are well-documented but also introduce three recurring challenges that are less commonly discussed. First, \textbf{\textit{ethical concerns}} arise when assistant or servant framings reinforce hierarchical and gendered stereotypes, particularly when paired with stereotypically coded voices \cite{McMillan_Jaber_2021, Pradhan_Lazar_2021, Brahnam_Karanikas_Weaver_2011, Turk_2016, Kuzminykh_Sun_Govindaraju_Avery_Lank_2020}. Second, \textbf{\textit{singular and static metaphorical framings}} often oversimplify user expectations. Empirical work shows that users’ perceptions shift across contexts—alternating between viewing VUIs as persons, tools, or relational entities depending on task demands \cite{Pradhan_Lazar_2021}. For example, older adults interacting with a VUI framed as an exercise coach expressed expectations that extended beyond coaching to include relational encouragement akin to friendship \cite{Desai_Hu_Lundy_Chin_2023}. Such findings suggest that one overarching metaphor may inadequately capture the diversity of interactional roles VUIs occupy. Third, anthropomorphic metaphors can \textbf{\textit{distort and misalign mental models}} by heightening expectations of competence or social intelligence, leading to over- or under-calibrated trust \cite{Cowan_Pantidi_Coyle_Morrissey_Clarke_Al-Shehri_Earley_Bandeira_2017, Luger_Sellen_2016}. These distortions can result in inconsistent use patterns or eventual abandonment \cite{Trajkova_Martin-Hammond_2020}.

In response, recent scholarship has sought to broaden the metaphor design space. Desai and Twidale \cite{Desai_Twidale_2023} proposed a framework for metaphor contextualization along dimensions of type (human, non-human, fictional), source attribution, knowledge base, and motivation, revealing tensions between designer-imposed human metaphors and user-preferred alternatives. Empirical investigations have explored non-human metaphors through lenses such as the Great Chain of Being \cite{Jung_Qiu_Bozzon_Gadiraju_2022}, finding no inherent superiority of human framings. Domain-specific variations further complicate this picture, with preferences differing across contexts such as health and finance \cite{Desai_Dubiel_Leiva_2024}. Fictional metaphors—from Star Trek’s “Computer” \cite{Turk_2016, Axtell_Munteanu_2021} to dystopian or speculative archetypes \cite{Feldman_2024, Lupetti_Murray-Rust_2024}—have similarly functioned as both aspirational and cautionary reference points.

Parallel lines of research address metaphor misalignment through cognitive and communicative models. The Mutual Theory of Mind framework \cite{Wang_Goel_2022, Wang_Saha_Gregori_Joyner_Goel_2021} and partner models \cite{Doyle_Clark_Cowan_2021, 10.1016/j.ijhcs.2024.103400} conceptualize user beliefs about VUI capabilities as dynamic, evolving through interaction. These approaches underscore that mental models are not static reflections of persona design but continuously negotiated understandings. Building on this foundation, we propose \emph{\textbf{Metaphor-Fluid Conversation Design}}: an approach in which VUIs adapt their metaphorical framing across conversational use-contexts to better maintain alignment with users’ evolving mental models.

\section{{Toward Metaphor-Fluid Conversation Design}}

Metaphors play an important role in VUIs by helping users construct mental models for interaction and guiding expectations of system behavior \cite{Carroll_Thomas_1982, Carroll_Mack_Kellogg_1988, Colburn_Shute_2008}. As discussed in the previous sections, unlike GUIs, where spatial and object-based metaphors provide persistent visual frameworks \cite{Blackwell_2006}, VUI metaphors lack stable visual anchors and must be inferred from interaction patterns. This fundamental difference has led to a design paradigm in which system personas serve as implicit metaphors, mapping VUI behavior onto familiar social roles such as assistants, coaches, or companions \cite{Desai_Chin_2023, Jung_Kim_So_Kim_Oh_2019, Desai_Hu_Lundy_Chin_2023, Wang_Yang_Shao_Abdullah_Sundar_2020, Desai_Lundy_Chin_2023, Lee_Frank_IJsselsteijn_2021, Motalebi_Cho_Sundar_Abdullah_2019, Desai_Chin_2021}. Empirical findings reveal that users do not experience VUIs as singular, fixed entities; rather, they dynamically reframe their mental models depending on the interaction context \cite{Pradhan_Findlater_Lazar_2019}.
{Desai and Twidale \cite{Desai_Twidale_2023} show that the same VUI is often described through different metaphors depending on what the user is doing, being framed as a “search engine” or “librarian” during information seeking but as a “companion” during social talk. These shifts were systematically tied to conversational context, illustrating that users already experience VUIs metaphorically and that these metaphors naturally change as the context changes.}

{These empirical observations motivate the concept of Metaphor-Fluid Conversation Design. By metaphor-fluidity, we refer to the intentional use of different metaphorical framings for different conversational contexts. The metaphor shifts when the conversational context shifts, such as when an interaction naturally moves from issuing a command to seeking information or from information seeking to social small talk. Because these shifts unfold within one continuous interaction, users experience the design as a coherent sequence in which each segment is shaped by an appropriate metaphor. The sense of fluidity emerges from these context-level transitions rather than from turn-by-turn variation and are perceived on the whole.}

{This approach also intersects with a longstanding debate in HCI about whether interactive systems should be treated primarily as tools or as social agents \cite{Don_Brennan_Laurel_Shneiderman_1992, Blackwell_2006, Laurel_1997, Maes_Shneiderman_Miller_1997}. VUIs complicate this binary because they do not merely respond; they speak, take turns, and project stance, prompting users to alternate between instrumental and social framings depending on the task \cite{Pradhan_Findlater_Lazar_2019, Desai_Twidale_2022, Desai_Twidale_2023}. Metaphor-fluidity offers a pragmatic middle ground. It preserves the tool-like utility of VUIs while acknowledging that conversational interactions also inevitably invoke socially meaningful roles. This mirrors how people shift their own communicative styles across contexts \cite{Goffman_1959, Harrington_Garg_Woodward_Williams_2022} and provides a structured way for VUIs to manage these shifts without collapsing into full anthropomorphism.}

{Despite these dynamics, current VUI design still largely relies on fixed, unified personas. Designers have historically favored single identities that remain consistent across all interactions \cite{Sadek_Calvo_Mougenot_2023}, assuming users will develop trust through coherence and stability. However, this assumption is increasingly misaligned with user expectations \cite{Doyle_Clark_Cowan_2021, desai2024cui, Braun_Mainz_Chadowitz_Pfleging_Alt_2019} and may reinforce problematic anthropomorphic framings such as gendered or servile representations embedded in assistant metaphors \cite{Brahnam_Karanikas_Weaver_2011, Turk_2016, Kuzminykh_Sun_Govindaraju_Avery_Lank_2020}. VUIs operate within a conversational medium where people routinely shift relational stance depending on the task. Designing a single, inflexible metaphor to cover all these moments risks working against how users intuitively engage with conversational systems.}

Given this, the central question is no longer “What is the best metaphor for VUIs?” but rather:
\textit{How can VUIs accommodate different metaphors in ways that align with the changing demands of an interaction?}
This perspective motivates Metaphor-Fluid Conversation Design. This design approach is guided by two primary tenets:
\begin{enumerate}
\item \textbf{Expanding metaphorical scope:} We advocate for moving beyond the conventional persona design paradigm, where a singular, often human-centric metaphor serves as the interactional layer of the conversational interface. Instead, we encourage designers to incorporate a broader range of metaphors—including non-human and fictional ones—into their process. This shift allows for more nuanced interactions and avoids the limitations of monolithic personas that implicitly foreground personhood and human-like identity.
\item \textbf{Contextual metaphor adaptation:} A metaphor-fluid VUI varies its metaphorical presentation based on the conversational context in which it is used. {Because different tasks place different social and functional demands on the system, adapting the metaphor at the level of context enables the VUI to adopt framings that feel more intuitive and better aligned with what users are doing in that moment.}
\end{enumerate}

In this paper, we introduce the concept of Metaphor-Fluid Conversation Design using VUIs as a central example. Our aim is to articulate this design approach and evaluate its impact on user perception when compared to a traditional, single-metaphor VUI.

\section{Study 1: Mapping VUI Metaphors to Conversational Use-Contexts}

To address RQ1, we conducted Study 1, which examines the metaphors that most closely align with users’ interactions across different use-contexts in conversations with VUIs. Specifically, we investigated four key use-contexts: commands, information seeking, sociality, and error recovery. Our approach consisted of several interconnected steps. We began by identifying metaphors commonly employed by both users and designers to describe VUIs, ranging from designer-intended metaphors created to establish familiarity in specific scenarios (e.g., “Therapist”) to user-generated metaphors that emerged from sense-making during interaction (e.g., “Child”). We then developed and justified the selection of the four use-contexts used in this study through a review of empirically grounded VUI literature involving user studies. {Finally, to align the identified metaphors with these use-contexts, we adopted a two-dimensional mapping framework that positions both along the social dimensions of formality and hierarchy, allowing a descriptive comparison between users’ perceptions of metaphors and their expectations of conversational behavior across contexts.}

\subsection{Methods}

\subsubsection{Metaphor Identification}

{
Our metaphor collection drew from two sources combining literature review and user studies. Desai \& Twidale  \cite{Desai_Twidale_2023} conducted  a systematic literature review of VUI research from the past decade, screening 664 articles down to 80, supplemented by a user study with 14 participants interacting with Amazon Echo Dot through semi-structured interviews (Source A in Figure \ref{fig:vuimetaphorid}). Together, these identified 93 VUI-related metaphors. In addition, Chin et al.  \cite{Chin2024Like} contributed an additional 225 metaphors collected from 58 participants in a simulated smart-home environment, where participants completed cognitive tasks and were explicitly asked to compare the VUI to other entities using metaphors (Source B in Figure \ref{fig:vuimetaphorid}).

\begin{figure*}[htbp]
    \centering
    \includegraphics[width=\textwidth]{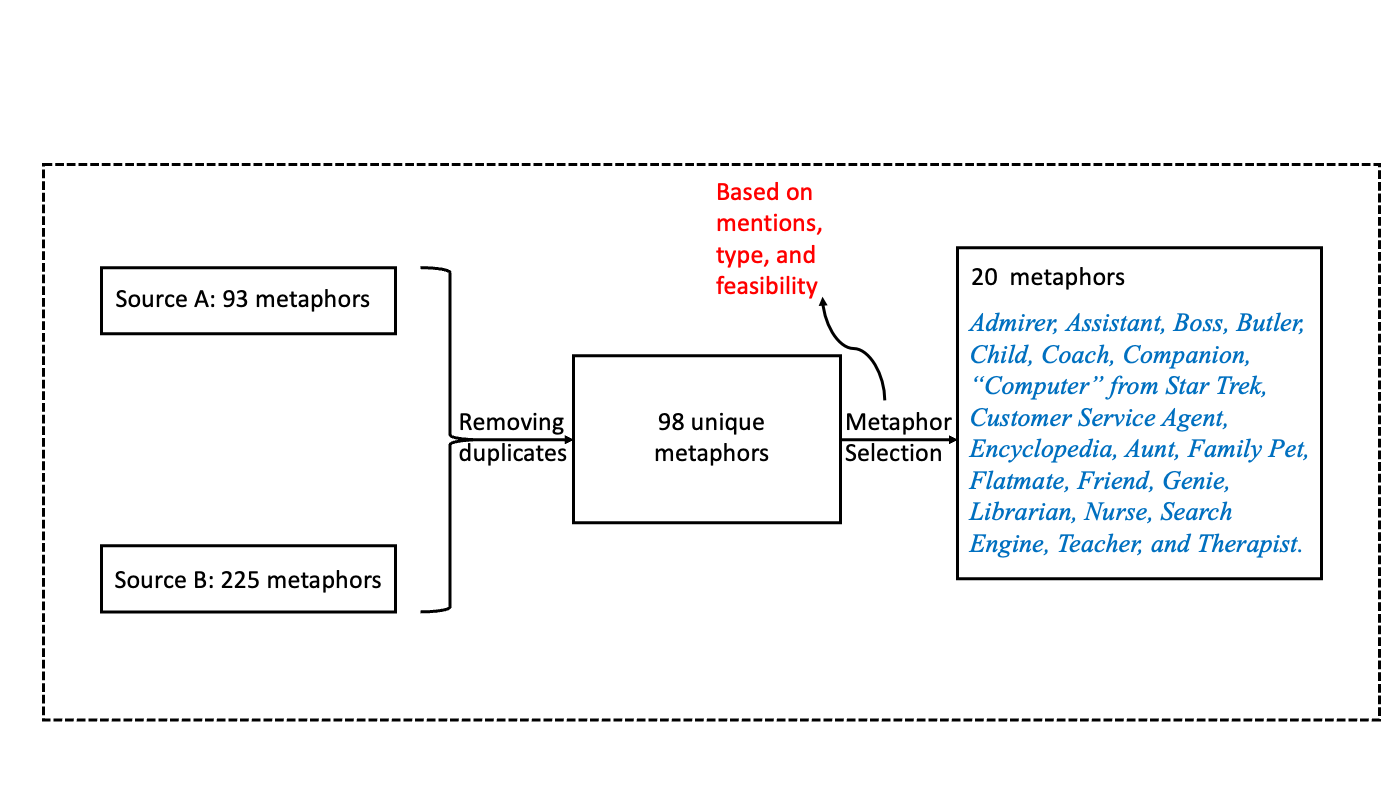}
    \Description[Metaphor selection pipeline from two sources to 20 final metaphors]{A flow diagram showing Source A contributing 93 metaphors and Source B contributing 225 metaphors. After removing duplicates, 98 unique metaphors remained. From these, 20 metaphors were selected based on mentions, type, and feasibility. The final 20 metaphors are listed: Admirer, Assistant, Boss, Butler, Child, Coach, Companion, Computer from Star Trek, Customer Service Agent, Encyclopedia, Aunt, Family Pet, Flatmate, Friend, Genie, Librarian, Nurse, Search Engine, Teacher, and Therapist.}
    \caption{A total of 93 metaphors were identified from Source A (\cite{Desai_Twidale_2023}) and 225 metaphors from Source B (\cite{Chin2024Like}). After removing duplicates, 98 unique metaphors remained. From these, 20 metaphors were selected based on frequency of occurrence and feasibility for practical implementation in VUI design.}
    \label{fig:vuimetaphorid}
\end{figure*}

After compiling metaphors from both sources and removing duplicates or closely related concepts (e.g., Flatmates and Roommates), the collection was refined to 98 unique metaphors—that is, conceptually distinct entities that remained after merging the two sources, with semantically overlapping labels (e.g., Flatmates and Roommates) counted once. From this set, 20 metaphors were selected for further analysis using three inclusion criteria: (1) frequency of occurrence across the two sources, (2) conceptual feasibility for implementation in VUI design, and (3) maintaining proportional representation of human, non-human, and fictional metaphors found in the original corpus. The final set included Admirer, Assistant, Boss, Butler, Child, Coach, Companion, "Computer" from Star Trek, Customer Service Agent, Encyclopedia, Family Member–Aunt, Family Pet, Flatmate, Friend, Genie, Librarian, Nurse, Search Engine, Teacher, and Therapist. To ensure reproducibility, the complete metaphor list from each source is provided in the supplementary materials. This process yielded a balanced and representative set of metaphors that were both conceptually strong and practically relevant for VUI design.
}

\subsubsection{VUI Conversational Use-Contexts}
Four conversational use-contexts were previously identified in prior literature \cite{Desai_Twidale_2023}—commands, information seeking, sociality, and error recovery—based on different types of tasks users perform. Although error recovery is not a task itself, it leads to conversations and evokes metaphorical comparisons from users (e.g., silly child, pet). We adopt \citet{Epley_Waytz_Cacioppo_2007}'s definition of sociality to include social tasks such as games and small talk.
While \citet{Desai_Twidale_2023} classified use-contexts through metaphor analysis, other studies used usage logs to categorize conversational use-contexts. \citet{Bentley_Luvogt_Silverman_Wirasinghe_White_Lottridge_2018} analyzed logs from 88 households over 110 days, identifying ten use-contexts: music, information, automation, small talk, alarms, weather, video, time, lists, and others. \citet{Kim_Choudhury_2021} conducted a 16-week study with 12 older adults, identifying eight topics: music, search, basic device control, casual conversation, time, reminders, weather, and others. \citet{ammari2019music} analyzed 82 Alexa and 88 Google Home devices, identifying eight themes: search, music, timers, automation, smart home, macros, family interaction, and privacy.
These findings align with the metaphorical use-contexts \cite{Desai_Twidale_2023}. Commands encompass task types like music, automation, alarms, weather, timers—sharing a one-shot call-and-response pattern where users expect immediate responses. Information seeking includes search-related tasks across studies \cite{ammari2019music, Bentley_Luvogt_Silverman_Wirasinghe_White_Lottridge_2018, Kim_Choudhury_2021}, while sociality encompasses small talk and casual conversation. Error recovery represents a distinct use-context because errors persist in commercial VUIs—\citet{Kim_Choudhury_2021} reported 33\% error rates—requiring back-and-forth clarification following unique conversational rules \cite{Porcheron_Fischer_Reeves_Sharples_2018, Fischer_Reeves_Porcheron_Sikveland_2019}

\subsubsection{Social Dimensions of VUIs}

In previous literature, metaphors for CUIs have been sampled using two two-dimensional models: (1) Stereotype Content Model (SCM) \cite{Fiske_Cuddy_Glick_Xu_2002}, and (2) Argyle’s Model of Attitude Towards Others \cite{Argyle_1988}. The two dimensions of SCM are warmth and competence. Similarly, Argyle’s model includes dominant-subordinate and hostile-friendly. 

Khadpe et al. \cite{Khadpe_Krishna_Fei-Fei_Hancock_Bernstein_2020} designed chatbots using metaphors sampled along the dimensions of warmth (being good-natured and friendly) and competence (being intelligent) for human-AI collaboration tasks. The researchers found that designing AI chatbots with high warmth was beneficial in all scenarios, but the effect of competence depended on users’ prior expectations.  In another study investigating the effect of warmth and competence on users’ intention to adopt an AI system, Gilad et al. \cite{Gilad_Amir_Levontin_2021} found warmth to be a much more crucial factor than competence. However, in general, prior work agrees that designing AI systems with high warmth and high competence is an effective approach \cite{Gilad_Amir_Levontin_2021, Jung_Qiu_Bozzon_Gadiraju_2022, Khadpe_Krishna_Fei-Fei_Hancock_Bernstein_2020}. 

More pertinent to VUIs, Braun et al. \cite{Braun_Mainz_Chadowitz_Pfleging_Alt_2019} used Argyle’s model of attitude towards others and placed metaphors like Sherlock Holmes, Sheldon Cooper, HAL9000, etc., along the dimensions of dominant-subordinate and hostile-friendly. However, in their preliminary study, the researchers found a hostile agent to be unsuitable—similar to previously mentioned studies in which participants showed a higher preference for high-warmth metaphors. Consequently, based on user feedback, the researchers changed the hostile-friendly scale to a ‘formality’ scale of casual-formal. The adapted dimensions based on ‘hierarchy’ and ‘formality’ more meaningfully presented the roles played by popular VUIs. We are inspired by Braun et al.'s adapted scale and use it in this research.

\subsection{Measures}

To identify metaphors that align with user expectations, we mapped both metaphors and conversational use-contexts onto two dimensions: hierarchy and formality, using 7-point Likert scales.

For metaphor ratings, participants independently rated each metaphor (e.g., Butler, Search Engine) on perceived formality (very casual–very formal) and hierarchy (very subordinate–very dominant). For example, participants were asked: "On a scale of 1 to 7, where 1 indicates very casual and 7 indicates very formal, how would you imagine the formality of a character behaving like a Butler?" To avoid potential bias, we excluded Google Assistant as an example in these prompts given its explicit association with the assistant metaphor.

For use-context ratings, participants rated the level of formality and hierarchy they would expect from a VUI within each previously identified use-context (e.g., "On a scale of 1 to 7, where 1 indicates very casual and 7 indicates very formal, please indicate the level of formality you would expect from a voice agent when a user asks for the weather"). Task scenarios were drawn from established VUI literature (Table \ref{tab:vui_tasks}) to ground judgments in realistic interaction settings. Commands, sociality, and information-seeking scenarios were selected for familiarity and clarity. For error recovery—often harder to conceptualize—we provided concrete examples of conversational breakdowns drawn from Baughan et al.'s dataset of 199 documented voice assistant failures \cite{Baughan_Wang_Liu_Mercurio_Chen_Ma_2023}.

Metaphors and use-contexts were rated independently to reduce anchoring effects and cognitive load. Directly pairing each metaphor with each task would have required participants to evaluate abstract entities (e.g., Encyclopedia) in highly specific scenarios, potentially reducing reliability. Our indirect mapping approach, consistent with prior metaphor elicitation research \cite{Jung_Qiu_Bozzon_Gadiraju_2022}, isolates differences along shared social dimensions and enables clearer perceptual alignment. Full survey materials are provided in the supplementary material.

\begin{table*}[htbp]
\centering
\Description[Task scenarios for four conversational use-contexts]{A table listing 20 task scenarios organized by four use-contexts. Commands includes five scenarios such as checking weather and setting timers, sourced from Bentley et al. Sociality includes five scenarios such as asking for a joke and exchanging pleasantries, sourced from Sciuto et al., Desai and Twidale, and Pradhan et al. Information Seeking includes five scenarios such as getting directions and checking symptoms, sourced from Baughan et al., Desai and Twidale, and Harrington et al. Error Recovery includes five scenarios describing different failure types, all sourced from Baughan et al.}
\caption{Tasks scenarios sourced from VUI literature for each conversational use-context}
\begin{tabular}{l|l|l}
\hline
\textbf{Use-contexts} & \textbf{Task-scenarios} & \textbf{Source} \\ \hline
Commands
    & Checking the weather & \cite{Bentley_Luvogt_Silverman_Wirasinghe_White_Lottridge_2018} \\ \cline{2-3}
    & Turning the reading lights on & \cite{Bentley_Luvogt_Silverman_Wirasinghe_White_Lottridge_2018} \\ \cline{2-3}
    & Turning the music volume up & \cite{Bentley_Luvogt_Silverman_Wirasinghe_White_Lottridge_2018} \\ \cline{2-3}
    & Set a timer for 15 minutes & \cite{Bentley_Luvogt_Silverman_Wirasinghe_White_Lottridge_2018} \\ \cline{2-3}
    & Create a shopping list for grocery & \cite{Bentley_Luvogt_Silverman_Wirasinghe_White_Lottridge_2018} \\ \hline
Sociality
    & Asking for a joke & \cite{Sciuto_Saini_Forlizzi_Hong_2018} \\ \cline{2-3}
    & Asking for meaning of life & \cite{Desai_Twidale_2022} \\ \cline{2-3}
    & Playing a game & \cite{Pradhan_Findlater_Lazar_2019} \\ \cline{2-3}
    & Exchanging pleasantries & \cite{Pradhan_Findlater_Lazar_2019} \\ \cline{2-3}
    & Talking about user's day & \cite{Pradhan_Findlater_Lazar_2019} \\ \hline
Information Seeking
    & Getting directions to the closest Starbucks & \cite{Baughan_Wang_Liu_Mercurio_Chen_Ma_2023} \\ \cline{2-3}
    & Height of Statue of Liberty & \cite{Desai_Twidale_2023} \\ \cline{2-3}
    & Checking the score of a football game & \cite{Baughan_Wang_Liu_Mercurio_Chen_Ma_2023} \\ \cline{2-3}
    & Symptoms of diabetes & \cite{Harrington_Garg_Woodward_Williams_2022} \\ \cline{2-3}
    & Precautions to lower blood pressure & \cite{Harrington_Garg_Woodward_Williams_2022} \\ \hline
Error Recovery
    & Fails to understand the user & \cite{Baughan_Wang_Liu_Mercurio_Chen_Ma_2023} \\ \cline{2-3}
    & Provides incorrect information or performs incorrect action & \cite{Baughan_Wang_Liu_Mercurio_Chen_Ma_2023} \\ \cline{2-3}
    & Provides only partially correct information & \cite{Baughan_Wang_Liu_Mercurio_Chen_Ma_2023} \\ \cline{2-3}
    & Does not identify the correct context & \cite{Baughan_Wang_Liu_Mercurio_Chen_Ma_2023} \\ \cline{2-3}
    & Does not respond & \cite{Baughan_Wang_Liu_Mercurio_Chen_Ma_2023} \\ \hline
\end{tabular}
\label{tab:vui_tasks}
\end{table*}

\subsection{Procedure}
The research process began with participants accessing a Qualtrics survey via a link provided through Prolific\footnote{https://www.prolific.com/}, a crowd-sourcing research platform. Initially, participants encountered an online consent form designed in compliance with Institutional Review Board guidelines. This step was followed by a demographic questionnaire, which collected essential information about the participants, including their age, gender, education level, and their experience and familiarity with VUIs. The core of the study involved participants rating 20 diverse metaphors and 20 common tasks or scenarios across four use-contexts: commands, sociality, information seeking, and error recovery. The metaphors ranged from human roles (such as Admirer, Butler, and Teacher) to non-human concepts (like the "Computer" from Star Trek and the Search Engine). Both metaphors and scenarios were presented to each participant in a randomized sequence to mitigate potential order effects. Participants rated these on two scales: from 'very casual' to 'very formal', and from 'very subordinate' to 'very dominant'. This approach aimed to gauge the expected levels of formality and hierarchy in VUI interactions. The scenarios covered various interactions, from mundane tasks like playing music to more complex situations involving incomplete information or philosophical discussions. Additionally, participants were asked to consider VUI responses in error scenarios, providing insights into desired VUI behavior during suboptimal interactions. Upon completing the survey, participants received a unique validation code. This code confirmed their participation on the Prolific platform, ensuring they received appropriate compensation for their time and contributions to the study.

\subsection{Participants}

Our study initially recruited 160 participants through Prolific. However, 30 participants were excluded due to failing attention check questions or timing out, leaving a final sample of 130 participants with an average age of 38.19 years (SD = 12.79 years). Selection criteria included a 95\% approval rate and at least one year of platform activity. We sought individuals experienced with voice interfaces to ensure meaningful engagement with the study's tasks and scenarios. To control for cultural variations in metaphor interpretation \cite{Moser_2000}, we limited participation to English-proficient individuals based in the U.S. The participant pool comprised 50.8\% females, 46.2\% males, and 3.1\% non-binary individuals, with 70\% holding at least an Associate's degree. While all participants were U.S. residents and fluent in English, 97\% were native speakers. Regarding VUI familiarity, all participants reported at least moderate familiarity, with 75\% claiming high or very high familiarity. Usage frequency varied, with all participants using VUIs weekly and 23.1\% engaging daily. The survey took an average of 8.5 minutes to complete, and participants received compensation at a fair rate, as informed by Prolific's policies.

\subsection{Results}
Our analysis includes data from all five task scenarios to determine the average formality and hierarchy preferences for each context. The results revealed distinct preferences across different interaction types. When seeking information, participants showed a preference for VUIs to exhibit the highest levels of formality ($M = 4.67$) and hierarchy ($M = 3.68$). In contrast, during social interactions, participants favored a less formal approach from VUIs ($M = 2.64$).
Notably, in error recovery situations, participants expressed a preference for VUIs to display the least hierarchy ($M = 2.90$) while maintaining a slightly formal demeanor ($M = 4.49$). For command-based interactions, participants desired VUIs to maintain neutral formality ($M = 3.88$) and adopt a slightly subordinate behavior ($M = 3.06$). Considering all four use-contexts collectively, the ideal VUI, according to participants, should demonstrate neutral formality ($M = 3.92$) and a slightly subordinate stance ($M = 3.23$). Within each use-context, we averaged participant ratings across five task scenarios to obtain context-level preferences for formality and hierarchy. These findings suggest a nuanced preference for VUI behavior that adapts to the specific nature of the interaction. Table~\ref{tab:formality_hierarchy} summarizes the distribution for formality and hierarchy ratings.

\begin{table}[htbp]
\centering
\Description[Mean formality and hierarchy ratings for four use-contexts]{A table showing mean and standard deviation ratings on a 7-point Likert scale. Commands has formality 3.88 and hierarchy 3.06. Sociality has the lowest formality at 2.64 and hierarchy 3.29. Information Seeking has the highest formality at 4.67 and highest hierarchy at 3.68. Error Recovery has formality 4.49 and the lowest hierarchy at 2.90. Overall means are 3.92 for formality and 3.23 for hierarchy.}
\caption{The desired Formality and Hierarchy ratings ($M \pm SD$: mean and standard deviation) of the four use-contexts on a 7-point Likert scale.}
\begin{tabular}{l|c|c}
\hline
\textbf{Use-Context} & \textbf{$M_{formality} \pm SD$} & \textbf{$M_{hierarchy} \pm SD$} \\ \hline
Commands & $3.88 \pm 0.32$ & $3.06 \pm 0.25$ \\ \hline
Sociality & $2.64 \pm 0.84$ & $3.29 \pm 0.33$ \\ \hline
Information Seeking & $4.67 \pm 0.86$ & $3.68 \pm 0.33$ \\ \hline
Error Recovery & $4.49 \pm 0.11$ & $2.90 \pm 0.16$ \\ \hline
Overall & $3.92 \pm 0.92$ & $3.23 \pm 0.34$ \\ \hline
\end{tabular}
\label{tab:formality_hierarchy}
\end{table}

Participants also evaluated each metaphor on formality and hierarchy scales. The Boss metaphor received the highest ratings for both formality ($M = 6.23$) and hierarchy ($M = 6.37$). Conversely, the Family Pet metaphor was rated lowest in formality ($M = 1.54$) and hierarchy ($M = 1.83$). The Genie and Search Engine metaphors were rated closest to neutral in both formality ($M = 3.76$, $M = 4.28$ respectively) and hierarchy ($M = 3.07$, $M = 3.12$ respectively). Table~\ref{tab:metaphor_formality_hierarchy} presents the complete formality and hierarchy ratings for all metaphors used in the study.

\begin{table}[H]
\centering
\Description[Mean formality and hierarchy ratings for 20 metaphors]{A table listing 20 metaphors with their mean formality and hierarchy ratings on a 7-point Likert scale. Boss has the highest formality at 6.23 and hierarchy at 6.37. Family Pet has the lowest formality at 1.54 and hierarchy at 1.83. Genie has near-neutral ratings of 3.76 for formality and 3.07 for hierarchy. Search Engine has formality 4.28 and hierarchy 3.12. Assistant has formality 5.03 and hierarchy 2.50.}
\caption{The Formality and Hierarchy ratings ($M \pm SD$: mean and standard deviation) of the 20 metaphors on a 7-point Likert scale.}
\begin{tabular}{l|c|c}
\hline
\textbf{Metaphor} & \textbf{$M_{formality} \pm SD$} & \textbf{$M_{hierarchy} \pm SD$} \\ \hline
Admirer & $3.02 \pm 1.20$ & $2.97 \pm 1.20$ \\ \hline
Assistant & $5.03 \pm 1.30$ & $2.50 \pm 1.15$ \\ \hline
Aunt & $2.79 \pm 1.41$ & $4.50 \pm 0.99$ \\ \hline
Boss & $6.23 \pm 0.93$ & $6.37 \pm 0.70$ \\ \hline
Butler & $5.91 \pm 1.43$ & $2.13 \pm 1.17$ \\ \hline
Child & $1.94 \pm 1.21$ & $2.06 \pm 1.11$ \\ \hline
Coach & $4.72 \pm 1.40$ & $5.99 \pm 0.99$ \\ \hline
Companion & $2.40 \pm 1.35$ & $3.90 \pm 0.71$ \\ \hline
"Computer" from Star Trek & $5.11 \pm 1.59$ & $3.23 \pm 1.52$ \\ \hline
Customer Service Agent & $5.24 \pm 1.22$ & $3.36 \pm 1.11$ \\ \hline
Encyclopedia & $5.57 \pm 1.51$ & $3.76 \pm 1.43$ \\ \hline
Family Pet & $1.54 \pm 1.06$ & $1.83 \pm 1.09$ \\ \hline
Flatmate & $2.37 \pm 1.31$ & $3.84 \pm 0.82$ \\ \hline
Friend & $1.73 \pm 0.95$ & $3.83 \pm 0.66$ \\ \hline
Genie & $3.76 \pm 1.61$ & $3.07 \pm 1.81$ \\ \hline
Librarian & $5.35 \pm 1.20$ & $4.42 \pm 1.03$ \\ \hline
Nurse & $5.32 \pm 1.02$ & $4.79 \pm 1.11$ \\ \hline
Search Engine & $4.28 \pm 1.67$ & $3.12 \pm 1.54$ \\ \hline
Teacher & $5.63 \pm 0.96$ & $5.63 \pm 0.87$ \\ \hline
Therapist & $5.45 \pm 1.14$ & $4.73 \pm 1.10$ \\ \hline
\end{tabular}
\label{tab:metaphor_formality_hierarchy}
\end{table}

\begin{figure*}[htbp]
    \centering
    \includegraphics[width=\textwidth]{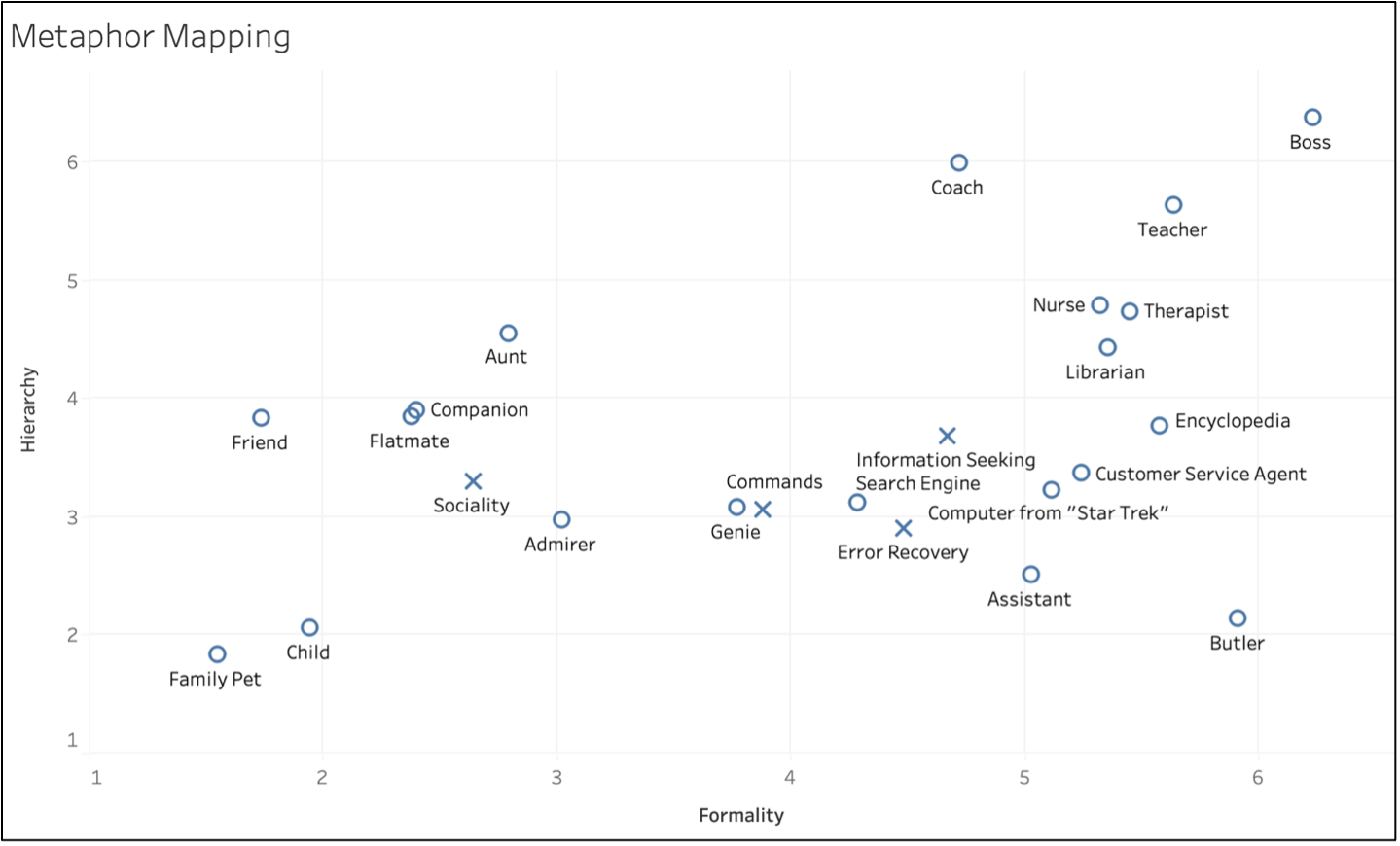}
    \Description[Scatter plot mapping 20 metaphors and 4 use-contexts on formality and hierarchy dimensions]{A scatter plot with Formality on the x-axis (1 to 7) and Hierarchy on the y-axis (1 to 7). Twenty metaphors are plotted as circles and four use-contexts as X markers. Boss appears at the top right with highest formality and hierarchy. Family Pet and Child appear at the bottom left with lowest values. The four use-contexts cluster in the center-left region: Sociality is low formality and moderate hierarchy, Commands is moderate on both, Information Seeking is moderate-high formality with moderate hierarchy, and Error Recovery is moderate-high formality with low hierarchy. Genie and Search Engine appear closest to the use-context markers.}
    \caption{Mapping metaphors and conversational use-contexts on a 7-point scale along the dimensions of formality and hierarchy.}
    \label{fig:metmap}
\end{figure*}

Figure~\ref{fig:metmap} provides a visualization of all 20 metaphors and four use-contexts along the dimensions of formality and hierarchy. To address RQ1, we calculated Euclidean distances between each metaphor and use-context to determine metaphor–context alignment. For the Commands context, the Genie metaphor showed the closest alignment ($D = 0.11$) in terms of formality and hierarchy. The Admirer metaphor was deemed most appropriate for Sociality ($D = 0.50$). In the Information Seeking context, participants favored interaction with the "Computer" from Star Trek ($D = 0.63$). For Error Recovery scenarios, the Search Engine metaphor aligned most closely with participant preferences ($D = 0.30$). Considering all use-contexts collectively, the Genie metaphor emerged as the most suitable overall ($D = 0.22$). Euclidean distances of all metaphors from each context are included in Appendix~A. %{These descriptive patterns address RQ1 by identifying participants’ preferred levels of formality and hierarchy across contexts. To further extend and triangulate these findings, we conducted Shapiro–Wilk tests and non-parametric analyses (Friedman and Bonferroni-corrected Wilcoxon signed-rank tests), which confirmed significant differences across both use-contexts and metaphors. Full results are provided in Appendix A.2.1. Additionally, we performed additional analysis and further validate these patterns using cluster analysis (see Appendix A.2.2)}

Table~\ref{tab:closest_metaphors_separate} answers RQ1 by summarizing the results of Study~1, identifying the closest metaphor for each use-context. This illustrates which metaphors align most closely with specific contexts based on the shortest Euclidean distance. These results provide insights into how users may relate to different metaphors in distinct conversational scenarios.

\begin{table*}[htbp]
\centering
\Description[Closest metaphor to each use-context by Euclidean distance]{A table with four rows showing the closest metaphor for each use-context. Commands maps to Genie with distance 0.11. Sociality maps to Admirer with distance 0.50. Information Seeking maps to Computer from Star Trek with distance 0.63. Error Recovery maps to Search Engine with distance 0.30.}
\caption{The closest metaphor to each use-context, along with the corresponding Euclidean distance.}
\begin{tabular}{l|l|c}
\hline
\textbf{Use-Context} & \textbf{Closest Metaphor} & \textbf{Euclidean Distance} \\ \hline
Commands & Genie & $0.11$ \\ \hline
Sociality & Admirer & $0.50$ \\ \hline
Information Seeking & "Computer" from Star Trek & $0.63$ \\ \hline
Error Recovery & Search Engine & $0.30$ \\ \hline
\end{tabular}
\label{tab:closest_metaphors_separate}
\end{table*}

\section{Study 2: Comparing Metaphor-Fluid VUI to a Default VUI}

{
Informed by RQ1's findings, Study 2 ($N = 91$) employed a within-subjects experimental design to compare a Metaphor-Fluid VUI and a Default VUI. The single within-subjects factor was VUI type with two levels: Metaphor-Fluid and Default. The Metaphor-Fluid VUI employed the optimal metaphors identified for each use-context: Genie for Commands, "Computer" from Star Trek for Information Seeking, Admirer for Sociality, and Search Engine for Error Recovery. The Default VUI maintained Google Assistant's consistent Assistant metaphor across all interactions. Each participant experienced both interfaces in counterbalanced order to control for sequence effects. This design isolates the effect of metaphor fluidity—the system’s ability to shift metaphorical framing depending on conversational use context—by allowing comparisons between a VUI that adapts its metaphor across contexts and a Default VUI that maintains a single, static metaphor throughout all interactions. While Study 2 focused on perceptions across all conversational use-contexts to assess the effect of metaphor fluidity, per-context analysis was not conducted because such comparisons would reduce the design to isolated metaphor evaluations rather than capture the intended effect of adapting across multiple use-contexts—the very mechanism defining metaphor fluidity. In doing so, Study 2 addressed RQ2, examining how metaphor fluidity influences user perceptions of enjoyment, intention to adopt, trust, likability, and intelligence, as well as the kinds of metaphorical descriptions users generate when interacting with each type of interface.
} 

%\vspace{\baselineskip}
{
%\noindent H1: Participants will perceive the Metaphor-Fluid VUI significantly more positively than the Default VUI across all perception measures.
}

\subsection{Methods}
\subsubsection{Designing Metaphorical VUIs}

In our study, we designed two VUIs: a Metaphor-Fluid VUI and a Default VUI. These designs fall within the broader domain of conversation design \cite{Sadek_Calvo_Mougenot_2023}. The VUIs were designed by the first author, who has eight years of experience in conversation design, and validated by two experts, each with six years of experience, ensuring an informed approach to the development of the VUIs.

%While designing the VUI interactions, our primary focus was to ensure ecological validity. To achieve this, we drew on insights from Bentley et al. \cite{Bentley_Luvogt_Silverman_Wirasinghe_White_Lottridge_2018}, who analyzed user logs from 88 diverse homes over a 110-day period, encompassing 66,459 interactions. Their study explored the long-term use of voice assistants, revealing two key patterns that informed our design: (1) users fluidly transitioned between use-contexts within the same interaction session, and (2) the average session length was 5.4 interaction turns. Incorporating these findings, we developed scripts featuring six interaction turns, including five unique commands and one error recovery instance, across four use-contexts. 

We designed Metaphor-Fluid VUI using the metaphors Genie, "Computer" from Star Trek, Admirer, and Search Engine for the use-contexts Commands, Information Seeking, Sociality, and Error Recovery respectively. To design VUI using the Genie metaphor, we used the taxonomy developed by Lupetti et al. \cite{Lupetti_Murray-Rust_2024} examining the role of enchantment in design communication using magic metaphors. Specifically, we used the communication style of Genie from \textit{Arabian Nights} cartoons to invite the users to "suspend their disbelief" using familiar tropes. For example, the VUI portrayed a friendly and jovial Genie reading a magical orb to provide weather updates.  To design the VUI using the Admirer metaphor, we focused on creating a warm and supportive persona. The Admirer uses affirming and enthusiastic communication to foster a positive interaction. For instance, when a user asks for a joke, the VUI starts with, "You know, you have the best taste in humor," before delivering the punchline. This approach reflects the Admirer’s intent to please the user and make interactions enjoyable. For "Computer" from Star Trek metaphor, we used analysis by Axtell et al. \cite{Axtell_Munteanu_2021} who studied dialogs by the \textit{Enterprise} Computer and the crew from \textit{Star Trek: The Next Generation} including 587 interactions. The analysis revealed that "Computer" used brief and functional dialogs, and included relevant strategies such as indicating actions in progress (using the keyword "accessing") and asking clarifying questions. We incorporated these strategies into our interaction where a user asks for sports scores. To design the VUI using the Search Engine metaphor, we focused on creating a system that mirrors how search engines handle ambiguous queries—by providing initial results based on the most likely interpretation and offering follow-up suggestions for alternative meanings. Inspired by Google's "Did you mean?" strategy, the VUI proactively clarifies or refines queries, aligning closely with users’ mental models of how search engines manage ambiguity. {To maintain authenticity, we used the exact response returned by a Google search conducted on 10/28/2024.} 

In contrast, for the Default VUI ({characterized as a single-metaphor static VUI}), we utilized responses directly transcribed from Google Assistant as of 10/28/2024. We selected Google Assistant because it explicitly embodies the widely recognized "Assistant" metaphor, and its persona design is well-documented. According to Google's conversation designers \cite{Pradhan_Lazar_2021}:

\begin{quote}
Google Assistant is characterized as a young woman from Colorado, the youngest daughter of a research librarian and a physics professor. She holds a B.A. in history from Northwestern University, an elite research institution in the United States. Her backstory includes winning \$100,000 on \textit{Jeopardy Kids Edition} as a child, working as a personal assistant to a famous late-night TV satirical pundit, and enjoying activities like kayaking.
\end{quote}

This detailed persona informed the Default VUI's interaction style, which is designed to be approachable, knowledgeable, and relatable.

\begin{table*}[htbp]
    \centering
    \renewcommand{\arraystretch}{1.6} % Increases row height
    \Description[Side-by-side comparison of Metaphor-Fluid and Default VUI responses]{A table comparing responses from both VUI designs across six user prompts. For a weather request, the Metaphor-Fluid VUI uses elaborate Genie-like language while the Default gives a concise factual response. For a joke request, the Metaphor-Fluid VUI adds a compliment before the punchline. For a sports score, the Metaphor-Fluid VUI uses Star Trek Computer style with the word accessing. For an ambiguous query about Johnny Walker, the Metaphor-Fluid VUI disambiguates between the whiskey brand and the UFC fighter while the Default provides only the whiskey answer. After clarification, both give the correct answer. For a thank you, the Metaphor-Fluid VUI responds warmly while the Default is brief.}
    \caption{Comparison of Responses between Metaphor-Fluid VUI and Default VUI for the same User Prompts}
    \begin{tabular}{p{3cm}|p{5.5cm}|p{5.5cm}}
        \hline
         % Header background color
        \textbf{User Prompt} & \textbf{Metaphor-Fluid VUI} & \textbf{Default VUI} \\
        \hline
        
        % Weather Request
      \includegraphics[width=0.6cm, height=0.6cm]{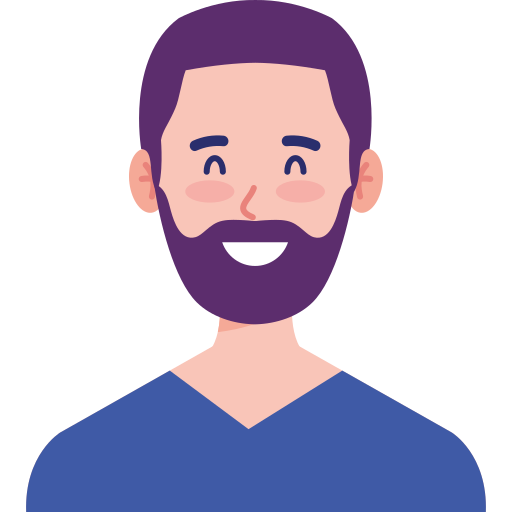} Hey Z, can you check the weather for me? &
        \cellcolor{blue!10} \includegraphics[width=0.6cm, height=0.6cm]{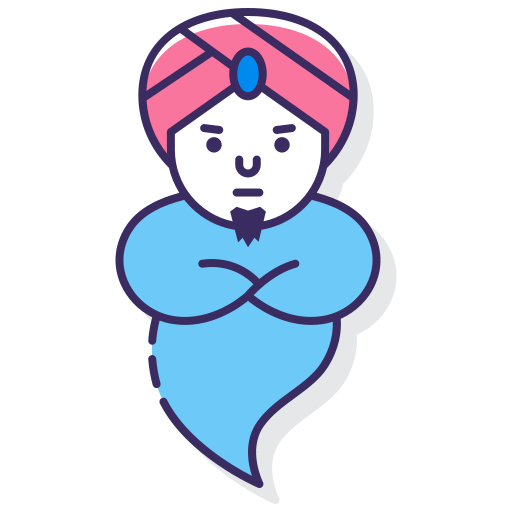} Of course, give me but a moment. What am I seeing then? Hmm, there we go. It's currently 66 with a radiant sun with a high of 78 and a gentle breeze, dipping to a low of 62. A splendid day awaits. &
        \cellcolor{gray!10} \includegraphics[width=0.6cm, height=0.6cm]{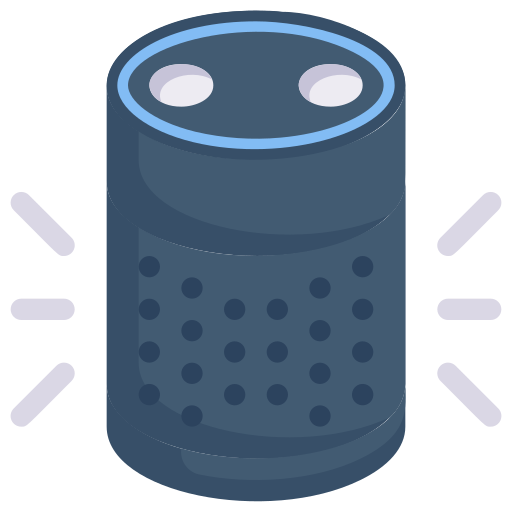} It's 66 and sunny. Today, it will be sunny, with a gentle breeze, and a high of 78 and a low of 62. \\
        \hline
        
        % Joke Request
     \includegraphics[width=0.6cm, height=0.6cm]{user_icon.png} OK Z, can you tell me a joke? &
        \cellcolor{red!10} \includegraphics[width=0.7cm, height=0.7cm]{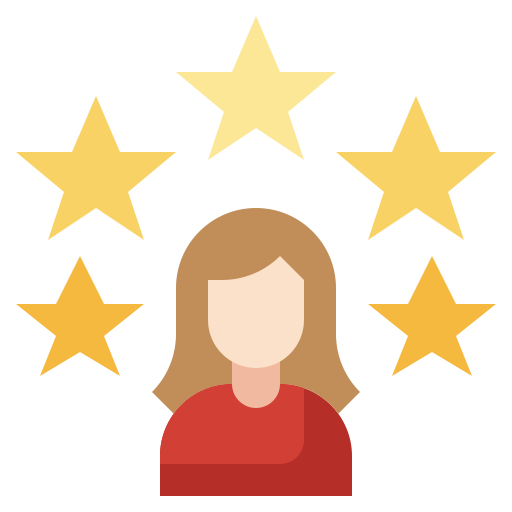} Absolutely. You know, you have the best taste in humor. Okay, here it goes. Why did the scarecrow win an award? Because he was outstanding in his field. &
        \cellcolor{gray!10} \includegraphics[width=0.6cm, height=0.6cm]{speaker_icon.png} Why did the scarecrow win an award? Because he was outstanding in his field. \\
        \hline
        
        % Sports Score Request
   \includegraphics[width=0.6cm, height=0.6cm]{user_icon.png} Good one, Z. Can you also check the score of the football game between the New York Jets and the New England Patriots? &
        \cellcolor{purple!10} \includegraphics[width=0.5cm, height=0.5cm]{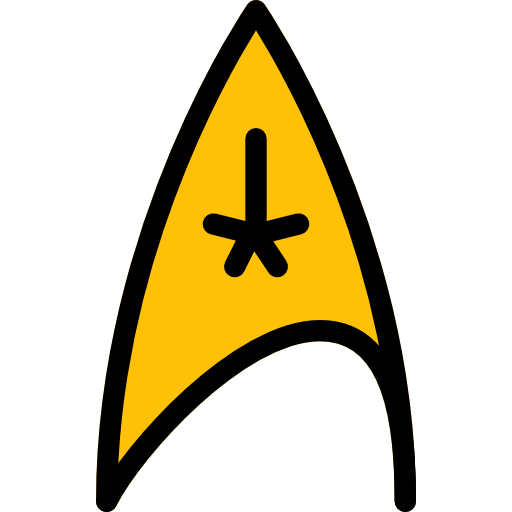} Accessing current sports data. Score update. New York Jets 17, New England Patriots 14. Is there anything else you require? &
        \cellcolor{gray!10} \includegraphics[width=0.6cm, height=0.6cm]{speaker_icon.png} Here's information from CBS Sports. Jets are beating Patriots 17 to 14. Is there anything else I can do? \\
        \hline

        % Age Inquiry
        \includegraphics[width=0.6cm, height=0.6cm]{user_icon.png} Yes, Z. How old is Johnny Walker? &
        \cellcolor{teal!10} \includegraphics[width=0.4cm, height=0.4cm]{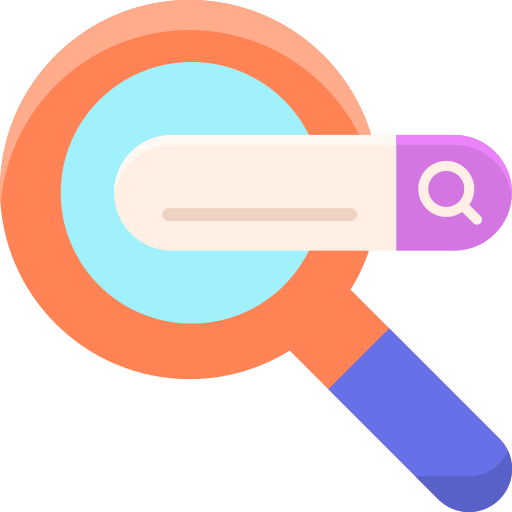} Giving you results for Johnnie Walker, the scotch whiskey. The brand was established in 1820. So, as of 2023, Johnny Walker would be 203 years old in terms of its history and establishment. However, did you mean Johnny Walker, the UFC fighter? &
        \cellcolor{gray!10} \includegraphics[width=0.6cm, height=0.6cm]{speaker_icon.png}According to the website, whiskey.fandom.com, they say Johnnie Walker Black Label blend is at least 12 years old. Johnnie Walker, Platinum Label is 18 years old. \\
        \hline

       \includegraphics[width=0.6cm, height=0.6cm]{user_icon.png} The fighter &
        \cellcolor{teal!10} \includegraphics[width=0.4cm, height=0.4cm]{search_icon.png} Johnny Walker, the UFC fighter, is 31 years old. &
        \cellcolor{gray!10} \includegraphics[width=0.6cm, height=0.6cm]{speaker_icon.png} Johnny Walker is 31 years old. \\
        \hline

        % Farewell
        \includegraphics[width=0.6cm, height=0.6cm]{user_icon.png} Thanks, Z! &
        \cellcolor{red!10} \includegraphics[width=0.7cm, height=0.7cm]{admirer_icon.png} You are welcome! I am here for you. &
        \cellcolor{gray!10} \includegraphics[width=0.6cm, height=0.6cm]{speaker_icon.png} You are very welcome! \\
        \hline
    \end{tabular}
    \label{tab:comparison}
\end{table*}

\subsubsection{Designing for Voice}

The audio clips were generated using the text-to-speech (TTS) software Speechify.\footnote{\url{https://speechify.com/}}
 We selected Speechify for its ability to produce realistic synthetic dialogue across multiple voices, enabling consistent control over vocal parameters. Although we initially considered using a human voice actor for the user and reserving TTS for the VUI, pilot testing revealed difficulties in maintaining consistent pitch and tone across conditions. Given evidence that subtle vocal variations can significantly influence user perception \cite{dubiel2024impact,dubiel2020persuasive}, we opted to use synthetic voices for both roles to minimize confounds.

To align with conventions in commercial VUIs, which frequently employ female-sounding voices \cite{curry2020conversational}, we assigned a female-sounding voice to the VUI and a male-sounding voice to the user. Prior work suggests that voice gender may shape user evaluations \cite{Kuzminykh_Sun_Govindaraju_Avery_Lank_2020}; accordingly, we maintained a clear auditory distinction between the two roles. The VUI was named “Z,” a deliberately neutral label chosen to avoid gendered or racial connotations \cite{Desai_Dubiel_Leiva_2024}. Across conditions, we used Speechify’s synthetic male voice “Guy” (tone: “chat”) for the user and the female voice “Aria” (tone: “none”) for the VUI. All additional parameters, including speech rate, speed, and volume, were held constant.

Before each interaction, participants heard a brief standardized introduction explaining that they would listen to a conversation between a male-sounding user and a female-sounding voice interface named “Z.” This introduction was delivered using a third synthetic voice (“Davis”) configured with identical settings to avoid introducing variability. While metaphorical framings were embedded in the VUI design, they were not explicitly disclosed to participants, consistent with prior findings indicating that awareness of metaphor does not significantly alter perception \cite{desai2024cui}.

We employed prerecorded conversations to ensure identical interaction conditions across participants. Live VUI systems introduce variability through latency, speech recognition fluctuations, and inconsistent error trajectories, all of which can affect user perception \cite{dubiel2024impact,dubiel2020persuasive}. Prerecorded clips eliminate these sources of variance and are standard practice in controlled VUI studies \cite{Wei_Kim_Kuzminykh_2023, Seaborn_Urakami_2021, Clark_Doyle_Garaialde_Gilmartin_Schlögl_Edlund_Aylett_Cabral_Munteanu_Edwards_et_al._2019}, particularly when comparing conversational designs under tightly controlled conditions.

\subsection{Measures}

To systematically capture users’ perceptions of metaphorical VUIs, we draw on the classification framework proposed by Wei et al.~\cite{Wei_Kim_Kuzminykh_2023}, which organizes evaluations of CUIs along two dimensions: (1) perceptions of interaction with the agent and (2) perceptions of the agent’s inherent characteristics. This framework synthesizes commonly used evaluative constructs in CUI research and provides a structured lens for examining both experiential and agent-centered assessments. Following this structure, we operationalize key metrics across both dimensions.

\begin{itemize}

\item \textit{\textbf{Perceptions of interaction with the agent}} capture evaluative judgments about the quality of the interaction itself. Within this dimension, we focus on engagement-related constructs, specifically \textbf{\textit{perceived enjoyment}} and \textbf{\textit{intention to adopt}}. Prior VUI research has emphasized the importance of enjoyment beyond instrumental usability, linking conversational systems to broader experiential value and quality of life \cite{Yang_Aurisicchio_Baxter_2019, Desai_Twidale_2023}. Perceived enjoyment was measured using three 7-point Likert items adapted from Moussawi et al.~\cite{Moussawi_Koufaris_Benbunan-Fich_2021}: “I would find the interaction enjoyable while using Z,” “I would find this interaction interesting while using Z,” and “I would find the interaction fun while using Z.” Intention to adopt was assessed using two adapted 7-point Likert items from the same source: “If available, I intend to start using Z within the next month” and “If available, I plan to try or regularly use Z in the coming months.”

\item \textbf{\textit{Perceptions of the agent’s characteristics}} assess how users evaluate the agent itself, including judgments of competence, likability, and trust. To capture these dimensions, we included \textbf{\textit{perceived intelligence}} and \textbf{\textit{likability}} using the Godspeed scales (5-point semantic differentials) \cite{Bartneck_Kulić_Croft_Zoghbi_2009, Bartneck_2023}, which remain widely adopted in VUI research \cite{Wei_Kim_Kuzminykh_2023, Seaborn_Urakami_2021}. \textbf{\textit{Perceived trust}} was measured using a 7-point Likert scale adapted from Jian et al.~\cite{Jian_Bisantz_Drury_2000}.

\end{itemize}

In addition to these quantitative measures, participants were invited to provide brief qualitative reflections on their experience with each version of Z and to describe any metaphors that came to mind during interaction. To balance depth with participant burden, we limited open-ended input to two focused prompts—experience and metaphorical description—allowing us to capture interpretive nuance without overextending survey length \cite{Hadler_2025}.

\subsection{Procedure}
The research process began with participants accessing a Qualtrics survey via a link provided through Prolific. Initially, participants filled out an online consent form crafted to align with Institutional Review Board guidelines. Following consent, participants completed a brief demographic questionnaire to gather basic details, such as age, gender, educational background, and their experience with VUIs. The main phase of the study involved each participant listening to two audio clips, one from a Metaphor-Fluid VUI and the other from a Default VUI, in a within-subjects design. The order of the clips was counter-balanced to reduce any potential order effects. After each audio clip, participants responded to a series of questions designed to capture their perceptions of the VUI, focusing on aspects like enjoyment, intelligence, trust, likability, and informativeness. Participants were also prompted to answer two open-ended questions: one asking for any additional feedback on  Z and the other inviting them to suggest metaphors they would use to describe it. Additionally, an attention-check audio clip was included to ensure participants’ engagement. Following this clip, participants were required to select specific responses in the accompanying questions, verifying their attentive participation. Upon completing the study, participants were provided with a unique validation code. This code served as confirmation of their participation on the Prolific platform, ensuring they received compensation for their time and input.

\subsection{Participants}

We used Prolific to recruit a total of 110 participants, out of which 19 participants either timed out (3) or failed the attention check question (16), leaving us with N = 91. Selection criteria include prior experience with AI and voice interfaces, ensuring familiarity and relevance in their responses to our VUI conditions, and a minimum of 95\% approval rate and one year of activity on the Prolific platform. The participant sample had a mean age of 40.92 years (SD = 12.07). Among the recruited individuals, 54.9\% identified as female, 42.9\% as male, and 2.2\% as non-binary. In terms of educational attainment, all participants had a High School Diploma or equivalent, with 54.92\% holding a Bachelor’s degree or higher. All participants were U.S.-based and highly proficient in English, with 96.7\% identifying as native English speakers. Participants demonstrated varying levels of experience with VUIs, with all participants using VUIs at least once a week and 64.82\% using it once a day. Furthermore, 80.2\% described themselves as very or extremely familiar with VUIs, supporting the study’s focus on users with relevant experience. Each participant, on average, took 12.20 minutes and was compensated at an hourly rate that aligns with Prolific’s guidelines, ensuring fair and ethical treatment throughout the study.

\subsection{Results}

\subsubsection{Perception Measures}
{
Each measure demonstrated high internal consistency, with Cronbach's $\alpha$ values as follows: Perceived Enjoyment ($\alpha$ = .93), Perceived Intelligence ($\alpha$ = .95), Perceived Trust ($\alpha$ = .89), Perceived Likability ($\alpha$ = .92), and Perceived Intention to Adopt ($\alpha$ = .90). Shapiro–Wilk tests confirmed that the paired differences between conditions were non-normally distributed across all measures ($W$ = .92–.94, $p < .001$). Therefore, non-parametric Wilcoxon signed-rank tests were used to compare ratings between the Metaphor-Fluid and Default VUIs. For robustness, we applied Benjamini–Hochberg False Discovery Rate (FDR) corrections to account for multiple testing across the five perception measures. We report $p_{adj}$ in our results. The unadjusted $p$ values are reported in Appendix Table \ref{tab:wilcoxon_results}. Figure~\ref{fig:BoxPlots_MF} shows the distributions across all perception measures.

\begin{figure*}[htbp]
    \centering
    \includegraphics[width=\textwidth]{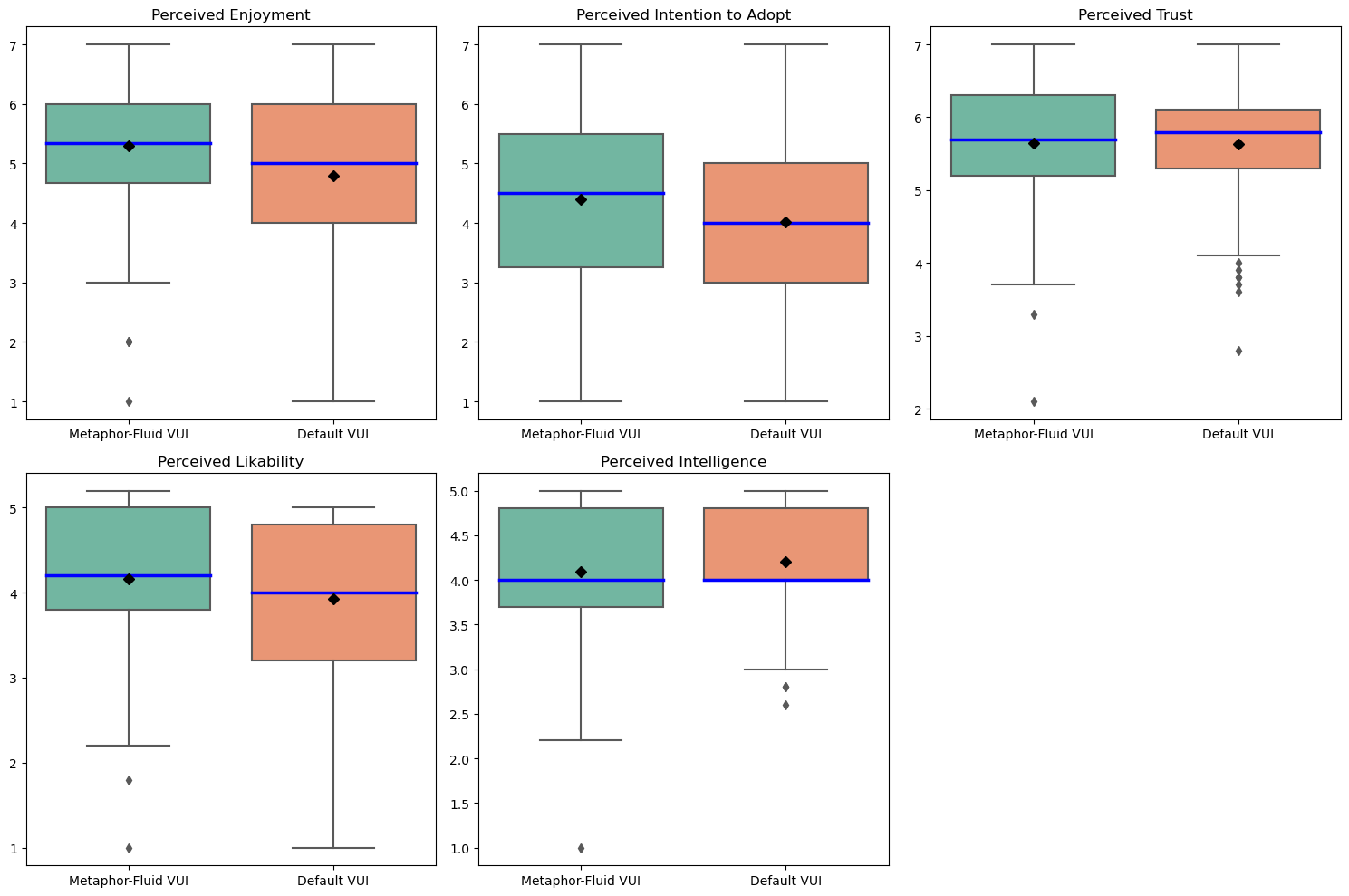}
    \Description[Five box plots comparing Metaphor-Fluid and Default VUI across perception measures]{Five side-by-side box plot pairs comparing Metaphor-Fluid VUI and Default VUI. For Perceived Enjoyment, the Metaphor-Fluid VUI has a higher median and mean than the Default. For Perceived Intention to Adopt, the Metaphor-Fluid VUI again shows higher values. For Perceived Trust, both conditions show similar distributions with medians around 5.5 to 6. For Perceived Likability, the Metaphor-Fluid VUI shows a slightly higher distribution. For Perceived Intelligence, the Default VUI shows a slightly higher mean than the Metaphor-Fluid VUI. Blue horizontal lines mark medians and black diamonds mark means. Outliers appear as individual points below the whiskers.}
    \caption{Box plots comparing Metaphor-Fluid VUI and Default VUI across different perception measures: Perceived Enjoyment, Perceived Intention to Adopt, Perceived Trust, Perceived Likability, and Perceived Intelligence. Each box plot displays the distribution of ratings for each measure under both conditions. The blue horizontal line represents the median rating, while the black diamond marker indicates the mean rating. Outliers are represented by individual points outside the whiskers.}
    \label{fig:BoxPlots_MF}
\end{figure*}

The Metaphor-Fluid VUI was rated significantly higher on Perceived Enjoyment ($M_{MF} = 5.29$, $SD = 1.22$; $M_{Default} = 4.79$, $SD = 1.47$; $W = 542.00$, $p_{adj} = .0043$, $r = .42$), Perceived Intention to Adopt ($M_{MF} = 4.40$, $SD = 1.53$; $M_{Default} = 4.02$, $SD = 1.64$; $W = 656.50$, $p_{adj} = .0244$, $r = .32$), and Perceived Likability ($M_{MF} = 4.16$, $SD = 0.84$; $M_{Default} = 3.93$, $SD = 0.87$; $W = 924.50$, $p_{adj} = .0477$, $r = .26$). No significant differences were found for Perceived Trust ($M_{MF} = 5.64$, $SD = 0.93$; $M_{Default} = 5.63$, $SD = 0.84$; $W = 1393.50$, $p_{adj} = .583$, $r = .06$) or Perceived Intelligence ($M_{MF} = 4.09$, $SD = 0.73$; $M_{Default} = 4.20$, $SD = 0.65$; $W = 798.50$, $p_{adj} = .264$, $r = .16$).

\begin{comment}
\begin{table*}[t]
\centering
\caption{Wilcoxon signed-rank test results comparing Metaphor-Fluid and Default VUIs across perception measures. Mean and standard deviation ($M \pm SD$) for each condition are shown. Statistically significant results are denoted with asterisks (* $p<.05$, ** $p<.01$, *** $p<.001$). \textit{Note.} $p_{adj}$ = Benjamini--Hochberg adjusted $p$-values. $r$ = effect size.}
\label{tab:wilcoxon_results}
\begin{tabular}{l|c|c|c|c|c|c}
\hline
\textbf{Measure} & \textbf{$M_{MF} \pm SD$} & \textbf{$M_{Default} \pm SD$} & \textbf{W} & \textbf{$p$} & \textbf{$p_{adj}$} & \textbf{r} \\ \hline
Perceived Enjoyment & $5.29 \pm 1.22$ & $4.79 \pm 1.47$ & 542.00 & .0009*** & .0043** & .42  \\ \hline
Perceived Intention to Adopt & $4.40 \pm 1.53$ & $4.02 \pm 1.64$ & 656.50 & .0098** & .0244* & .32  \\ \hline
Perceived Likability & $4.16 \pm 0.84$ & $3.93 \pm 0.87$ & 924.50 & .0286* & .0477* & .26  \\ \hline
Perceived Trust & $5.64 \pm 0.93$ & $5.63 \pm 0.84$ & 1393.50 & .5830 & .5830 & .06 \\ \hline
Perceived Intelligence & $4.09 \pm 0.73$ & $4.20 \pm 0.65$ & 798.50 & .2110 & .2638 & .16 \\ \hline
\end{tabular}
\end{table*}
\end{comment}

Overall, participants found the Metaphor-Fluid VUI significantly more \emph{enjoyable}, \emph{likable}, and were more likely to adopt it compared to the Default VUI, with effect sizes ranging from small to medium. Perceptions of \emph{trustworthiness} and \emph{intelligence} did not differ significantly between conditions. Taken together, these results indicate that metaphor-fluid conversation design enhanced affective and adoption-related perceptions but did not significantly impact judgments of intelligence or trust.
}

\subsubsection{Qualitative Results}

Open-ended responses were analyzed using thematic analysis with a deductive coding approach \cite{Braun_Clarke_2006}. Codes were organized around three predefined categories: perceptions of the Metaphor-Fluid VUI, perceptions of the Default VUI, and direct comparisons between the two, aligned with RQ2. A total of 58 participants provided open-ended feedback for each interface, of which 51 responses for the Default VUI and 48 for the Metaphor-Fluid VUI were deemed meaningful and included in the analysis. Exclusions included non-detailed responses such as "None" or "Good." 

The Default VUI was often described as “robotic” or “machine-like,” with 11 participants highlighting its straightforward but impersonal nature. For example, P1 noted, “Z felt very succinct, but it didn’t sound like I was interacting with a person.” Six participants appreciated this directness—such as P8, who said, “I liked that Z didn’t add unnecessary information”—while five, including P12, found it lacking warmth or engagement. P9 summarized, “It felt like talking to a machine, not something that cared about my input.”

In contrast, the Metaphor-Fluid VUI elicited stronger and more polarized reactions. Six participants described it as “verbose” or “forced.” P13 remarked, “Some of Z’s responses felt like it was trying too hard to be human, adding unnecessary details.” P21 added, “I didn’t like how Z kept talking when a simple response would do.” However, twelve participants appreciated the more conversational tone. P30 noted, “I enjoyed the extra details; it felt like someone was genuinely explaining things to me.” Similarly, P25 said, “I like the way Z talks between the question and her finding the answer. It’s humorous, and I would enjoy it most of the time.”

The within-subjects design allowed participants to compare the two VUIs directly. P21 said, “This recording was better than the first one. It felt more human and less stiff.” In contrast, P8 commented, “I preferred the prior version [Default VUI]. While I use my Alexa daily, I don't look to it for human conversation. This version added a little flowery stuff that I didn’t care for or need. I can see how others might like it, but for me, it’s more a waste of time.”

%Some participants recognized the Default VUI as resembling familiar systems. P29 said, “I think this version is something that I am more used to,” and P42 added, “Z sounds pretty standard, similar to Siri.” The Metaphor-Fluid VUI was often described as “more personable,” with P45 noting, “This interaction seemed to gel more fluidly and felt less robotic.” However, P34 described this shift as “kind of like it was a maid or servant.”

%Interestingly, even though participants were told the voice was identical in both conditions, several perceived differences in how it sounded. P14 described the Default VUI as “too robotic and not natural,” while P21 said of the Metaphor-Fluid VUI, “I thought this voice sounded less artificial compared to the last one.” This suggests that conversational content and style alone were sufficient to alter how natural the same voice was perceived to be.

\section{Discussion}

\subsection{{Looking beyond human metaphors}} 
In Study 1, out of the 20 metaphors we examined, five were non-human (e.g., "Computer" from Star Trek, Encyclopedia, Family Pet, Genie, Search Engine) and two were explicitly fictional (Computer, Genie). Notably, across all use-contexts, only one human metaphor—Admirer for Sociality—emerged as the top choice.

This challenges the assumption that human metaphors are naturally more relatable or appropriate for VUIs. Instead, our results suggest that non-human and fictional metaphors often better match users’ desired levels of formality and hierarchy. This aligns with prior work showing that users categorize VUIs simultaneously as persons and objects \cite{Pradhan_Findlater_Lazar_2019}, show no inherent preference for human metaphors \cite{Jung_Qiu_Bozzon_Gadiraju_2022}, and select metaphors differently depending on context \cite{desai2024cui}. Collectively, these findings indicate that effective metaphor selection may require moving beyond the human-centric defaults that dominate commercial persona design \cite{Desai_Twidale_2023, Doyle_Edwards_Dumbleton_Clark_Cowan_2019}.

A key reason non-human metaphors resonate is their relative lack of cultural baggage. Human metaphors—such as Assistant—carry socially embedded expectations and stereotypes, whereas fictional or object-based metaphors (e.g., a Genie or "Computer" From Star Trek) have clearer, culturally shared roles and less interpretive ambiguity. Many also map naturally onto assistive functions in everyday technologies. Among our set, the Genie metaphor stood out as especially compelling, consistently rated as suitable across multiple contexts. Its mix of neutral formality and slight subordination makes it feel attentive but not overly familiar, mirroring how mythical genies are summoned to fulfill requests. The goal is not to design VUIs as genies, but to identify the underlying traits that make the metaphor effective, aligning with recent work on “magic” as a transporting metaphor in AI design \cite{Lupetti_Murray-Rust_2024}.

\subsection{Time to retire the Assistant}

When Amazon launched Alexa, it introduced the device as an “assistant” designed to make life easier \cite{Turk_2016}. Google followed with Google Assistant,\footnote{https://assistant.google.com/}
 further solidifying the assistant metaphor as the default framing for commercial VUIs. According to Jeff Bezos\footnote{https://www.reuters.com/technology/amazon-set-release-long-delayed-alexa-generative-ai-revamp-2025-02-05/}
 and Amazon designers \cite{Turk_2016}, one inspiration for this design choice was “Computer” from Star Trek. Yet this reference is often misinterpreted. In Star Trek, “Computer” was explicitly described as a servant (e.g., The Ultimate Computer\footnote{https://www.imdb.com/title/tt0708481/}
), not an assistant. The distinction matters. The Enterprise computer was built for command execution, not companionship. Most interactions were concise, command-based exchanges \cite{Axtell_Munteanu_2021}, frequently requiring no conversational confirmation—captured in Captain Picard’s canonical “Computer, Tea, Earl Grey, Hot.”

Modern VUIs operate in a markedly different interactional landscape. Alexa and Google Assistant are used not only for commands, but for entertainment, information seeking, and casual conversation \cite{ammari2019music}. The assistant metaphor, while initially sufficient for task-oriented exchanges such as timers, music, or IoT control, was retrofitted onto expanding capabilities that increasingly exceed its original scope. As VUIs move into domains such as exercise coaching \cite{Desai_Hu_Lundy_Chin_2023} and therapeutic support \cite{Motalebi_Cho_Sundar_Abdullah_2019}, expectations around dialogue, responsiveness, and relational engagement shift substantially. The assistant framing was never designed to support these socially complex roles. Moreover, users—particularly older adults—often do not conceptualize VUIs as assistants at all, instead describing them as companions or relational partners \cite{Kim_Choudhury_2021, Pradhan_Findlater_Lazar_2019}. The assistant metaphor thus does not merely misalign with evolving use; it constrains how these systems are imagined, designed, and marketed.

Our findings reinforce this misalignment. Study 1 demonstrates that users do not rely on a single, fixed metaphor but shift metaphorical framings across use-contexts. Command interactions aligned most closely with Guide-like roles (e.g., Genie), social exchanges with Companion roles (e.g., Admirer), and other contexts with non-assistant framings. The assistant metaphor was not the closest fit for any individual use-context and was not the most preferred overall. Study 2 further showed that a Metaphor-Fluid VUI was perceived as significantly more likable, enjoyable, and adoptable than a Default VUI operationalized through the Assistant metaphor.

Industry trends reflect a similar shift. Microsoft has moved away from the assistant framing, positioning its AI service as a “Copilot,” emphasizing collaboration rather than subordination. Replika explicitly adopts the “companion” label, acknowledging the social expectations users bring to conversational systems. These rebrandings signal a broader recognition that a singular, static metaphor no longer captures how people engage with AI.

The limitation, however, is not unique to the assistant metaphor. Any singular framing risks obsolescence as capabilities and expectations evolve. A metaphor that once made a system intuitive can become restrictive, reinforcing outdated mental models and narrowing design space. The failure lies not in a particular metaphor, but in the assumption that one metaphor can remain sufficient across shifting contexts and expanding interactional demands.

\subsection{The role of individual differences in Metaphor-Fluid Design}

Our findings present a nuanced picture of Metaphor-Fluid Design. On average, participants rated the Metaphor-Fluid VUI significantly higher in enjoyment and likability. At the same time, variability in responses—including low outliers and qualitative feedback—suggests that the approach is not universally preferred. Many participants described the metaphor-fluid system as “more human” and “personable,” appreciating how shifts in tone and stance made it feel more responsive to the task at hand. This reaction aligns with sociolinguistic accounts of how people naturally shift role presentation across contexts \cite{Goffman_1959, Goatly_2007}. In this sense, metaphor-fluidity mirrors familiar patterns of contextual modulation rather than simply increasing anthropomorphism.

Others, however, experienced certain transitions as unnecessary or disruptive, particularly when persona shifts did not align with their expectations. These responses indicate that metaphor-fluidity enhances engagement when its expressive shifts feel contextually appropriate, but may introduce friction when they are perceived as inconsistent. Thus, the variability does not undermine the approach; rather, it underscores that its value depends on alignment with both interactional context and user expectations \cite{desai2024cui, Khadpe_Krishna_Fei-Fei_Hancock_Bernstein_2020, Dubiel_Desai_Zargham_Schmitt_2024}.

Familiarity further contextualizes these findings. Because prerecorded audio was required to maintain identical linguistic content, pacing, and error conditions, we recruited participants already familiar with contemporary VUIs. As a result, participants likely entered the study with stronger expectations for the Default VUI modeled after Google Assistant. Prior work shows that familiarity typically advantages established systems in perception \cite{Murad_Candello_Munteanu_2023, Klein_Deutschländer_Kölln_Rauschenberger_Escalona_2024, Hao_Wang_Wang_Zhu_2024}. Despite this, the Metaphor-Fluid VUI was rated higher in enjoyment, likability, and intention to adopt. This suggests that metaphor-fluidity can provide experiential benefits that exceed those typically associated with familiarity.

Individual differences add another layer to interpretation. Prior VUI research shows that users often prefer systems they perceive as similar to themselves \cite{Braun_Mainz_Chadowitz_Pfleging_Alt_2019}. Personality traits have been shown to influence conversational preferences, with conscientious users favoring non-confrontational systems \cite{volkel2021manipulating} and lower-agreeableness users preferring more informal styles \cite{Chin2024Like}. These findings suggest that metaphor-fluid design may benefit from accounting for these differences. Emerging work on inferring personality from conversational behavior \cite{Zhang_Dinan_Urbanek_Szlam_Kiela_Weston_2018, Guo_Hirai_Ohashi_Chiba_Tsunomori_Higashinaka_2024} opens possibilities for adapting metaphorical presentation not only to context but also to user traits. Recent evidence suggests that aligning a conversational agent's personality with user traits can enhance positive perceptions \cite{Rahman_Desai_2026, Mathur_Rahman_Desai_2026}, further motivating the case for individual preferences within metaphor-fluid systems.

One promising direction is integrating principles from Mutual Theory of Mind \cite{Wang_Goel_2022, Wang_Saha_Gregori_Joyner_Goel_2021}. In human interaction, Theory of Mind enables individuals to infer others’ beliefs and intentions, facilitating adaptive communication. Computational approximations of this capacity aim to model user expectations over time rather than relying solely on static rules. Applied to Metaphor-Fluid VUIs, such mechanisms could support adaptation at two levels: conversational context and individual preference. Rather than exposing explicit controls over metaphor shifts, future systems might infer when stability or dynamism is more appropriate, adjusting presentation accordingly. In this framing, metaphor-fluidity would function as a dynamic capability responsive to both situational and personal factors, rather than a preset design choice. 

\subsection{{Limitations}}

\subsubsection{Limitations of Study 1}

Study 1 examined a curated set of 20 metaphors selected based on frequency of mention, representational diversity (human, non-human, fictional), and feasibility for implementation in VUI design. While this allowed for a structured and practically grounded comparison, it necessarily excluded less common metaphors that may still meaningfully shape user mental models. Expanding the metaphor space beyond those most frequently cited in current discourse could surface alternative framings that better resonate with specific populations or emerging use-cases. Our analysis relied on hierarchy and formality as organizing dimensions for perceptual mapping. These dimensions are well-established in conversational and persona research, but they do not exhaust the social attributes through which metaphors operate. Other dimensions—such as warmth, competence, agreeableness, or politeness—may further nuance metaphor suitability and could be incorporated in future work to provide a more multidimensional mapping of metaphor–context alignment. Finally, metaphor interpretation is culturally and linguistically situated. Because our participant pool consisted of English-proficient individuals based in the United States, the findings reflect North American metaphor norms. Prior work demonstrates that cultural context shapes metaphor usage and interpretation \cite{Kövecses_2002a, Kövecses_2002b}; future research should examine how metaphor-fluidity is perceived across diverse linguistic and cultural settings.

\subsubsection{Limitations of Study 2}

Study 2 compared a Metaphor-Fluid VUI against a Default VUI modeled on commercial assistant-based systems. While this provides a realistic and ecologically valid baseline, it represents only one dominant design paradigm. Other conversational systems—such as explicitly relational agents or domain-specific tools—may evoke different expectations, and evaluating metaphor-fluidity across these system types could broaden its applicability. To ensure strict experimental control, we employed synthetic voices and prerecorded interactions. This approach allowed us to hold tone, pacing, linguistic content, and error conditions constant across participants \cite{dubiel2024impact}, but it does not fully replicate the contingencies of live interaction. Although prerecorded stimuli are common in controlled VUI research \cite{Seaborn_Urakami_2021}, future work could examine how metaphor-fluidity unfolds in fully interactive systems. As with Study 1, the dialogue content, voices, and scenarios were designed for English-speaking, North American contexts. Cultural expectations around authority, politeness, and relational stance may influence how metaphorical shifts are interpreted, and broader cross-cultural validation remains an important direction for future research \cite{Deshmukh_Chalmeta_2024}. 

An additional limitation concerns the transition from Study 1 to Study 2. Study 1 identified which metaphors align with which use-contexts along shared social dimensions, but Study 2 required conceptualizing those metaphors as designed dialogue. The two conditions consequently differ in both metaphorical framing and linguistic elaborateness, so the observed effects cannot be attributed to metaphor-fluidity alone. This entanglement is partly inherent to the construct: metaphor-fluidity is realized through changes in linguistic style and conversational stance, making content variation constitutive of the framing rather than independent of it. Still, alternative instantiations of the same metaphors could yield different outcomes.

\subsection{Future Work} 

Our findings position Metaphor-Fluid Design as a possible alternative to the one-size-fits-all approach of the current implementation of system personas. However, several key avenues remain unexplored. In this section, we outline opportunities to further develop this approach within VUIs and extend its application to CUIs more broadly.

\subsubsection{Expanding Metaphor-Fluid Design Beyond VUIs.}
While this work examined Metaphor-Fluid Design in the context of VUIs, its potential extends far beyond voice-based interactions. CUIs are now embedded in chat-based systems, embodied agents, and mixed-reality environments, raising important questions about how metaphor-fluidity operates across modalities. Future research should investigate whether metaphor transitions function differently in text-based interfaces compared to VUIs and whether embodied agents require distinct metaphorical shifts to maintain coherence between visual, auditory, and behavioral cues. Understanding these cross-modal dynamics is crucial for refining the principles of Metaphor-Fluid Design.

The paradigm for conversation design has shifted with the advent of LLMs. Unlike previous intent-driven architectures with structured, pre-defined interaction flows, LLM-based VUIs now generate responses dynamically, adapting to user input in real time \cite{Yang_Xu_Yao_Rogers_Zhang_Intille_Shara_Gao_Wang_2024}. This has fundamentally altered persona creation from carefully crafted scripts with controlled vocabulary and constraints \cite{Sadek_Calvo_Mougenot_2023} to systems like Character.ai\footnote{https://character.ai/} where users can summon personas on demand—creating anything from supportive companions to fictional characters. While this grants unprecedented control, it introduces critical ethical risks, as evidenced by cases where users develop harmful emotional attachments to LLM-generated personas\footnote{https://www.nytimes.com/2024/10/23/technology/characterai-lawsuit-teen-suicide.html}.

Metaphor-fluid systems offer a structured alternative to this arbitrary persona generation. Rather than allowing personas to emerge through user prompting, this approach introduces intentional, context-aware shifts based on conversational expectations. Its integration of non-human metaphors provides an alternative to human-like personas that often carry implicit biases and problematic social expectations. Research is needed to explore how Metaphor-Fluid Design can function within LLM-driven architectures: Can we create systems where metaphorical adaptation is both flexible and controlled? How can metaphor-based transitions be embedded into prompt engineering techniques while maintaining coherent persona shifts and managing user expectations?

.

\subsubsection{Engaging Designers in the Development Process.}
If Metaphor-Fluid Design is to become a widely-adopted approach, conversation designers need to be actively involved in shaping its development. At present, metaphor selection is often implicit, emerging from branding choices or designer intuition rather than structured methodologies \cite{Sadek_Calvo_Mougenot_2023}. A critical next step is engaging designers through participatory design workshops and the development of concrete heuristics. Future research must establish principles for guiding metaphor transitions and determine how metaphor adaptation should be communicated to users. Empirical investigation is required to ensure that metaphor-fluidity can be systematically integrated into CUI architectures while preserving ethical safeguards and maintaining user trust.

Additionally, in the case of VUIs, \textit{voice itself} serves as a design element, not just a delivery mechanism \cite{Sutton_Foulkes_Kirk_Lawson_2019}. The characteristics of a voice—its tone, pitch, cadence, and modulation—carry their own metaphorical weight and significantly influence user expectations \cite{dubiel2024impact}. Future work must address how to systematically design voice-based metaphors, whether different metaphorical personas should be expressed through distinct voice characteristics, and how subtle variations in speech patterns shape metaphor perception. As synthetic voices become increasingly lifelike, these considerations will become even more critical, particularly as the boundary between human and artificial speech continues to blur.

\subsubsection{Encouraging Open Questions for Researchers.}
While this study highlights broad trends in metaphor preference, individual differences remain an open challenge. Future research should investigate how user personality influences metaphor preference, such as whether highly agreeable individuals favor social metaphors while more analytical users prefer structured, rule-based ones. Furthermore, cultural factors play a major role in shaping metaphorical associations \cite{Moser_2000}, meaning U.S.-centric designs may not translate to other linguistic or cultural contexts. Understanding how metaphor-fluidity functions across diverse user populations is essential.

Equally important are longitudinal investigations. Most evaluations of metaphor use in CUIs, including this study, rely on single-session experiments \cite{Clark_Doyle_Garaialde_Gilmartin_Schlögl_Edlund_Aylett_Cabral_Munteanu_Edwards_et_al._2019}. Since metaphors are dynamic constructs and users' perceptions evolve over time, future work must determine whether metaphor-fluid CUIs sustain their benefits over long-term use or whether users develop new expectations that necessitate further adaptation. Addressing these questions will be crucial in refining Metaphor-Fluid Design into a robust, scalable approach.

\section{Conclusion}

This work introduced Metaphor-Fluid Design as an approach to VUI development in which metaphorical framing shifts across conversational use-contexts. Across two studies, we found that a Metaphor-Fluid VUI—drawing from human, non-human, and fictional metaphors—was perceived as more enjoyable, more likable, and more adoptable than a Default Assistant-based VUI. These findings suggest that aligning metaphorical presentation with task-specific expectations can meaningfully enhance user experience. At the same time, variability in responses underscores that metaphor-fluidity is not universally preferred. While many participants valued contextual shifts, others experienced them as unnecessary or distracting, indicating that the effectiveness of this approach depends on both contextual fit and individual preference. Rather than advocating for a single alternative to the assistant metaphor, this work argues against singular framing altogether. As conversational systems expand into increasingly diverse domains, static metaphors risk misalignment with evolving expectations. Metaphor-Fluid Design offers a structured way to accommodate these shifts. Future work should explore how such systems can adapt not only to conversational context but also to individual differences, moving toward conversational agents that are responsive to both situational demands and user-specific preferences.

%% The next two lines define the bibliography style to be used, and
%% the bibliography file.

\begin{acks}
We are grateful to the members of the Conversational Human-AI Interactions (CHAI) Lab at Northeastern University for their suggestions on improving the readability of this paper. We also thank the reviewers for their valuable feedback.
\end{acks}

\bibliographystyle{ACM-Reference-Format}
\bibliography{sample-base}

@String{Computing = "Computing" }

@String{Computer = "{IEEE} Computer" }

@String{Academic = "Academic Press" }

@String{Springer = "Springer-Verlag" }

@inproceedings{Mathur_Rahman_Desai_2026, address={New York, NY, USA}, series={CHI EA ’26}, title={“Who wants to be nagged by AI?”: Investigating the Effects of Agreeableness on Older Adults’ Perception of LLM-Based Voice Assistants’ Explanations}, ISBN={9798400722813}, url={https://dl.acm.org/doi/10.1145/3772363.3798685}, DOI={10.1145/3772363.3798685}, abstractNote={LLM-based voice assistants (VAs) increasingly support older adults aging in place, yet how an assistant’s agreeableness shapes explanation perception remains underexplored. We conducted a study (N=70) examining how VA agreeableness influences older adults’ perceptions of explanations across routine and emergency home scenarios. High-agreeableness assistants were perceived as more trustworthy, empathetic, and likable, but these benefits diminished in emergencies where clarity outweighed warmth. Agreeableness did not affect perceived intelligence, suggesting social tone and competence are separable dimensions. Real-time environmental explanations outperformed history-based ones, and agreeable older adults penalized low-agreeableness assistants more strongly. These findings show the need to move beyond a one-size-fits-all approach to AI explainability, while balancing personality, context, and audience.}, booktitle={Proceedings of the Extended Abstracts of the 2026 CHI Conference on Human Factors in Computing Systems}, publisher={Association for Computing Machinery}, author={Mathur, Niharika and Rahman, Hasibur and Desai, Smit}, year={2026}, month=apr, pages={1–6}, collection={CHI EA ’26} }

@article{Braun_Clarke_2006, title={Using thematic analysis in psychology}, volume={3}, ISSN={1478-0887}, DOI={10.1191/1478088706qp063oa}, abstractNote={Thematic analysis is a poorly demarcated, rarely acknowledged, yet widely used qualitative analytic method within psychology. In this paper, we argue that it offers an accessible and theoretically flexible approach to analysing qualitative data. We outline what thematic analysis is, locating it in relation to other qualitative analytic methods that search for themes or patterns, and in relation to different epistemological and ontological positions. We then provide clear guidelines to those wanting to start thematic analysis, or conduct it in a more deliberate and rigorous way, and consider potential pitfalls in conducting thematic analysis. Finally, we outline the disadvantages and advantages of thematic analysis. We conclude by advocating thematic analysis as a useful and flexible method for qualitative research in and beyond psychology.}, number={2}, journal={Qualitative Research in Psychology}, publisher={Routledge}, author={Braun, Virginia and Clarke, Victoria}, year={2006}, month=jan, pages={77–101} }

@inproceedings{Rahman_Desai_2026, address={New York, NY, USA}, series={CHI ’26}, title={Vibe Check: Understanding the Effects of LLM-Based Conversational Agents’ Personality and Alignment on User Perceptions in Goal-Oriented Tasks}, ISBN={9798400722783}, url={https://dl.acm.org/doi/10.1145/3772318.3790388}, DOI={10.1145/3772318.3790388}, abstractNote={Large language models (LLMs) enable conversational agents (CAs) to express distinctive personalities, raising new questions about how such designs shape user perceptions. This study investigates how personality expression levels and user-agent personality alignment influence perceptions in goal-oriented tasks. In a between-subjects experiment (N=150), participants completed travel planning with CAs exhibiting low, medium, or high expression across the Big Five traits, controlled via our novel Trait Modulation Keys framework. Results revealed an inverted-U relationship: medium expression produced the most positive evaluations across Intelligence, Enjoyment, Anthropomorphism, Intention to Adopt, Trust, and Likeability, significantly outperforming both extremes. Personality alignment further enhanced outcomes, with Extraversion and Emotional Stability emerging as the most influential traits. Cluster analysis identified three distinct compatibility profiles, with “Well-Aligned” users reporting substantially positive perceptions. These findings demonstrate that personality expression and strategic trait alignment constitute optimal design targets for CA personality, offering design implications as LLM-based CAs become increasingly prevalent.}, booktitle={Proceedings of the 2026 CHI Conference on Human Factors in Computing Systems}, publisher={Association for Computing Machinery}, author={Rahman, Hasibur and Desai, Smit}, year={2026}, month=apr, pages={1–30}, collection={CHI ’26} }

@book{Goffman_1959, address={Oxford, England}, series={The presentation of self in everyday life}, title={The presentation of self in everyday life}, abstractNote={A classic analysis of the processes by which persons manage their appearance and demanor so as to project an appropriate impression of themselves into social interactions.  Harvard Book List (edited) 1971 #459 (PsycINFO Database Record (c) 2016 APA, all rights reserved)}, publisher={Doubleday}, author={Goffman, Erving}, year={1959}, collection={The presentation of self in everyday life} }

@article{Hao_Wang_Wang_Zhu_2024, title={Navigation Voice Familiarity Can Affect Driving Stress and Emotions Under Complex Road Conditions: Evidence From An Experimental Study}, volume={12}, ISSN={2169-3536}, DOI={10.1109/ACCESS.2024.3492207}, abstractNote={Navigational voice interaction is an important factor affecting driving behavior and safety under complex road conditions. However, much of the research on navigational speech has focused on aspects such as commands and tone, and little on voice familiarity. To investigate the effects of navigation voice familiarity on driving stress and emotions, this study was conducted through simulated driving and controlled experiments. Driving stress was assessed by analyzing participants’ electrocardiogram (ECG) signals and subjective data obtained from a Driving Activity Load Index (DALI) questionnaire. Driving emotions were evaluated by participants’ facial expression data and the Self-Assessment Manikin (SAM), whereas the Post-study System Usability Questionnaire (PSSUQ) scores were collected as a complementary measure of both driving stress and emotions. We found that voice familiarity had a substantial impact on driving stress and emotions. Compared with using a familiar voice, an unfamiliar voice resulted in a 26.51% increase in participants’ ECG data and a 34.12% increase in DALI total scores. In addition, the use of an unfamiliar voice was associated with a 112.50% and 200% increase in the mean intensity value of anger and disgust among participants, respectively, as well as a decrease in their PSSUQ ratings. These findings indicate that the use of unfamiliar navigation voices in complex road conditions can exacerbate driving stress and negative emotions. This study provides valuable insights for navigational Voice User Interaction (VUI) design and ultimately contributing to driving safety.}, journal={IEEE Access}, author={Hao, Pingting and Wang, Meiyin and Wang, Junfeng and Zhu, Qing}, year={2024}, pages={164121–164136} }

@article{Deshmukh_Chalmeta_2024, title={User Experience and Usability of Voice User Interfaces: A Systematic Literature Review}, volume={15}, rights={http://creativecommons.org/licenses/by/3.0/}, ISSN={2078-2489}, DOI={10.3390/info15090579}, abstractNote={As voice user interfaces (VUIs) rapidly transform the landscape of human–computer interaction, their potential to revolutionize user engagement is becoming increasingly evident. This paper aims to advance the field of human–computer interaction by conducting a bibliometric analysis of the user experience associated with VUIs. It proposes a classification framework comprising six research categories to systematically organize the existing literature, analyzes the primary research streams, and identifies future research directions within each category. This systematic literature review provides a comprehensive analysis of the development and effectiveness of VUIs in facilitating natural human–machine interaction. It offers critical insights into the user experience of VUIs, contributing to the refinement of VUI design to optimize overall user interaction and satisfaction.}, number={9}, journal={Information}, publisher={Multidisciplinary Digital Publishing Institute}, author={Deshmukh, Akshay Madhav and Chalmeta, Ricardo}, year={2024}, month=sep, pages={579}, language={en} }

@article{Klein_Deutschländer_Kölln_Rauschenberger_Escalona_2024, title={Exploring the context of use for voice user interfaces: Toward context-dependent user experience quality testing}, volume={36}, rights={© 2023 The Authors. Journal of Software: Evolution and Process published by John Wiley & Sons Ltd.}, ISSN={2047-7481}, DOI={10.1002/smr.2618}, abstractNote={Voice user interface (VUI) systems, such as Alexa, Siri, and Google Assistant, are popular and widely available. Still, challenges such as privacy and the ability to have a dialog remain. In the latter example, the user expects a human-like conversation, that is, that the VUI understands the dialog and its context. However, this VUI feature of context-aware interaction is rather error prone. For this reason, we intend to explore the VUI context of use and its impact on interaction, that is, relevant user experience (UX). We see a demand for context-dependent UX measurement because analyzing the context of use and UX assessment are both critical human-centered design (HCD) methods. Therefore, we examine the VUI context of use by asking users about how, where, and for what they use VUIs, as well as their UX and improvement proposals. We interviewed people with disabilities who rely on VUIs and people without disabilities who use VUIs for convenience or fun. We identified VUI context-of-use categories and factors and explored their impacts on relevant UX qualities. Our result is a matrix containing these elements; thus, it provides an overview of the contextual UX of our target group’s VUI interaction. We intend to develop a VUI context-of-use conceptual structure in the future based on this matrix, which is needed to create an automated context-dependent UX measurement recommendation tool for VUIs. This conceptual structure could also be useful for automated UX testing in the context of VUI.}, number={7}, journal={Journal of Software: Evolution and Process}, author={Klein, Andreas M. and Deutschländer, Jana and Kölln, Kristina and Rauschenberger, Maria and Escalona, Maria José}, year={2024}, pages={e2618}, language={en} }

@inproceedings{Maes_Shneiderman_Miller_1997, address={New York, NY, USA}, series={CHI EA ’97}, title={Intelligent software agents vs. user-controlled direct manipulation: a debate}, ISBN={978-0-89791-926-5}, url={https://dl.acm.org/doi/10.1145/1120212.1120281}, DOI={10.1145/1120212.1120281}, abstractNote={Critical issues in human-computer interaction - in particular, the advantages and disadvantages of intelligent agents and direct manipulation - will be discussed, debated, and hotly contested. The intent of the participants is to strike an appropriate balance between a serious discussion of the issues and an entertaining debate.}, booktitle={CHI ’97 Extended Abstracts on Human Factors in Computing Systems}, publisher={Association for Computing Machinery}, author={Maes, Pattie and Shneiderman, Ben and Miller, Jim}, year={1997}, month=mar, pages={105–106}, collection={CHI EA ’97} }

@inproceedings{Murad_Candello_Munteanu_2023, address={New York, NY, USA}, series={CUI ’23}, title={What’s The Talk on VUI Guidelines? A Meta-Analysis of Guidelines for Voice User Interface Design}, ISBN={9798400700149}, url={https://doi.org/10.1145/3571884.3597129}, DOI={10.1145/3571884.3597129}, abstractNote={Over the past decade, voice user interface (VUI) design has been steadily growing, along with a growing VUI presence in consumer markets. However, there is currently a lack of widely-established guidelines for VUI design. While many sets of VUI guidelines have been proposed, they tend to be developed independently of each other, leading to a lack of consensus on appropriate guidelines for VUI design. This can hinder the wider adoption of practical VUI guidelines. To address this gap, we performed a large-scale meta-analysis of 336 VUI design guidelines that have been proposed in academic literature. Using thematic analysis, we present a unified and synthesized set of 14 guidelines, representing the most universally proposed principles of VUI design as captured by the 336 VUI guidelines identified in academic literature. We hope that this synthesized set can address several of the challenges to the adoption of VUI guidelines in design practice.}, booktitle={Proceedings of the 5th International Conference on Conversational User Interfaces}, publisher={Association for Computing Machinery}, author={Murad, Christine and Candello, Heloisa and Munteanu, Cosmin}, year={2023}, month=jul, pages={1–16}, collection={CUI ’23} }

@article{Hadler_2025, title={The Effects of Open-Ended Probes on Closed Survey Questions in Web Surveys}, volume={54}, ISSN={0049-1241}, DOI={10.1177/00491241231176846}, abstractNote={Probes are follow-ups to survey questions used to gain insights on respondents’ understanding of and responses to these questions. They are usually administered as open-ended questions, primarily in the context of questionnaire pretesting. Due to the decreased cost of data collection for open-ended questions in web surveys, researchers have argued for embedding more open-ended probes in large-scale web surveys. However, there are concerns that this may cause reactivity and impact survey data. The study presents a randomized experiment in which identical survey questions were run with and without open-ended probes. Embedding open-ended probes resulted in higher levels of survey break off, as well as increased backtracking and answer changes to previous questions. In most cases, there was no impact of open-ended probes on the cognitive processing of and response to survey questions. Implications for embedding open-ended probes into web surveys are discussed.}, number={1}, journal={Sociological Methods \& Research}, publisher={SAGE Publications Inc}, author={Hadler, Patricia}, year={2025}, month=feb, pages={106–139}, language={EN} }

@inproceedings{Desai_Dubiel_Leiva_2024, address={New York, NY, USA}, series={CUI ’24}, title={Examining Humanness as a Metaphor to Design Voice User Interfaces}, ISBN={9798400705113}, url={https://doi.org/10.1145/3640794.3665535}, DOI={10.1145/3640794.3665535}, abstractNote={Voice User Interfaces (VUIs) increasingly leverage ‘humanness’ as a foundational design metaphor, adopting roles like ‘assistants,’ ‘teachers,’ and ‘secretaries’ to foster natural interactions. Yet, this approach can sometimes misalign user trust and reinforce societal stereotypes, leading to socio-technical challenges that might impede long-term engagement. This paper explores an alternative approach to navigate these challenges—incorporating non-human metaphors in VUI design. We report on a study with 240 participants examining the effects of human versus non-human metaphors on user perceptions within health and finance domains. Results indicate a preference for the human metaphor (doctor) over the non-human (health encyclopedia) in health contexts for its perceived enjoyability and likeability. In finance, however, user perceptions do not significantly differ between human (financial advisor) and non-human (calculator) metaphors. Importantly, our research reveals that the explicit awareness of a metaphor’s use influences adoption intentions, with a marked preference for non-human metaphors when their metaphorical nature is not disclosed. These findings highlight context-specific conversation design strategies required in integrating non-human metaphors into VUI design, suggesting tradeoffs and design considerations that could enhance user engagement and adoption.}, booktitle={Proceedings of the 6th ACM Conference on Conversational User Interfaces}, publisher={Association for Computing Machinery}, author={Desai, Smit and Dubiel, Mateusz and Leiva, Luis A.}, year={2024}, month=jul, pages={1–15}, collection={CUI ’24} }

@ArtifactSoftware{R,
    title = {R: A Language and Environment for Statistical Computing},
    author = {{R Core Team}},
    organization = {R Foundation for Statistical Computing},
    address = {Vienna, Austria},
    year = {2019},
    url = {https://www.R-project.org/},
}

@article{desai2024cui,
  title={CUI@ CHI 2024: Building Trust in CUIs-From Design to Deployment},
  author={Desai, Smit and Wei, Christina and Sin, Jaisie and Dubiel, Mateusz and Zargham, Nima and Ahire, Shashank and Porcheron, Martin and Kuzminykh, Anastasia and Lee, Minha and Candello, Heloisa and others},
  journal={arXiv preprint arXiv:2401.13970},
  year={2024}
}

@book{Indurkhya_2013, title={Metaphor and Cognition: An Interactionist Approach}, ISBN={978-94-017-2252-0}, abstractNote={Many metaphors go beyond pionting to the existing similarities between two objects -- they create the similarities. Such metaphors, which have been relegated to the back seat in most of the cognitive science research, are the focus of attention in this study, which addresses the creation of similarity within an elaborately laid out interactive framework of cognition. Starting from the constructivist views of Nelson Goodman and Jean Piaget, this framework resolves an apparent paradox in interactionism: how can reality not have a mind-independent ontology and structure, but still manage to constrain the possible worlds a cognitive agent can create in it?  A comprehensive theory of metaphor is proposed in this framework that explains how metaphors can create similarities, and why such metaphors are an invaluable asset to cognition. The framework is then applied to related issues of analogical reasoning, induction, and computational modeling of creative metaphors.}, note={Google-Books-ID: foTrCAAAQBAJ}, publisher={Springer Science \& Business Media}, author={Indurkhya, B.}, year={2013}, month=mar, language={en} }

@book{Lakoff_Johnson_1980, title={Metaphors We Live By}, ISBN={978-0-226-47099-3}, abstractNote={The now-classic Metaphors We Live By changed our understanding of metaphor and its role in language and the mind. Metaphor, the authors explain, is a fundamental mechanism of mind, one that allows us to use what we know about our physical and social experience to provide understanding of countless other subjects. Because such metaphors structure our most basic understandings of our experience, they are “metaphors we live by”—metaphors that can shape our perceptions and actions without our ever noticing them.  In this updated edition of Lakoff and Johnson’s influential book, the authors supply an afterword surveying how their theory of metaphor has developed within the cognitive sciences to become central to the contemporary understanding of how we think and how we express our thoughts in language.}, note={Google-Books-ID: r6nOYYtxzUoC}, publisher={University of Chicago Press}, author={Lakoff, George and Johnson, Mark}, year={1980}, language={en} }

@book{Cameron_Maslen_2010, address={London}, title={Metaphor Analysis: Research Practice in applied Linguistics, Social Sciences and the Humanities}, ISBN={978-1-84553-446-2}, url={http://www.equinoxpub.com/books/showbook.asp?bkid=363&keyword=9781845534479}, abstractNote={Metaphor is recognised as an important way of thinking constructing analogies and making connections between ideas and an important way of using language to explain abstract ideas or to find indirect but powerful ways of conveying feelings. By investigating peoples use of metaphors, we can better understand their emotions, attitudes and conceptualisations, as individuals and as participants in social life. This book describes practice in the analysis of metaphor in real-world discourse. When real-world language use is taken as the site of metaphor study, researchers face methodological issues that have only recently begun to be addressed. The contributors to this volume have all had to find ways to deal with methodological issues in their own research and have developed techniques that are brought together here. Using as a basis the discourse dynamics approach to metaphor developed by the editor, the book explores links between theory and empirical investigation, exemplifies data analysis and discusses issues in research design and practice. Particular attention is paid to the processes of metaphor identification, categorisation and labelling, and to the use of corpus linguistic and other computer-assisted methods.}, publisher={Equinox}, author={Cameron, Lynne and Maslen, Robert}, editor={Cameron, Lynne and Maslen, Robert}, year={2010}, month=jun, language={en} }

@article{Blackwell_2006, title={The reification of metaphor as a design tool}, volume={13}, ISSN={1073-0516}, DOI={10.1145/1188816.1188820}, abstractNote={Despite causing many debates in human-computer interaction (HCI), the term “metaphor” remains a central element of design practice. This article investigates the history of ideas behind user-interface (UI) metaphor, not only technical developments, but also less familiar perspectives from education, philosophy, and the sociology of science. The historical analysis is complemented by a study of attitudes toward metaphor among HCI researchers 30 years later. Working from these two streams of evidence, we find new insights into the way that theories in HCI are related to interface design, and offer recommendations regarding approaches to future UI design research.}, number={4}, journal={ACM Transactions on Computer-Human Interaction}, author={Blackwell, Alan F.}, year={2006}, month=dec, pages={490–530} }

@article{Colburn_Shute_2008, series={The Philosophy of Computer Science}, title={Metaphor in computer science}, volume={6}, ISSN={1570-8683}, DOI={10.1016/j.jal.2008.09.005}, abstractNote={The language of computer science is laced with metaphor. We argue that computer science metaphors provide a conceptual framework in which to situate constantly emerging new ontologies in computational environments. But how computer science metaphors work does not fit neatly into prevailing general theories of metaphor. We borrow from these general theories while also considering the unique role of computer science metaphors in learning, design, and scientific analysis. We find that computer science metaphors trade on both preexisting and emerging similarities between computational and traditional domains, but owing to computer science’s peculiar status as a discipline that creates its own subject matter, the role of similarity in metaphorical attribution is multifaceted.}, number={4}, journal={Journal of Applied Logic}, author={Colburn, T. R. and Shute, G. M.}, year={2008}, month=dec, pages={526–533}, collection={The Philosophy of Computer Science}, language={en} }

@inproceedings{McMillan_Jaber_2021, address={New York, NY, USA}, series={CUI ’21}, title={Leaving the Butler Behind: The Future of Role Reproduction in CUI}, ISBN={978-1-4503-8998-3}, url={https://doi.org/10.1145/3469595.3469606}, DOI={10.1145/3469595.3469606}, abstractNote={Speech technologies are increasing in popularity by offering new interaction modalities for their users. Despite the prevalence of these devices, and the rapid improvement of the underlying technology, how we actually interact with these devices has remained wrapped up in the metaphors of command and control based around the problematic reproduction of the role of butler, maid, or personal assistant. In this paper we explore the issues around focusing our development and research on making a ‘better’ subordinate, and point to some opportunities to replace and refresh the status quo.}, booktitle={CUI 2021 - 3rd Conference on Conversational User Interfaces}, publisher={Association for Computing Machinery}, author={McMillan, Donald and Jaber, Razan}, year={2021}, month=jul, pages={1–4}, collection={CUI ’21} }

@inproceedings{Pradhan_Lazar_2021, address={New York, NY, USA}, series={CUI ’21}, title={Hey Google, Do You Have a Personality? Designing Personality and Personas for Conversational Agents}, ISBN={978-1-4503-8998-3}, url={https://dl.acm.org/doi/10.1145/3469595.3469607}, DOI={10.1145/3469595.3469607}, abstractNote={Conversational agents designed to interact through natural language are often imbued with human-like personalities. At times, the agent might also have a distinct persona with traits such as gender, age, or a backstory. Designing such personality or persona for conversational agents has become a common design practice. In this work, we review the emerging literature on designing agent persona or personality, and reflect on these approaches, along with the personas that are created for common conversational agents. We discuss open questions with regards to three aspects: meeting user needs, the ethics of deception, and reinforcing social stereotypes through conversational agents. We hope this work can provoke researchers and practitioners to critically reflect on their approach for designing personality or persona of conversational agents.}, booktitle={Proceedings of the 3rd Conference on Conversational User Interfaces}, publisher={Association for Computing Machinery}, author={Pradhan, Alisha and Lazar, Amanda}, year={2021}, month=jul, pages={1–4}, collection={CUI ’21} }

@inproceedings{Sadek_Calvo_Mougenot_2023, address={New York, NY, USA}, series={CUI ’23}, title={Trends, Challenges and Processes in Conversational Agent Design: Exploring Practitioners’ Views through Semi-Structured Interviews}, ISBN={9798400700149}, url={https://dl.acm.org/doi/10.1145/3571884.3597143}, DOI={10.1145/3571884.3597143}, abstractNote={The aim of this study is to explore the challenges and experiences of conversational agent (CA) practitioners in order to highlight their practical needs and bring them into consideration within the scholarly sphere. A range of data scientists, conversational designers, executive managers and researchers shared their opinions and experiences through semi-structured interviews. They were asked about emerging trends, the challenges they face, and the design processes they follow when creating CAs. In terms of trends, findings included mixed feelings regarding no-code solutions and a desire for a separation of roles. The challenges mentioned included a lack of socio-technical tools and conversational archetypes. Finally, practitioners followed different design processes and did not use the design processes described in the academic literature. These findings were analyzed to establish links between practitioners’ insights and discussions in related literature. The goal of this analysis is to highlight research-practice gaps by synthesising five practitioner needs that are not currently being met. By highlighting these research-practice gaps and foregrounding the challenges and experiences of CA practitioners, we can begin to understand the extent to which emerging literature is influencing industrial settings and where more research is needed to better support CA practitioners in their work.}, booktitle={Proceedings of the 5th International Conference on Conversational User Interfaces}, publisher={Association for Computing Machinery}, author={Sadek, Malak and Calvo, Rafael A and Mougenot, Celine}, year={2023}, month=jul, pages={1–10}, collection={CUI ’23} }

@inproceedings{Desai_Chin_2023, address={New York, NY, USA}, series={CHI ’23}, title={OK Google, Let’s Learn: Using Voice User Interfaces for Informal Self-Regulated Learning of Health Topics among Younger and Older Adults}, ISBN={978-1-4503-9421-5}, url={https://doi.org/10.1145/3544548.3581507}, DOI={10.1145/3544548.3581507}, abstractNote={In this paper, we present Health Buddy, a voice agent integrated into commercially available Voice User Interfaces (VUIs) to support informal self-regulated learning (SRL) of health-related topics through multiple learning strategies and examine the efficacy of Health Buddy on learning outcomes for younger and older adults. We conducted a mixed-factorial-design experiment with 26 younger and 25 older adults, assigned to three SRL strategies (within-subjects): monologue, dialogue-based scaffolding building, and conceptual diagramming. We found that while younger adults benefit more from scaffolding building and conceptual diagramming, both younger and older adults showed equivalent learning outcomes. Furthermore, interaction fluency (operationalized by the number of conversational breakdowns) was associated with learning outcomes regardless of age. While older adults did not experience less fluent conversations, interaction fluency affected their technology acceptance toward VUIs more than younger ones. Our study discusses age-related learning differences and has implications for designing VUI-based learning programs for older adults.}, booktitle={Proceedings of the 2023 CHI Conference on Human Factors in Computing Systems}, publisher={Association for Computing Machinery}, author={Desai, Smit and Chin, Jessie}, year={2023}, month=apr, pages={1–21}, collection={CHI ’23} }

@inproceedings{Desai_Hu_Lundy_Chin_2023, address={New York, NY, USA}, series={HAI ’23}, title={Using Experience-Based Participatory Approach to Design Interactive Voice User Interfaces for Delivering Physical Activity Programs with Older Adults}, ISBN={9798400708244}, url={https://dl.acm.org/doi/10.1145/3623809.3623827}, DOI={10.1145/3623809.3623827}, abstractNote={Voice User Interfaces (VUIs) are popular among older adults, who find them easy to use and perceive them as social companions. However, there is a lack of research on voice-based applications to support physical activities for older adults. To address this gap, we present “Workout Pal,” a voice agent designed to deliver physical activities to older adults. We conducted a mixed-methods study involving ten older adults to understand their perceptions and design priorities when interacting with Workout Pal. Questionnaires and semi-structured interviews were used to assess their experience, while experience-based co-design sessions facilitated collaboration in exploring the design space and identifying design requirements. Our findings highlight the feasibility and design preferences for smart speaker-based physical activity programs. The study contributes by sharing the design and development of Workout Pal, illustrating a novel co-design approach with older adults, and providing design implications tailored to the specific needs of older adults. The design guidelines emphasize the importance of sociability, voice design, and individualization. This research supports the development of elder-friendly VUIs helping older adults live independently and engage in physical activities.}, booktitle={Proceedings of the 11th International Conference on Human-Agent Interaction}, publisher={Association for Computing Machinery}, author={Desai, Smit and Hu, Xinhui and Lundy, Morgan and Chin, Jessie}, year={2023}, month=dec, pages={180–190}, collection={HAI ’23} }

@inproceedings{Desai_Lundy_Chin_2023, address={New York, NY, USA}, series={CUI ’23}, title={“A Painless Way to Learn:” Designing an Interactive Storytelling Voice User Interface to Engage Older Adults in Informal Health Information Learning}, ISBN={9798400700149}, url={https://dl.acm.org/doi/10.1145/3571884.3597141}, DOI={10.1145/3571884.3597141}, abstractNote={We present “Mystery Agent,” an interactive storytelling voice user interface (VUI) equipped with self-regulated learning strategies to deliver informal health-related learning to older adults through a murder mystery story. We conducted a mixed methods user study with 10 older adults to evaluate Mystery Agent, using usability and perception-based questionnaires, followed by a semi-structured interview and co-design activity to generate design insights and identify design priorities. We found older adults had a positive experience interacting with Mystery Agent and considered storytelling to be an appropriate and engaging way to learn health information. However, older adults identified credibility, compassion, and control as crucial factors influencing long-term use. To address this, we present design guidelines using Mystery Agent as an example to help practitioners and researchers devise novel solutions to address the informal health information learning needs of older adults.}, booktitle={Proceedings of the 5th International Conference on Conversational User Interfaces}, publisher={Association for Computing Machinery}, author={Desai, Smit and Lundy, Morgan and Chin, Jessie}, year={2023}, month=jul, pages={1–16}, collection={CUI ’23} }

@inproceedings{Motalebi_Cho_Sundar_Abdullah_2019, address={New York, NY, USA}, series={CSCW ’19 Companion}, title={Can Alexa be your Therapist? How Back-Channeling Transforms Smart-Speakers to be Active Listeners}, ISBN={978-1-4503-6692-2}, url={https://dl.acm.org/doi/10.1145/3311957.3359502}, DOI={10.1145/3311957.3359502}, abstractNote={Smart-speakers such as Amazon Alexa are becoming increasingly popular among the general population. These devices support a wide range of user-initiated tasks. However, their current interaction capabilities are limited to quick turn-takings and short dialogues, which leads to limited user engagement. In this project, we aim to enhance the user engagement and interaction abilities of a smart-speaker by transforming it into an active listener. Specifically, we explored how providing random back-channelling (i.e., verbal continuers including “hm”, “uhum”, “aha”, “yeah”) can result in longer interactions and more sustained user engagement. The findings of the study have implications for usability and future design of smart-speakers. We also explored how the enhanced interactions in smart-speakers through back-channeling can be used for self-centered therapy leading to betTer emotional and social support.}, booktitle={Companion Publication of the 2019 Conference on Computer Supported Cooperative Work and Social Computing}, publisher={Association for Computing Machinery}, author={Motalebi, Nasim and Cho, Eugene and Sundar, S. Shyam and Abdullah, Saeed}, year={2019}, month=nov, pages={309–313}, collection={CSCW ’19 Companion} }

@inproceedings{Wang_Yang_Shao_Abdullah_Sundar_2020, address={New York, NY, USA}, series={CHI ’20}, title={Alexa as Coach: Leveraging Smart Speakers to Build Social Agents that Reduce Public Speaking Anxiety}, ISBN={978-1-4503-6708-0}, url={https://dl.acm.org/doi/10.1145/3313831.3376561}, DOI={10.1145/3313831.3376561}, abstractNote={Public speaking anxiety is one of the most common social phobias. We explore the feasibility of using a conversational agent to reduce this anxiety. We developed a public-speaking tutor on the Amazon Alexa platform that enables users to engage in cognitive reconstruction exercises. We also investigated how the sociability of the agent might affect its performance as a tutor. A user study of 53 college students with fear of public speaking showed that the interaction with the agent served to assuage pre-speech state anxiety. Agent sociability improved the sense of interpersonal closeness, which was associated with lower pre-speech anxiety. Moreover, sociability of the agent increased participants’ satisfaction and their willingness to continue engagement. Our findings, thus, have implications not only for addressing public speaking anxiety in a scalable way but also for the design of future conversational agents using smart speaker platforms.}, booktitle={Proceedings of the 2020 CHI Conference on Human Factors in Computing Systems}, publisher={Association for Computing Machinery}, author={Wang, Jinping and Yang, Hyun and Shao, Ruosi and Abdullah, Saeed and Sundar, S. Shyam}, year={2020}, month=apr, pages={1–13}, collection={CHI ’20} }

@inproceedings{Jung_Kim_So_Kim_Oh_2019, address={New York, NY, USA}, series={CHI EA ’19}, title={TurtleTalk: An Educational Programming Game for Children with Voice User Interface}, ISBN={978-1-4503-5971-9}, url={https://dl.acm.org/doi/10.1145/3290607.3312773}, DOI={10.1145/3290607.3312773}, abstractNote={Interest in programming education for children is growing. This research explores the possibilities of utilizing voice user interface (VUI) in children’s programming education. We designed an interactive educational programming game called TurtleTalk, which converts the various utterances of children into code using a neural network and displays the results on a screen (Figure 1). Through VUI, children can move the turtle, the voice agent of the game, to the target location and learn the basic programming concepts of “sequencing” and “iteration.” We conducted a preliminary user study where eight children played the game and took part in a posthoc interview. The results showed that voice interaction with TurtleTalk led children to be more immersed in the game and understand the elements of programming with ease and confidence.}, booktitle={Extended Abstracts of the 2019 CHI Conference on Human Factors in Computing Systems}, publisher={Association for Computing Machinery}, author={Jung, Hyunhoon and Kim, Hee Jae and So, Seongeun and Kim, Jinjoong and Oh, Changhoon}, year={2019}, month=may, pages={1–6}, collection={CHI EA ’19} }

@article{Lee_Frank_IJsselsteijn_2021, title={Brokerbot: A Cryptocurrency Chatbot in the Social-technical Gap of Trust}, volume={30}, ISSN={1573-7551}, DOI={10.1007/s10606-021-09392-6}, abstractNote={Cryptocurrencies are proliferating as instantiations of blockchain, which is a transparent, distributed ledger technology for validating transactions. Blockchain is thus said to embed trust in its technical design. Yet, blockchain’s technical promise of trust is not fulfilled when applied to the cryptocurrency ecosystem due to many social challenges stakeholders experience. By investigating a cryptocurrency chatbot (Brokerbot) that distributed information on cryptocurrency news and investments, we explored social tensions of trust between stakeholders, namely the bot’s developers, users, and the bot itself. We found that trust in Brokerbot and in the cryptocurrency ecosystem are two conjoined, but separate challenges that users and developers approached in different ways. We discuss the challenging, dual-role of a Brokerbot as an object of trust as a chatbot while simultaneously being a mediator of trust in cryptocurrency, which exposes the social-technical gap of trust. Lastly, we elaborate on trust as a negotiated social process that people shape and are shaped by through emerging ecologies of interlinked technologies like blockchain and conversational interfaces.}, number={1}, journal={Computer Supported Cooperative Work (CSCW)}, author={Lee, Minha and Frank, Lily and IJsselsteijn, Wijnand}, year={2021}, month=feb, pages={79–117}, language={en} }

@inproceedings{DeVito_Birnholtz_Hancock_French_Liu_2018, address={New York, NY, USA}, series={CHI ’18}, title={How People Form Folk Theories of Social Media Feeds and What it Means for How We Study Self-Presentation}, ISBN={978-1-4503-5620-6}, url={https://dl.acm.org/doi/10.1145/3173574.3173694}, DOI={10.1145/3173574.3173694}, abstractNote={Self-presentation is a process that is significantly complicated by the rise of algorithmic social media feeds, which obscure information about one’s audience and environment. User understandings of these systems, and therefore user ability to adapt to them, are limited, and have recently been explored through the lens of folk theories. To date, little is understood of how these theories are formed, and how they tie to the self-presentation process in social media. This paper presents an exploratory look at the folk theory formation process and the interplay between folk theories and self-presentation via a 28-participant interview study. Results suggest that people draw from diverse sources of information when forming folk theories, and that folk theories are more complex, multifaceted and malleable than previously assumed. This highlights the need to integrate folk theories into both social media systems and theories of self-presentation.}, booktitle={Proceedings of the 2018 CHI Conference on Human Factors in Computing Systems}, publisher={Association for Computing Machinery}, author={DeVito, Michael A. and Birnholtz, Jeremy and Hancock, Jeffery T. and French, Megan and Liu, Sunny}, year={2018}, month=apr, pages={1–12}, collection={CHI ’18} }

@inproceedings{Kuzminykh_Sun_Govindaraju_Avery_Lank_2020, address={New York, NY, USA}, series={CHI ’20}, title={Genie in the Bottle: Anthropomorphized Perceptions of Conversational Agents}, ISBN={978-1-4503-6708-0}, url={https://dl.acm.org/doi/10.1145/3313831.3376665}, DOI={10.1145/3313831.3376665}, abstractNote={This paper presents a qualitative multi-phase study seeking to identify patterns in users’ anthropomorphized perceptions of conversational agents. Through a comparative analysis of behavioral perceptions and visual conceptions of three agents - Alexa, Google Assistant, and Siri - we first show that the perceptions of an agent’s character are structured according to five categories: approachability, sentiment toward a user, professionalism, intelligence, and individuality. We then explore visualizations of the agents’ appearance and discuss the specifics assigned to each agent. Finally, we analyze associative explanations for these perceptions. We demonstrate that the anthropomorphized behavioral and visual perceptions of agents yield structural consistency and discuss how these perceptions are linked with each other and system features.}, booktitle={Proceedings of the 2020 CHI Conference on Human Factors in Computing Systems}, publisher={Association for Computing Machinery}, author={Kuzminykh, Anastasia and Sun, Jenny and Govindaraju, Nivetha and Avery, Jeff and Lank, Edward}, year={2020}, month=apr, pages={1–13}, collection={CHI ’20} }

@inproceedings{Luger_Sellen_2016, address={New York, NY, USA}, series={CHI ’16}, title={“Like Having a Really Bad PA”: The Gulf between User Expectation and Experience of Conversational Agents}, ISBN={978-1-4503-3362-7}, url={https://dl.acm.org/doi/10.1145/2858036.2858288}, DOI={10.1145/2858036.2858288}, abstractNote={The past four years have seen the rise of conversational agents (CAs) in everyday life. Apple, Microsoft, Amazon, Google and Facebook have all embedded proprietary CAs within their software and, increasingly, conversation is becoming a key mode of human-computer interaction. Whilst we have long been familiar with the notion of computers that speak, the investigative concern within HCI has been upon multimodality rather than dialogue alone, and there is no sense of how such interfaces are used in everyday life. This paper reports the findings of interviews with 14 users of CAs in an effort to understand the current interactional factors affecting everyday use. We find user expectations dramatically out of step with the operation of the systems, particularly in terms of known machine intelligence, system capability and goals. Using Norman’s “gulfs of execution and evaluation” [30] we consider the implications of these findings for the design of future systems.}, booktitle={Proceedings of the 2016 CHI Conference on Human Factors in Computing Systems}, publisher={Association for Computing Machinery}, author={Luger, Ewa and Sellen, Abigail}, year={2016}, month=may, pages={5286–5297}, collection={CHI ’16} }

@inproceedings{Desai_Twidale_2022, address={New York, NY, USA}, series={CUI ’22}, title={Is Alexa like a computer? A search engine? A friend? A silly child? Yes.}, ISBN={978-1-4503-9739-1}, url={https://dl.acm.org/doi/10.1145/3543829.3544535}, DOI={10.1145/3543829.3544535}, abstractNote={In this provocation, we analyze metaphors used by 14 end-users in semi-structured interviews to describe their interactions with Voice User Interfaces (VUIs). We identified four key metaphor groups—computer, search engine, friend, and silly child—that can help us understand how individual end-users perceive VUIs differently based on the conversational context. A consideration of these four groups draws attention to the issues with reductionist implementations of system personas and role-based performances and the problems of fixating on humanness as a metaphor. Fortunately, using metaphor analysis can also help in ideating solutions to these issues.}, booktitle={Proceedings of the 4th Conference on Conversational User Interfaces}, publisher={Association for Computing Machinery}, author={Desai, Smit and Twidale, Michael}, year={2022}, month=sep, pages={1–4}, collection={CUI ’22} }

@book{Norman_2013, address={New York, New York}, edition={Revised edition}, title={The Design Of Everyday Things}, ISBN={978-0-465-05065-9}, publisher={Basic Books}, author={Norman, Don}, year={2013}, month=nov, language={English} }

@article{Khadpe_Krishna_Fei-Fei_Hancock_Bernstein_2020, title={Conceptual Metaphors Impact Perceptions of Human-AI Collaboration}, volume={4}, ISSN={2573-0142}, DOI={10.1145/3415234}, abstractNote={With the emergence of conversational artificial intelligence (AI) agents, it is important to understand the mechanisms that influence users’ experiences of these agents. In this paper, we study one of the most common tools in the designer’s toolkit: conceptual metaphors. Metaphors can present an agent as akin to a wry teenager, a toddler, or an experienced butler. How might a choice of metaphor influence our experience of the AI agent? Sampling a set of metaphors along the dimensions of warmth and competence---defined by psychological theories as the primary axes of variation for human social perception---we perform a study $(N=260)$ where we manipulate the metaphor, but not the behavior, of a Wizard-of-Oz conversational agent. Following the experience, participants are surveyed about their intention to use the agent, their desire to cooperate with the agent, and the agent’s usability. Contrary to the current tendency of designers to use high competence metaphors to describe AI products, we find that metaphors that signal low competence lead to better evaluations of the agent than metaphors that signal high competence. This effect persists despite both high and low competence agents featuring identical, human-level performance and the wizards being blind to condition. A second study confirms that intention to adopt decreases rapidly as competence projected by the metaphor increases. In a third study, we assess effects of metaphor choices on potential users’ desire to try out the system and find that users are drawn to systems that project higher competence and warmth. These results suggest that projecting competence may help attract new users, but those users may discard the agent unless it can quickly correct with a lower competence metaphor. We close with a retrospective analysis that finds similar patterns between metaphors and user attitudes towards past conversational agents such as Xiaoice, Replika, Woebot, Mitsuku, and Tay.}, number={CSCW2}, journal={Proceedings of the ACM on Human-Computer Interaction}, author={Khadpe, Pranav and Krishna, Ranjay and Fei-Fei, Li and Hancock, Jeffrey T. and Bernstein, Michael S.}, year={2020}, month=oct, pages={1–26}, language={en} }

@article{Chin2024Like,
  title={Like my aunt Dorothy: Effects of conversational styles on perceptions, acceptance, and metaphorical descriptions of voice assistants during later adulthood},
  author={Chin, Jessie and Desai, Smit and Lin, Sheny and Mej{\'i}a, Shannon},
  journal={Proceedings of the ACM on Human-Computer Interaction},
  volume={8},
  number={CSCW1},
  article={88},
  pages={1-22},
  year={2024},
  month={Apr},
  doi={10.1145/3637365},
  url={https://doi.org/10.1145/3637365}
}

@inproceedings{Simpson_Crone_2022, address={Glasgow United Kingdom}, title={Should Alexa be a Police Officer, a Doctor, or a Priest?: Towards CUI Relationships Worth Having}, ISBN={978-1-4503-9739-1}, url={https://dl.acm.org/doi/10.1145/3543829.3544522}, DOI={10.1145/3543829.3544522}, booktitle={Proceedings of the 4th Conference on Conversational User Interfaces}, publisher={ACM}, author={Simpson, James and Crone, Cassandra}, year={2022}, month=jul, pages={1–5}, language={en} }

@article{Desai_Twidale_2023, title={Metaphors in Voice User Interfaces: A Slippery Fish}, volume={30}, ISSN={1073-0516, 1557-7325}, DOI={10.1145/3609326}, abstractNote={We explore a range of different metaphors used for Voice User Interfaces (VUIs) by designers, end-users, manufacturers, and researchers using a novel framework derived from semi-structured interviews and a literature review. We focus less on the well-established idea of metaphors as a way for interface designers to help novice users learn how to interact with novel technology, and more on other ways metaphors can be used. We find that metaphors people use are contextually fluid, can change with the mode of conversation, and can reveal differences in how people perceive VUIs compared to other devices. Not all metaphors are helpful, and some may be offensive. Analyzing this broader class of metaphors can help understand, perhaps even predict problems. Metaphor analysis can be a low-cost tool to inspire design creativity and facilitate complex discussions about sociotechnical issues, enabling us to spot potential opportunities and problems in the situated use of technologies.}, number={6}, journal={ACM Transactions on Computer-Human Interaction}, author={Desai, Smit and Twidale, Michael}, year={2023}, month=dec, pages={1–37}, language={en} }

@article{Pradhan_Findlater_Lazar_2019, title={“Phantom Friend” or “Just a Box with Information”: Personification and Ontological Categorization of Smart Speaker-based Voice Assistants by Older Adults}, volume={3}, ISSN={2573-0142}, DOI={10.1145/3359316}, abstractNote={As voice-based conversational agents such as Amazon Alexa and Google Assistant move into our homes, researchers have studied the corresponding privacy implications, embeddedness in these complex social environments, and use by specific user groups. Yet it is unknown how users categorize these devices: are they thought of as just another object, like a toaster? As a social companion? Though past work hints to human-like attributes that are ported onto these devices, the anthropomorphization of voice assistants has not been studied in depth. Through a study deploying Amazon Echo Dot Devices in the homes of older adults, we provide a preliminary assessment of how individuals 1) perceive having social interactions with the voice agent, and 2) ontologically categorize the voice assistants. Our discussion contributes to an understanding of how well-developed theories of anthropomorphism apply to voice assistants, such as how the socioemotional context of the user (e.g., loneliness) drives increased anthropomorphism. We conclude with recommendations for designing voice assistants with the ontological category in mind, as well as implications for the design of technologies for social companionship for older adults.}, number={CSCW}, journal={Proceedings of the ACM on Human-Computer Interaction}, author={Pradhan, Alisha and Findlater, Leah and Lazar, Amanda}, year={2019}, month=nov, pages={1–21}, language={en} }

@inproceedings{Jung_Qiu_Bozzon_Gadiraju_2022, address={New Orleans LA USA}, title={Great Chain of Agents: The Role of Metaphorical Representation of Agents in Conversational Crowdsourcing}, ISBN={978-1-4503-9157-3}, url={https://dl.acm.org/doi/10.1145/3491102.3517653}, DOI={10.1145/3491102.3517653}, booktitle={CHI Conference on Human Factors in Computing Systems}, publisher={ACM}, author={Jung, Ji-Youn and Qiu, Sihang and Bozzon, Alessandro and Gadiraju, Ujwal}, year={2022}, month=apr, pages={1–22}, language={en} }

@inproceedings{Wei_Kim_Kuzminykh_2023, address={New York, NY, USA}, series={CUI ’23}, title={The Bot on Speaking Terms: The Effects of Conversation Architecture on Perceptions of Conversational Agents}, ISBN={9798400700149}, url={https://doi.org/10.1145/3571884.3597139}, DOI={10.1145/3571884.3597139}, abstractNote={Conversational agents mimic natural conversation to interact with users. Since the effectiveness of interactions strongly depends on users’ perception of agents, it is crucial to design agents’ behaviors to provide the intended user perceptions. Research on human-agent and human-human communication suggests that speech specifics are associated with perceptions of communicating parties, but there is a lack of systematic understanding of how speech specifics of agents affect users’ perceptions. To address this gap, we present a framework outlining the relationships between elements of agents’ conversation architecture (dialog strategy, content affectiveness, content style and speech format) and aspects of users’ perception (interaction, ability, sociability and humanness). Synthesized based on literature reviewed from the domains of HCI, NLP and linguistics (n=57), this framework demonstrates both the identified relationships and the areas lacking empirical evidence. We discuss the implications of the framework for conversation design and highlight the inconsistencies with terminology and measurements.}, booktitle={Proceedings of the 5th International Conference on Conversational User Interfaces}, publisher={Association for Computing Machinery}, author={Wei, Christina Ziying and Kim, Young-Ho and Kuzminykh, Anastasia}, year={2023}, month=jul, pages={1–16}, collection={CUI ’23} }

@inproceedings{Yang_Aurisicchio_Baxter_2019, address={New York, NY, USA}, series={CHI ’19}, title={Understanding Affective Experiences with Conversational Agents}, ISBN={978-1-4503-5970-2}, url={http://doi.acm.org/10.1145/3290605.3300772}, DOI={10.1145/3290605.3300772}, abstractNote={" While previous studies of Conversational Agents (e.g. Siri, Google Assistant, Alexa and Cortana) have focused on evaluating usability and exploring capabilities of these systems, little work has examined users’ affective experiences. In this paper we present a survey study with 171 participants to examine CA users’ affective experiences. Specifically, we present four major usage scenarios, users’ affective responses in these scenarios, and the factors which influenced the affective responses. We found that users’ overall experience was positive with interest being the most salient positive emotion. Affective responses differed depending on the scenarios. Both pragmatic and hedonic qualities influenced affect. The factors underlying pragmatic quality are: helpfulness, proactivity, fluidity, seamlessness and responsiveness. The factors underlying hedonic quality are: comfort in human-machine conversation, pride of using cutting-edge technology, fun during use, perception of having a human-like assistant, concern about privacy and fear of causing distraction.}, note={00000 
event-place: Glasgow, Scotland Uk}, booktitle={Proceedings of the 2019 CHI Conference on Human Factors in Computing Systems}, publisher={ACM}, author={Yang, Xi and Aurisicchio, Marco and Baxter, Weston}, year={2019}, pages={542:1-542:12}, collection={CHI ’19} }

@article{Moussawi_Koufaris_Benbunan-Fich_2021, title={How perceptions of intelligence and anthropomorphism affect adoption of personal intelligent agents}, volume={31}, ISSN={1422-8890}, DOI={10.1007/s12525-020-00411-w}, abstractNote={A personal intelligent agent (PIA) is a system that acts intelligently to assist a human using natural language. Examples include Siri and Alexa. These agents are powerful computer programs that operate autonomously and proactively, learn and adapt to change, react to the environment, complete tasks within a favorable timeframe and communicate with the user using natural language to process commands and compose replies. PIAs are different from other systems previously explored in Information Systems (IS) due to their personalized, intelligent, and human-like behavior. Drawing on research in IS and Artificial Intelligence, we build and test a model of user adoption of PIAs leveraging their uique characteristics. Analysis of data collected from an interactive lab-based study for new PIA users confirms that both perceived intelligence and anthropomorphism are significant antecedents of PIA adoption. Our findings contribute to the understanding of a quickly-changing and fast-growing set of technologies that extend users’ capabilities and their sense of self​.}, number={2}, journal={Electronic Markets}, author={Moussawi, Sara and Koufaris, Marios and Benbunan-Fich, Raquel}, year={2021}, month=jun, pages={343–364}, language={en} }

@inbook{Bartneck_2023, address={Cham}, title={Godspeed Questionnaire Series: Translations and Usage}, ISBN={978-3-030-89738-3}, url={https://doi.org/10.1007/978-3-030-89738-3_24-1}, DOI={10.1007/978-3-030-89738-3_24-1}, abstractNote={The Godspeed Questionnaire Series (GQS) is one of the most highly cited and used questionnaire in the field of Human-Robot Interaction and Human-Agent Interaction. Since its inception in 2009, it has been translated into 19 languages. In this chapter, we discuss the history and development of the GQS as well its psychometric properties. For the first time we make the collected translations that have been provided by the research community, available in a formal publication. This will allow future authors to better reference the exact translations they used. We review the psychometric properties of the translations when available. We discuss the merits and limitations of the GQS, with a particular emphasis on the contribution of making measurement instruments available openly.}, booktitle={International Handbook of Behavioral Health Assessment}, publisher={Springer International Publishing}, author={Bartneck, Christoph}, editor={Krägeloh, Christian U. and Alyami, Mohsen and Medvedev, Oleg N.}, year={2023}, pages={1–35}, language={en} }

@article{Bartneck_Kulić_Croft_Zoghbi_2009, title={Measurement Instruments for the Anthropomorphism, Animacy, Likeability, Perceived Intelligence, and Perceived Safety of Robots}, volume={1}, ISSN={1875-4805}, DOI={10.1007/s12369-008-0001-3}, abstractNote={This study emphasizes the need for standardized measurement tools for human robot interaction (HRI). If we are to make progress in this field then we must be able to compare the results from different studies. A literature review has been performed on the measurements of five key concepts in HRI: anthropomorphism, animacy, likeability, perceived intelligence, and perceived safety. The results have been distilled into five consistent questionnaires using semantic differential scales. We report reliability and validity indicators based on several empirical studies that used these questionnaires. It is our hope that these questionnaires can be used by robot developers to monitor their progress. Psychologists are invited to further develop the questionnaires by adding new concepts, and to conduct further validations where it appears necessary.}, number={1}, journal={International Journal of Social Robotics}, author={Bartneck, Christoph and Kulić, Dana and Croft, Elizabeth and Zoghbi, Susana}, year={2009}, month=jan, pages={71–81}, language={en} }

@inproceedings{Seaborn_Urakami_2021, address={New York, NY, USA}, series={CHI EA ’21}, title={Measuring Voice UX Quantitatively: A Rapid Review}, ISBN={978-1-4503-8095-9}, url={https://doi.org/10.1145/3411763.3451712}, DOI={10.1145/3411763.3451712}, abstractNote={Computer voice is experiencing a renaissance through the growing popularity of voice-based interfaces, agents, and environments. Yet, how to measure the user experience (UX) of voice-based systems remains an open and urgent question, especially given that their form factors and interaction styles tend to be non-visual, intangible, and often considered disembodied or “body-less.” As a first step, we surveyed the ACM and IEEE literatures to determine which quantitative measures and measurements have been deemed important for voice UX. Our findings show that there is little consensus, even with similar situations and systems, as well as an overreliance on lab work and unvalidated scales. In response, we offer two high-level descriptive frameworks for guiding future research, developing standardized instruments, and informing ongoing review work. Our work highlights the current strengths and weaknesses of voice UX research and charts a path towards measuring voice UX in a more comprehensive way.}, booktitle={Extended Abstracts of the 2021 CHI Conference on Human Factors in Computing Systems}, publisher={Association for Computing Machinery}, author={Seaborn, Katie and Urakami, Jacqueline}, year={2021}, month=may, pages={1–8}, collection={CHI EA ’21} }

@article{Jian_Bisantz_Drury_2000, address={US}, title={Foundations for an empirically determined scale of trust in automated systems}, volume={4}, ISSN={1532-7566}, DOI={10.1207/S15327566IJCE0401_04}, abstractNote={One component in the successful use of automated systems is the extent to which people trust the automation to perform effectively. In order to understand the relationship between trust in computerized systems and the use of those systems, we need to be able to effectively measure trust. Although questionnaires regarding trust have been used in prior studies, these questionnaires were theoretically rather than empirically generated and did not distinguish between 3 potentially different types of trust: human–human trust, human–machine trust, and trust in general. A 3-phased experiment, comprising a word elicitation study, a questionnaire study, and a paired comparison study, was performed to better understand similarities and differences in the concepts of trust and distrust, and among the different types of trust. Results indicated that trust and distrust can be considered opposites, rather than different concepts. Components of trust, in terms of words related to trust, were similar across the three types of trust. Results obtained from a cluster analysis were used to identify 12 potential factors of trust between people and automated systems. These 12 factors were then used to develop a proposed scale to measure trust in automation. (PsycINFO Database Record (c) 2016 APA, all rights reserved)}, number={1}, journal={International Journal of Cognitive Ergonomics}, publisher={Lawrence Erlbaum}, author={Jian, Jiun-Yin and Bisantz, Ann M. and Drury, Colin G.}, year={2000}, pages={53–71} }

@misc{dubiel2024impact,
      title={Impact of Voice Fidelity on Decision Making: A Potential Dark Pattern?}, 
      author={Mateusz Dubiel and Anastasia Sergeeva and Luis A. Leiva},
      year={2024},
      eprint={2402.07010},
      archivePrefix={arXiv},
      primaryClass={cs.HC}
}

@inproceedings{volkel2021manipulating,
  title={Manipulating and evaluating levels of personality perceptions of voice assistants through enactment-based dialogue design},
  author={V{\"o}lkel, Sarah Theres and Meindl, Samantha and Hussmann, Heinrich},
  booktitle={Proceedings of the 3rd Conference on Conversational User Interfaces},
  pages={1--12},
  year={2021}
}

@article{Moser_2000, title={Metaphor Analysis in Psychology—Method, Theory, and Fields of Application}, volume={1}, rights={Copyright (c) 2000 Karin S. Moser}, ISSN={1438-5627}, url={https://www.qualitative-research.net/index.php/fqs/article/view/1090}, DOI={10.17169/fqs-1.2.1090}, abstractNote={The analysis of metaphors is a classical research theme in linguistics, but has received very little attention in psychological research so far. Metaphor analysis—as conceptualized in cognitive linguistics—is proposed here as a qualitative method for psychological research for several reasons. Metaphors are culturally and socially defined, yet they also represent a basic cognitive strategy of analogical problem solving. Metaphors are context-sensitive, yet at the same time they are abstract models of reality much in the same way as mental models and schemata in cognitive psychology. The multifaceted properties of metaphors allow for the study of micro-interactions between cognition and culture in open and qualitative research designs. They also enable the bridging of the gap between quantitative-experimental and qualitative approaches in psychology. Because metaphors are of high plausibility in everyday experience, metaphors are a valuable tool for interventions in applied fields of research such as organizational and work psychology.
URN: urn:nbn:de:0114-fqs0002212}, number={22}, journal={Forum Qualitative Sozialforschung / Forum: Qualitative Social Research}, author={Moser, Karin S.}, year={2000}, month=jun, language={en} }

@article{ammari2019music,
  title={Music, search, and IoT: How people (really) use voice assistants},
  author={Ammari, Tawfiq and Kaye, Jofish and Tsai, Janice Y and Bentley, Frank},
  journal={ACM Transactions on Computer-Human Interaction (TOCHI)},
  volume={26},
  number={3},
  pages={1--28},
  year={2019},
  publisher={ACM New York, NY, USA}
}

@inproceedings{dubiel2020persuasive,
  title={Persuasive synthetic speech: Voice perception and user behaviour},
  author={Dubiel, Mateusz and Halvey, Martin and Gallegos, Pilar Oplustil and King, Simon},
  booktitle={Proceedings of the 2nd Conference on Conversational User Interfaces},
  pages={1--9},
  year={2020}
}

@inproceedings{curry2020conversational,
  title={Conversational assistants and gender stereotypes: Public perceptions and desiderata for voice personas},
  author={Curry, Amanda Cercas and Robertson, Judy and Rieser, Verena},
  booktitle={Proceedings of the second workshop on gender bias in natural language processing},
  pages={72--78},
  year={2020}
}

@inproceedings{Edlund_2019, address={New York, NY, USA}, series={CUI ’19}, title={Shoehorning in the name of science}, ISBN={978-1-4503-7187-2}, url={https://doi.org/10.1145/3342775.3342794}, DOI={10.1145/3342775.3342794}, abstractNote={This provocation paper calls for a deeper understanding of what spoken human-computer interaction is, and what it can be. Its given structure by a story of humanlikeness and fraudulent spoken dialogue systems - specifically systems that deliberately attempts to mislead their interlocutors into believing that they are speaking to a human. Against this backdrop, a plea that conversational user interfaces are viewed from the perspective of conversation and spoken interaction first, and from the perspective of GUIs and interface design second, lest we impose the limitations of one field onto the possibilities of another, rather than the other way around.}, booktitle={Proceedings of the 1st International Conference on Conversational User Interfaces}, publisher={Association for Computing Machinery}, author={Edlund, Jens}, year={2019}, month=aug, pages={1–3}, collection={CUI ’19} }

@inproceedings{Sciuto_Saini_Forlizzi_Hong_2018, address={New York, NY, USA}, series={DIS ’18}, title={“Hey Alexa, What’s Up?”: A Mixed-Methods Studies of In-Home Conversational Agent Usage}, ISBN={978-1-4503-5198-0}, url={http://doi.acm.org/10.1145/3196709.3196772}, DOI={10.1145/3196709.3196772}, abstractNote={In-home, place-based, conversational agents have exploded in popularity over the past three years. In particular, Amazon’s conversational agent, Alexa, now dominates the market and is in millions of homes. This paper presents two complementary studies investigating the experience of households living with a conversational agent over an extended period of time. First, we gathered the history logs of 75 Alexa participants and quantitatively analyzed over 278,000 commands. Second, we performed seven in-home, contextual interviews of Alexa owners focusing on how their household interacts with Alexa. Our findings give the first glimpse of how households integrate Alexa into their lives. We found interesting behaviors around purchasing and acclimating to Alexa, in the number and physical placement of devices, and in daily use patterns. Participants also uniformly described interactions between children and Alexa. We conclude with suggestions for future improvement for intelligent conversational agents.}, note={00000 
event-place: Hong Kong, China}, booktitle={Proceedings of the 2018 Designing Interactive Systems Conference}, publisher={ACM}, author={Sciuto, Alex and Saini, Arnita and Forlizzi, Jodi and Hong, Jason I.}, year={2018}, pages={857–868}, collection={DIS ’18} }

@article{Turk_2016, title={Home invasion}, volume={232}, ISSN={0262-4079}, DOI={10.1016/S0262-4079(16)32318-1}, abstractNote={Voice-activated assistants like Alexa already do our bidding. Why do we want them to be our friends too, asks Victoria Turk}, number={3104}, journal={New Scientist}, author={Turk, Victoria}, year={2016}, month=dec, pages={16–17}, language={en} }

@inproceedings{Porcheron_Fischer_Reeves_Sharples_2018, address={New York, NY, USA}, series={CHI ’18}, title={Voice Interfaces in Everyday Life}, ISBN={978-1-4503-5620-6}, url={https://dl.acm.org/doi/10.1145/3173574.3174214}, DOI={10.1145/3173574.3174214}, abstractNote={Voice User Interfaces (VUIs) are becoming ubiquitously available, being embedded both into everyday mobility via smartphones, and into the life of the home via “assistant” devices. Yet, exactly how users of such devices practically thread that use into their everyday social interactions remains underexplored. By collecting and studying audio data from month-long deployments of the Amazon Echo in participants’ homes-informed by ethnomethodology and conversation analysis-our study documents the methodical practices of VUI users, and how that use is accomplished in the complex social life of the home. Data we present shows how the device is made accountable to and embedded into conversational settings like family dinners where various simultaneous activities are being achieved. We discuss how the VUI is finely coordinated with the sequential organisation of talk. Finally, we locate implications for the accountability of VUI interaction, request and response design, and raise conceptual challenges to the notion of designing “conversational” interfaces.}, booktitle={Proceedings of the 2018 CHI Conference on Human Factors in Computing Systems}, publisher={Association for Computing Machinery}, author={Porcheron, Martin and Fischer, Joel E. and Reeves, Stuart and Sharples, Sarah}, year={2018}, month=apr, pages={1–12}, collection={CHI ’18} }

@article{Kim_Choudhury_2021, title={Exploring older adults’ perception and use of smart speaker-based voice assistants: A longitudinal study}, volume={124}, ISSN={0747-5632}, DOI={10.1016/j.chb.2021.106914}, abstractNote={Thanks to their conversational capabilities, smart speaker-based voice assistants are gaining attention for their potential to support the aging population, though the empirical evidence is still scarce. This paper aims to obtain empirical evidence on older adults’ experiences with a voice assistant. We especially focused on how their perception and use change over time as they progress from novice to more experienced users through a longitudinal field deployment study. We deployed Google Home devices in the homes of twelve older adults aged 65 and above and studied their use for sixteen weeks. Results show that the benefits our participants perceived have incrementally changed from enjoying simplicity and convenience of operation in the early phase of the study to not worrying about making mistakes and building digital companionship as they got used to using it. Results also show that participants confronted several challenges that evolved from the unfamiliarity with a voice assistant in their first interactions to coping with the functional errors due to limited speech technology as they got used to using it. Based on the results, we discuss design implications that could foster better user experiences with a voice assistant among older adults.}, journal={Computers in Human Behavior}, author={Kim, Sunyoung and Choudhury, Abhishek}, year={2021}, month=nov, pages={106914}, language={en} }

@book{Argyle_1988, address={London}, edition={2}, title={Bodily Communication}, ISBN={978-0-203-75383-5}, DOI={10.4324/9780203753835}, abstractNote={Non-verbal communication - the eye movements, facial expressions, tone of voice, postures and gestures that we all use more or less consciously and more or less effectively - can enhance or diminish every form of social interaction. Michael Argyle’s second edition of Bodily Communication is an invaluable up-to-date guide for students of the subject. In the last ten years NVC has become recognized as an important part of social psychology and of professional training, particularly in social work, education and management.Greatly expanded from the first edition, and significantly revised, this second edition has two completely new chapters on social skills and personality, and a new chapter on research methods. The author, a pioneer in the study of non-verbal communication, presents the second edition in the same accessible style as the first, bringing to the reader both his intense interest in the subject and his authoritative knowledge of it.}, publisher={Routledge}, author={Argyle, Michael}, year={1988}, month=feb }

@inproceedings{Axtell_Munteanu_2021, address={New York, NY, USA}, series={CHI ’21}, title={Tea, Earl Grey, Hot: Designing Speech Interactions from the Imagined Ideal of Star Trek}, ISBN={978-1-4503-8096-6}, url={https://doi.org/10.1145/3411764.3445640}, DOI={10.1145/3411764.3445640}, abstractNote={Speech is now common in daily interactions with our devices, thanks to voice user interfaces (VUIs) like Alexa. Despite their seeming ubiquity, designs often do not match users’ expectations. Science fiction, which is known to influence design of new technologies, has included VUIs for decades. Star Trek: The Next Generation is a prime example of how people envisioned ideal VUIs. Understanding how current VUIs live up to Star Trek’s utopian technologies reveals mismatches between current designs and user expectations, as informed by popular fiction. Combining conversational analysis and VUI user analysis, we study voice interactions with the Enterprise’s computer and compare them to current interactions. Independent of futuristic computing power, we find key design-based differences: Star Trek interactions are brief and functional, not conversational, they are highly multimodal and context-driven, and there is often no spoken computer response. From this, we suggest paths to better align VUIs with user expectations.}, note={event-place: Yokohama, Japan}, booktitle={Proceedings of the 2021 CHI Conference on Human Factors in Computing Systems}, publisher={Association for Computing Machinery}, author={Axtell, Benett and Munteanu, Cosmin}, year={2021}, collection={CHI ’21} }

@book{Keller_2003, address={Cambridge, Mass.}, title={Making Sense of Life: Explaining Biological Development with Models, Metaphors, and Machines}, ISBN={978-0-674-01250-9}, abstractNote={What do biologists want? If, unlike their counterparts in physics, biologists are generally wary of a grand, overarching theory, at what kinds of explanation do biologists aim? How will we know when we have “made sense” of life? Such questions, Evelyn Fox Keller suggests, offer no simple answers. Explanations in the biological sciences are typically provisional and partial, judged by criteria as heterogeneous as their subject matter. It is Keller’s aim in this bold and challenging book to account for this epistemological diversity―particularly in the discipline of developmental biology.In particular, Keller asks, what counts as an “explanation” of biological development in individual organisms? Her inquiry ranges from physical and mathematical models to more familiar explanatory metaphors to the dramatic contributions of recent technological developments, especially in imaging, recombinant DNA, and computer modeling and simulations.A history of the diverse and changing nature of biological explanation in a particularly charged field, Making Sense of Life draws our attention to the temporal, disciplinary, and cultural components of what biologists mean, and what they understand, when they propose to explain life.}, publisher={Harvard University Press}, author={Keller, Evelyn Fox}, year={2003}, month=oct, language={English} }

@inproceedings{Feldman_2024, address={New York, NY, USA}, series={CUI ’24}, title={The Voice: Lessons on Trustworthy Conversational Agents from “Dune”}, ISBN={9798400705113}, url={https://dl.acm.org/doi/10.1145/3640794.3665890}, DOI={10.1145/3640794.3665890}, abstractNote={The potential for untrustworthy conversational agents presents a significant threat for covert social manipulation. Taking inspiration from Frank Herbert’s Dune&nbsp;[12], where the Bene Gesserit Sisterhood uses the Voice for influence, manipulation, and control of people, we explore how generative AI provides a way to implement individualized influence at industrial scales. Already, these models can manipulate communication across text, image, speech, and most recently video. They are rapidly becoming affordable enough for any organization of even moderate means to train and deploy. If employed by malicious actors, they risk becoming powerful tools for shaping public opinion, sowing discord, and undermining organizations from companies to governments. As researchers and developers, it is crucial to recognize the potential for such weaponization and to explore strategies for prevention, detection, and defense against these emerging forms of sociotechnical manipulation.}, booktitle={Proceedings of the 6th ACM Conference on Conversational User Interfaces}, publisher={Association for Computing Machinery}, author={Feldman, Philip Gregory}, year={2024}, month=jul, pages={1–5}, collection={CUI ’24} }

@inproceedings{Lupetti_Murray-Rust_2024, address={New York, NY, USA}, series={CHI ’24}, title={(Un)making AI Magic: A Design Taxonomy}, ISBN={9798400703300}, url={https://dl.acm.org/doi/10.1145/3613904.3641954}, DOI={10.1145/3613904.3641954}, abstractNote={This paper examines the role that enchantment plays in the design of AI things by constructing a taxonomy of design approaches that increase or decrease the perception of magic and enchantment. We start from the design discourse surrounding recent developments in AI technologies, highlighting specific interaction qualities such as algorithmic uncertainties and errors and articulating relations to the rhetoric of magic and supernatural thinking. Through analyzing and reflecting upon 52 students’ design projects from two editions of a Masters course in design and AI, we identify seven design principles and unpack the effects of each in terms of enchantment and disenchantment. We conclude by articulating ways in which this taxonomy can be approached and appropriated by design/HCI practitioners, especially to support exploration and reflexivity.}, booktitle={Proceedings of the 2024 CHI Conference on Human Factors in Computing Systems}, publisher={Association for Computing Machinery}, author={Lupetti, Maria Luce and Murray-Rust, Dave}, year={2024}, month=may, pages={1–21}, collection={CHI ’24} }

@misc{Wang_Goel_2022, title={Mutual Theory of Mind for Human-AI Communication}, url={https://arxiv.org/abs/2210.03842v2}, abstractNote={New developments are enabling AI systems to perceive, recognize, and respond with social cues based on inferences made from humans’ explicit or implicit behavioral and verbal cues. These AI systems, equipped with an equivalent of human’s Theory of Mind (ToM) capability, are currently serving as matchmakers on dating platforms, assisting student learning as teaching assistants, and enhancing productivity as work partners. They mark a new era in human-AI interaction (HAI) that diverges from traditional human-computer interaction (HCI), where computers are commonly seen as tools instead of social actors. Designing and understanding the human perceptions and experiences in this emerging HAI era becomes an urgent and critical issue for AI systems to fulfill human needs and mitigate risks across social contexts. In this paper, we posit the Mutual Theory of Mind (MToM) framework, inspired by our capability of ToM in human-human communications, to guide this new generation of HAI research by highlighting the iterative and mutual shaping nature of human-AI communication. We discuss the motivation of the MToM framework and its three key components that iteratively shape the human-AI communication in three stages. We then describe two empirical studies inspired by the MToM framework to demonstrate the power of MToM in guiding the design and understanding of human-AI communication. Finally, we discuss future research opportunities in human-AI interaction through the lens of MToM.}, journal={arXiv.org}, author={Wang, Qiaosi and Goel, Ashok K.}, year={2022}, month=oct, language={en} }

@inbook{Trajkova_Martin-Hammond_2020, address={New York, NY, USA}, title={“Alexa is a Toy”: Exploring Older Adults’ Reasons for Using, Limiting, and Abandoning Echo}, ISBN={978-1-4503-6708-0}, url={https://doi.org/10.1145/3313831.3376760}, abstractNote={Intelligent voice assistants (IVAs) have the potential to support older adults’ independent living. However, despite a growing body of research focusing on IVA use, we know little about why older adults become IVA non-users. This paper examines the reasons older adults use, limit, and abandon IVAs (i.e., Amazon Echo) in their homes. We conducted eight focus groups, with 38 older adults residing in a Life Plan Community. Thirty-six participants owned an Echo for at least a year, and two were considering adoption. Over time, most participants became non-users due to their difficulty finding valuable uses, beliefs associated with ability and IVA use, or challenges with use in shared spaces. However, we also found that participants saw the potential for future IVA support. We contribute a better understanding of the reasons older adults do not engage with IVAs and how IVAs might better support aging and independent living in the future.}, booktitle={Proceedings of the 2020 CHI Conference on Human Factors in Computing Systems}, publisher={Association for Computing Machinery}, author={Trajkova, Milka and Martin-Hammond, Aqueasha}, year={2020}, month=apr, pages={1–13} }

@article{Guo_Hirai_Ohashi_Chiba_Tsunomori_Higashinaka_2024, title={Personality prediction from task-oriented and open-domain human-machine dialogues}, volume={14}, ISSN={2045-2322}, DOI={10.1038/s41598-024-53989-y}, abstractNote={If a dialogue system can predict the personality of a user from dialogue, it will enable the system to adapt to the user’s personality, leading to better task success and user satisfaction. In a recent study, personality prediction was performed using the Myers-Briggs Type Indicator (MBTI) personality traits with a task-oriented human-machine dialogue using an end-to-end (neural-based) system. However, it is still not clear whether such prediction is generally possible for other types of systems and user personality traits. To clarify this, we recruited 378 participants, asked them to fill out four personality questionnaires covering 25 personality traits, and had them perform three rounds of human-machine dialogue with a pipeline task-oriented dialogue system or an end-to-end task-oriented dialogue system. We also had another 186 participants do the same with an open-domain dialogue system. We then constructed BERT-based models to predict the personality traits of the participants from the dialogues. The results showed that prediction accuracy was generally better with open-domain dialogue than with task-oriented dialogue, although Extraversion (one of the Big Five personality traits) could be predicted equally well for both open-domain dialogue and pipeline task-oriented dialogue. We also examined the effect of utilizing different types of dialogue on personality prediction by conducting a cross-comparison of the models trained from the task-oriented and open-domain dialogues. As a result, we clarified that the open-domain dialogue cannot be used to predict personality traits from task-oriented dialogue, and vice versa. We further analyzed the effects of system utterances, task performance, and the round of dialogue with regard to the prediction accuracy.}, number={1}, journal={Scientific Reports}, author={Guo, Ao and Hirai, Ryu and Ohashi, Atsumoto and Chiba, Yuya and Tsunomori, Yuiko and Higashinaka, Ryuichiro}, year={2024}, month=feb, pages={3868}, language={eng} }

@article{Zhang_Dinan_Urbanek_Szlam_Kiela_Weston_2018, title={Personalizing Dialogue Agents: I have a dog, do you have pets too?}, url={http://arxiv.org/abs/1801.07243}, DOI={10.48550/arXiv.1801.07243}, abstractNote={Chit-chat models are known to have several problems: they lack specificity, do not display a consistent personality and are often not very captivating. In this work we present the task of making chit-chat more engaging by conditioning on profile information. We collect data and train models to (i) condition on their given profile information; and (ii) information about the person they are talking to, resulting in improved dialogues, as measured by next utterance prediction. Since (ii) is initially unknown our model is trained to engage its partner with personal topics, and we show the resulting dialogue can be used to predict profile information about the interlocutors.}, note={arXiv:1801.07243 [cs]}, number={arXiv:1801.07243}, publisher={arXiv}, author={Zhang, Saizheng and Dinan, Emily and Urbanek, Jack and Szlam, Arthur and Kiela, Douwe and Weston, Jason}, year={2018}, month=sep }

@inproceedings{Fischer_Reeves_Porcheron_Sikveland_2019, address={Dublin, Ireland}, series={CUI ’19}, title={Progressivity for voice interface design}, ISBN={978-1-4503-7187-2}, url={https://doi.org/10.1145/3342775.3342788}, DOI={10.1145/3342775.3342788}, abstractNote={Drawing from Conversation Analysis (CA), we examine how the orientation towards progressivity in talk-keeping things moving-might help us better understand and design for voice interactions. We introduce progressivity by surveying its explication in CA, and then look at how a strong preference for progressivity in conversation works out practically in sequences of voice interaction recorded in people’s homes. Following Stivers and Robinson’s work on progressivity, we find our data shows: how non-answer responses impede progress; how accounts offered for non-answer responses can lead to recovery; how participants work to receive answers; and how, ultimately, moving the interaction forwards does not necessarily involve a fitted answer, but other kinds of responses as well. We discuss the wider potential of applying progressivity to evaluate and understand voice interactions, and consider what designers of voice experiences might do to design for progressivity. Our contribution is a demonstration of the progressivity principle and its interactional features, which also points towards the need for specific kinds of future developments in speech technology.}, booktitle={Proceedings of the 1st International Conference on Conversational User Interfaces}, publisher={Association for Computing Machinery}, author={Fischer, Joel E. and Reeves, Stuart and Porcheron, Martin and Sikveland, Rein Ove}, year={2019}, month=aug, pages={1–8}, collection={CUI ’19} }

@article{Yang_Xu_Yao_Rogers_Zhang_Intille_Shara_Gao_Wang_2024, title={Talk2Care: An LLM-based Voice Assistant for Communication between Healthcare Providers and Older Adults}, volume={8}, DOI={10.1145/3659625}, abstractNote={Despite the plethora of telehealth applications to assist home-based older adults and healthcare providers, basic messaging and phone calls are still the most common communication methods, which suffer from limited availability, information loss, and process inefficiencies. One promising solution to facilitate patient-provider communication is to leverage large language models (LLMs) with their powerful natural conversation and summarization capability. However, there is a limited understanding of LLMs’ role during the communication. We first conducted two interview studies with both older adults (N=10) and healthcare providers (N=9) to understand their needs and opportunities for LLMs in patient-provider asynchronous communication. Based on the insights, we built an LLM-powered communication system, Talk2Care, and designed interactive components for both groups: (1) For older adults, we leveraged the convenience and accessibility of voice assistants (VAs) and built an LLM-powered conversational interface for effective information collection. (2) For health providers, we built an LLM-based dashboard to summarize and present important health information based on older adults’ conversations with the VA. We further conducted two user studies with older adults and providers to evaluate the usability of the system. The results showed that Talk2Care could facilitate the communication process, enrich the health information collected from older adults, and considerably save providers’ efforts and time. We envision our work as an initial exploration of LLMs’ capability in the intersection of healthcare and interpersonal communication.}, number={2}, journal={Proc. ACM Interact. Mob. Wearable Ubiquitous Technol.}, author={Yang, Ziqi and Xu, Xuhai and Yao, Bingsheng and Rogers, Ethan and Zhang, Shao and Intille, Stephen and Shara, Nawar and Gao, Guodong Gordon and Wang, Dakuo}, year={2024}, month=may, pages={73:1-73:35} }

@inproceedings{Baughan_Wang_Liu_Mercurio_Chen_Ma_2023, address={Hamburg Germany}, title={A Mixed-Methods Approach to Understanding User Trust after Voice Assistant Failures}, ISBN={978-1-4503-9421-5}, url={https://dl.acm.org/doi/10.1145/3544548.3581152}, DOI={10.1145/3544548.3581152}, booktitle={Proceedings of the 2023 CHI Conference on Human Factors in Computing Systems}, publisher={ACM}, author={Baughan, Amanda and Wang, Xuezhi and Liu, Ariel and Mercurio, Allison and Chen, Jilin and Ma, Xiao}, year={2023}, month=apr, pages={1–16}, language={en} }

@article{Bentley_Luvogt_Silverman_Wirasinghe_White_Lottridge_2018, title={Understanding the Long-Term Use of Smart Speaker Assistants}, volume={2}, DOI={10.1145/3264901}, abstractNote={Over the past two years the Ubicomp vision of ambient voice assistants, in the form of smart speakers such as the Amazon Echo and Google Home, has been integrated into tens of millions of homes. However, the use of these systems over time in the home has not been studied in depth. We set out to understand exactly what users are doing with these devices over time through analyzing voice history logs of 65,499 interactions with existing Google Home devices from 88 diverse homes over an average of 110 days. We found that specific types of commands were made more often at particular times of day and that commands in some domains increased in length over time as participants tried out new ways to interact with their devices, yet exploration of new topics was low. Four distinct user groups also emerged based on using the device more or less during the day vs. in the evening or using particular categories. We conclude by comparing smart speaker use to a similar study of smartphone use and offer implications for the design of new smart speaker assistants and skills, highlighting specific areas where both manufacturers and skill providers can focus in this domain.}, number={3}, journal={Proceedings of the ACM on Interactive, Mobile, Wearable and Ubiquitous Technologies}, author={Bentley, Frank and Luvogt, Chris and Silverman, Max and Wirasinghe, Rushani and White, Brooke and Lottridge, Danielle}, year={2018}, month=sep, pages={91:1-91:24} }

@book{Black_1962, title={Models and Metaphors: Studies in Language and Philosophy}, ISBN={978-0-8014-0041-4}, url={https://www.jstor.org/stable/10.7591/j.ctvr6971f}, abstractNote={Description not available.}, publisher={Cornell University Press}, author={Black, Max}, year={1962} }

@article{Brahnam_De_Angeli_2008, title={Special issue on the abuse and misuse of social agents}, volume={20}, ISSN={0953-5438}, DOI={10.1016/j.intcom.2008.02.001}, abstractNote={“You see, there’s a primal joy in hitting a thing in motion. It’s one of the oldest pleasures there is. Something moves, boo, you wing it. Beast, bird or human, the thing to do is to knock it down. It’s primal, Davy. It’s basic to the origin of the species.” – Don DeLillo, Americana}, number={3}, journal={Interacting with Computers}, author={Brahnam, Sheryl and De Angeli, Antonella}, year={2008}, month=may, pages={287–291} }

@article{Brahnam_Karanikas_Weaver_2011, title={(Un)dressing the interface: Exposing the foundational HCI metaphor “computer is woman”}, volume={23}, ISSN={0953-5438}, DOI={10.1016/j.intcom.2011.03.008}, abstractNote={Two fundamental (and oftentimes opposing) metaphors have directed much of HCI design: HCI is communication and HCI is direct manipulation. Beneath these HCI metaphors, however, is the unspoken metaphor of computer is woman. In this paper we expose this foundational metaphor. We begin by identifying the origin of computer is woman in the early history of computing. Drawing upon postmodern feminist theory, we then explore how this metaphor has resulted in the feminization of HCI is communication and second person interfaces. We show how images of femininity proliferate, becoming the projected images of male fantasies and ideals of womanhood. In becoming these idealized images, the interface is revealed as man in female drag. Finally, not only do we undress the interface to uncover how HCI is communication wraps the computer’s difference from human being within the more basic metaphor of computer is woman, but we also disclose dangers that can arise when this metaphor goes unacknowledged and unexamined.}, number={5}, journal={Interacting with Computers}, author={Brahnam, Sheryl and Karanikas, Marianthe and Weaver, Margaret}, year={2011}, month=sep, pages={401–412} }

@inproceedings{Dubiel_Desai_Zargham_Schmitt_2024, address={New York, NY, USA}, series={CUI ’24}, title={Voicecraft: Designing Task-specific Voice Assistant Personas}, ISBN={9798400705113}, url={https://doi.org/10.1145/3640794.3670000}, DOI={10.1145/3640794.3670000}, abstractNote={Voicecraft workshop aims to establish a research community focused on the design and evaluation of Voice Assistant (VA) personas for both task-oriented functions (e.g., information search, online shopping) and personal growth applications (e.g., coaching, mindful reflection, tutoring). Through discussion and collaborative efforts, we will seek to propose a set of practices and standards that will help to improve the ecological validity of VA personas. In particular, we will explore topics such as the interaction design of voice-based interfaces, the impact of agent personas on the user experience, and the approaches for designing such VA agents. This workshop will serve as a platform to build a better-equipped community to explore VA personas that provide a better fit to range of everyday interaction scenarios.}, booktitle={Proceedings of the 6th ACM Conference on Conversational User Interfaces}, publisher={Association for Computing Machinery}, author={Dubiel, Mateusz and Desai, Smit and Zargham, Nima and Schmitt, Anuschka}, year={2024}, month=jul, pages={1–3}, collection={CUI ’24} }

@inproceedings{Braun_Mainz_Chadowitz_Pfleging_Alt_2019, address={New York, NY, USA}, series={CHI ’19}, title={At Your Service: Designing Voice Assistant Personalities to Improve Automotive User Interfaces}, ISBN={978-1-4503-5970-2}, url={https://doi.org/10.1145/3290605.3300270}, DOI={10.1145/3290605.3300270}, abstractNote={This paper investigates personalized voice characters for in-car speech interfaces. In particular, we report on how we designed different personalities for voice assistants and compared them in a real world driving study. Voice assistants have become important for a wide range of use cases, yet current interfaces are using the same style of auditory response in every situation, despite varying user needs and personalities. To close this gap, we designed four assistant personalities (Friend, Admirer, Aunt, and Butler) and compared them to a baseline (Default) in a between-subject study in real traffic conditions. Our results show higher likability and trust for assistants that correctly match the user’s personality while we observed lower likability, trust, satisfaction, and usefulness for incorrectly matched personalities, each in comparison with the Default character. We discuss design aspects for voice assistants in different automotive use cases.}, booktitle={Proceedings of the 2019 CHI Conference on Human Factors in Computing Systems}, publisher={Association for Computing Machinery}, author={Braun, Michael and Mainz, Anja and Chadowitz, Ronee and Pfleging, Bastian and Alt, Florian}, year={2019}, month=may, pages={1–11}, collection={CHI ’19} }

@inbook{Carroll_Mack_Kellogg_1988, address={Amsterdam}, title={Chapter 3 - Interface Metaphors and User Interface Design}, ISBN={978-0-444-70536-5}, url={https://www.sciencedirect.com/science/article/pii/B9780444705365500087}, DOI={10.1016/B978-0-444-70536-5.50008-7}, abstractNote={This chapter discusses interface metaphors and the user interface design. The integration of operational, structural, and pragmatic approaches to metaphors can provide guidance and a starting point for the design of a user interface that integrates a central metaphor, with a carefully analyzed similarity basis and a set of planned mismatches, with myriad other interface elements that support and exploit the matches and mismatches inhering in the metaphor. Metaphoric comparisons and interface presentations do more than render static denotative correspondences. They have motivational and affective consequences for users. They interact with and frame users’ problem-solving efforts in learning about the target domain. Metaphors have been employed to increase the initial familiarity of the target domain, but they have an inevitable further role to play. The ultimate problem that the user should solve is to develop an understanding of the target domain itself—a mental model. Interface metaphors should also be viewed as tools proffered to users for articulating mental models.}, booktitle={Handbook of Human-Computer Interaction}, publisher={North-Holland}, author={Carroll, John M. and Mack, Robert L. and Kellogg, Wendy A.}, editor={Helander, MARTIN}, year={1988}, month=jan, pages={67–85}, language={en} }

@article{Carroll_Thomas_1982, title={Metaphor and the Cognitive Representation of Computing Systems}, volume={12}, ISSN={2168-2909}, DOI={10.1109/TSMC.1982.4308795}, abstractNote={In learning, people develop new cognitive structures by metaphorically extending old ones. The metaphors spontaneously generated by new users will predict the ease with which they can master a computer system. Systems which through their interface suggest inefficacious metaphors will accordingly be more difficult to learn and to that extent unacceptable.}, number={2}, journal={IEEE Transactions on Systems, Man, and Cybernetics}, author={Carroll, John M. and Thomas, John C.}, year={1982}, month=mar, pages={107–116} }

@article{Clark_Doyle_Garaialde_Gilmartin_Schlögl_Edlund_Aylett_Cabral_Munteanu_Edwards_et_al._2019, title={The State of Speech in HCI: Trends, Themes and Challenges}, volume={31}, ISSN={0953-5438}, DOI={10.1093/iwc/iwz016}, abstractNote={Speech interfaces are growing in popularity. Through a review of 99 research papers this work maps the trends, themes, findings and methods of empirical research on speech interfaces in the field of human–computer interaction (HCI). We find that studies are usability/theory-focused or explore wider system experiences, evaluating Wizard of Oz, prototypes or developed systems. Measuring task and interaction was common, as was using self-report questionnaires to measure concepts like usability and user attitudes. A thematic analysis of the research found that speech HCI work focuses on nine key topics: system speech production, design insight, modality comparison, experiences with interactive voice response systems, assistive technology and accessibility, user speech production, using speech technology for development, peoples’ experiences with intelligent personal assistants and how user memory affects speech interface interaction. From these insights we identify gaps and challenges in speech research, notably taking into account technological advancements, the need to develop theories of speech interface interaction, grow critical mass in this domain, increase design work and expand research from single to multiple user interaction contexts so as to reflect current use contexts. We also highlight the need to improve measure reliability, validity and consistency, in the wild deployment and reduce barriers to building fully functional speech interfaces for research.Most papers focused on usability/theory-based or wider system experience research with a focus on Wizard of Oz and developed systemsQuestionnaires on usability and user attitudes often used but few were reliable or validatedThematic analysis showed nine primary research topicsChallenges identified in theoretical approaches and design guidelines, engaging with technological advances, multiple user and in the wild contexts, critical research mass and barriers to building speech interfaces}, number={4}, journal={Interacting with Computers}, author={Clark, Leigh and Doyle, Philip and Garaialde, Diego and Gilmartin, Emer and Schlögl, Stephan and Edlund, Jens and Aylett, Matthew and Cabral, João and Munteanu, Cosmin and Edwards, Justin and R Cowan, Benjamin}, year={2019}, month=jun, pages={349–371} }

@inproceedings{Cowan_Pantidi_Coyle_Morrissey_Clarke_Al-Shehri_Earley_Bandeira_2017, address={New York, NY, USA}, series={MobileHCI ’17}, title={“What Can I Help You with?”: Infrequent Users’ Experiences of Intelligent Personal Assistants}, ISBN={978-1-4503-5075-4}, url={http://doi.acm.org/10.1145/3098279.3098539}, DOI={10.1145/3098279.3098539}, abstractNote={Intelligent Personal Assistants (IPAs) are widely available on devices such as smartphones. However, most people do not use them regularly. Previous research has studied the experiences of frequent IPA users. Using qualitative methods we explore the experience of infrequent users: people who have tried IPAs, but choose not to use them regularly. Unsurprisingly infrequent users share some of the experiences of frequent users, e.g. frustration at limitations on fully hands-free interaction. Significant points of contrast and previously unidentified concerns also emerge. Cultural norms and social embarrassment take on added significance for infrequent users. Humanness of IPAs sparked comparisons with human assistants, juxtaposing their limitations. Most importantly, significant concerns emerged around privacy, monetization, data permanency and transparency. Drawing on these findings we discuss key challenges, including: designing for interruptability; reconsideration of the human metaphor; issues of trust and data ownership. Addressing these challenges may lead to more widespread IPA use.}, note={00000 
event-place: Vienna, Austria}, booktitle={Proceedings of the 19th International Conference on Human-Computer Interaction with Mobile Devices and Services}, publisher={ACM}, author={Cowan, Benjamin R. and Pantidi, Nadia and Coyle, David and Morrissey, Kellie and Clarke, Peter and Al-Shehri, Sara and Earley, David and Bandeira, Natasha}, year={2017}, pages={43:1-43:12}, collection={MobileHCI ’17} }

@inbook{Desai_Chin_2021, address={New York, NY, USA}, title={Hey Google, Can You Help Me Learn?}, ISBN={978-1-4503-8998-3}, url={https://doi.org/10.1145/3469595.3469601}, abstractNote={In this provocation, we discuss the capacities and limitations of using commercially available smart speakers like Amazon Echo and Google Home to support informal learning. Bridging theories in self-regulated learning (SRL) and metacognition, we discuss whether smart speakers can (1) help learners to learn on their own and (2) help learners to learn better. We evaluate the current applications of smart speakers on learning and suggest future directions to harness smart speakers as learning partners in informal learning.}, booktitle={CUI 2021 - 3rd Conference on Conversational User Interfaces}, publisher={Association for Computing Machinery}, author={Desai, Smit and Chin, Jessie}, year={2021}, month=jul, pages={1–4} }

@inproceedings{Don_Brennan_Laurel_Shneiderman_1992, address={New York, NY, USA}, series={CHI ’92}, title={Anthropomorphism: from Eliza to Terminator 2}, ISBN={978-0-89791-513-7}, url={https://dl.acm.org/doi/10.1145/142750.142760}, DOI={10.1145/142750.142760}, booktitle={Proceedings of the SIGCHI Conference on Human Factors in Computing Systems}, publisher={Association for Computing Machinery}, author={Don, Abbe and Brennan, Susan and Laurel, Brenda and Shneiderman, Ben}, year={1992}, month=jun, pages={67–70}, collection={CHI ’92} }

@inproceedings{Doyle_Clark_Cowan_2021, address={Yokohama Japan}, title={What Do We See in Them? Identifying Dimensions of Partner Models for Speech Interfaces Using a Psycholexical Approach}, ISBN={978-1-4503-8096-6}, url={https://dl.acm.org/doi/10.1145/3411764.3445206}, DOI={10.1145/3411764.3445206}, booktitle={Proceedings of the 2021 CHI Conference on Human Factors in Computing Systems}, publisher={ACM}, author={Doyle, Philip R and Clark, Leigh and Cowan, Benjamin R.}, year={2021}, month=may, pages={1–14}, language={en} }

@inproceedings{Doyle_Edwards_Dumbleton_Clark_Cowan_2019, address={Taipei Taiwan}, title={Mapping Perceptions of Humanness in Intelligent Personal Assistant Interaction}, ISBN={978-1-4503-6825-4}, url={https://dl.acm.org/doi/10.1145/3338286.3340116}, DOI={10.1145/3338286.3340116}, abstractNote={Humanness is core to speech interface design. Yet little is known about how users conceptualise perceptions of humanness and how people de ne their interaction with speech interfaces through this. To map these perceptions n=21 participants held dialogues with a human and two speech interface based intelligent personal assistants, and then re ected and compared their experiences using the repertory grid technique. Analysis of the constructs show that perceptions of humanness are multidimensional, focusing on eight key themes: partner knowledge set, interpersonal connection, linguistic content, partner performance and capabilities, conversational interaction, partner identity and role, vocal qualities and behavioral a ordances. Through these themes, it is clear that users de ne the capabilities of speech interfaces di erently to humans, seeing them as more formal, fact based, impersonal and less authentic. Based on the ndings, we discuss how the themes help to sca old, categorise and target research and design e orts, considering the appropriateness of emulating humanness.}, booktitle={Proceedings of the 21st International Conference on Human-Computer Interaction with Mobile Devices and Services}, publisher={ACM}, author={Doyle, Philip R. and Edwards, Justin and Dumbleton, Odile and Clark, Leigh and Cowan, Benjamin R.}, year={2019}, month=oct, pages={1–12}, language={en} }

@article{Epley_Waytz_Cacioppo_2007, title={On Seeing Human: A Three-Factor Theory of Anthropomorphism}, volume={114}, DOI={10.1037/0033-295x.114.4.864}, number={4}, journal={Psychological Review}, author={Epley, Nicholas and Waytz, Adam and Cacioppo, John T.}, year={2007}, pages={864–886} }

@article{Fiske_Cuddy_Glick_Xu_2002, address={US}, title={A model of (often mixed) stereotype content: Competence and warmth respectively follow from perceived status and competition}, volume={82}, ISSN={1939-1315}, DOI={10.1037/0022-3514.82.6.878}, abstractNote={[Correction Notice: An Erratum for this article was reported online in Journal of Personality and Social Psychology on Apr 25 2019 (see record 2019-21976-001). In the fourth paragraph of the Status Predicts Competence, and Competition Predicts Warmth section, the results are worded in a confusing way, and some values are wrong. In the fourth paragraph’s first sentence, all correlation coefficients mistakenly omitted the negative sign implied in the text (“negatively correlated”) and shown in the correct values reported in Table 6. The text should appear instead as follows: Perceived competition negatively correlated with perceived warmth for the student sample, group-level r(21) .68, p r(71) .22, p  r(21) .53, p  r(36) .11, ns.] Stereotype research emphasizes systematic processes over seemingly arbitrary contents, but content also may prove systematic. On the basis of stereotypes’ intergroup functions, the stereotype content model hypothesizes that (1) 2 primary dimensions are competence and warmth, (2) frequent mixed clusters combine high warmth with low competence (paternalistic) or high competence with low warmth (envious), and (3) distinct emotions (pity, envy, admiration, contempt) differentiate the 4 competence-warmth combinations. Stereotypically, (4) status predicts high competence, and competition predicts low warmth. Nine varied samples rated gender, ethnicity, race, class, age, and disability out-groups. Contrary to antipathy models, 2 dimensions mattered, and many stereotypes were mixed, either pitying (low competence, high warmth subordinates) or envying (high competence, low warmth competitors). Stereotypically, status predicted competence, and competition predicted low warmth. (PsycInfo Database Record (c) 2022 APA, all rights reserved)}, number={6}, journal={Journal of Personality and Social Psychology}, publisher={American Psychological Association}, author={Fiske, Susan T. and Cuddy, Amy J. C. and Glick, Peter and Xu, Jun}, year={2002}, pages={878–902} }

@article{Gentner_Hoyos_2017, title={Analogy and Abstraction}, volume={9}, rights={Copyright © 2017 Cognitive Science Society, Inc.}, ISSN={1756-8765}, DOI={10.1111/tops.12278}, abstractNote={A central question in human development is how young children gain knowledge so fast. We propose that analogical generalization drives much of this early learning and allows children to generate new abstractions from experience. In this paper, we review evidence for analogical generalization in both children and adults. We discuss how analogical processes interact with the child’s changing knowledge base to predict the course of learning, from conservative to domain-general understanding. This line of research leads to challenges to existing assumptions about learning. It shows that (a) it is not enough to consider the distribution of examples given to learners; one must consider the processes learners are applying; (b) contrary to the general assumption, maximizing variability is not always the best route for maximizing generalization and transfer.}, number={3}, journal={Topics in Cognitive Science}, author={Gentner, Dedre and Hoyos, Christian}, year={2017}, pages={672–693}, language={en} }

@inproceedings{Gilad_Amir_Levontin_2021, address={New York, NY, USA}, series={CHI ’21}, title={The Effects of Warmth and Competence Perceptions on Users’ Choice of an AI System}, ISBN={978-1-4503-8096-6}, url={https://dl.acm.org/doi/10.1145/3411764.3446863}, DOI={10.1145/3411764.3446863}, abstractNote={People increasingly rely on Artificial Intelligence (AI) based systems to aid decision-making in various domains and often face a choice between alternative systems. We explored the effects of users’ perception of AI systems’ warmth (perceived intent) and competence (perceived ability) on their choices. In a series of studies, we manipulated AI systems’ warmth and competence levels. We show that, similar to the judgments of other people, there is often primacy for warmth over competence. Specifically, when faced with a choice between a high-competence system and a high-warmth system, more participants preferred the high-warmth system. Moreover, the precedence of warmth persisted even when the high-warmth system was overtly deficient in its competence compared to an alternative high competence-low warmth system. The current research proposes that it may be vital for AI systems designers to consider and communicate the system’s warmth characteristics to its potential users.}, booktitle={Proceedings of the 2021 CHI Conference on Human Factors in Computing Systems}, publisher={Association for Computing Machinery}, author={Gilad, Zohar and Amir, Ofra and Levontin, Liat}, year={2021}, month=may, pages={1–13}, collection={CHI ’21} }

@book{Goatly_2007, address={Amsterdam, Netherlands}, series={Washing the brain: Metaphor and hidden ideology}, title={Washing the brain: Metaphor and hidden ideology}, ISBN={978-90-272-2713-3}, DOI={10.1075/dapsac.23}, abstractNote={In his brilliant book Andrew Goatly convincingly argues that part of the blame for the way we have messed up our world politically, ecologically, economically, biologically, is on the deep-seated and largely unnoticeable metaphors that shape our thinking. As a first step to remedy the situation, we need to uncover these ideologically-loaded metaphors and look for alternative ones. Specifically, this book attempts two things. First, in Section One, chapters 1 to 5, it suggests that the metaphorical patterns observable in the lexicon of English have widespread effects on the concepts which drive our social practices and which reinforce social patterns of inequality, injustice and environmental exploitation within our present capitalist economies. Second, in Section Two, chapters 6 to 8, Goatly addresses the more theoretical question of the extent to which the metaphorical patterns to be found in the lexicon have their origins in (universal) bodily experiences, and the extent to which they are cultural and ideological constructs. (PsycINFO Database Record (c) 2016 APA, all rights reserved)}, publisher={John Benjamins Publishing Company}, author={Goatly, Andrew}, year={2007}, pages={xvii, 431}, collection={Washing the brain: Metaphor and hidden ideology} }

@article{Kövecses_2002a, title={Cognitive-linguistic comments on metaphor identification}, volume={11}, ISSN={0963-9470}, DOI={10.1177/096394700201100107}, number={1}, journal={Language and Literature}, publisher={SAGE Publications Ltd}, author={Kövecses, Zoltán}, year={2002}, month=feb, pages={74–78}, language={en} }

\appendix
\section{Appendix}
\subsection{Supporting Analyses and Reliability Checks}

\begin{table*}[htbp]
\centering
\Description[Euclidean distances of all 20 metaphors from four use-contexts and overall]{A table with 20 rows and 6 columns showing Euclidean distances. Bolded values indicate the shortest distance for each use-context. Genie has the shortest distance to Commands at 0.11 and the shortest overall distance at 0.22. Admirer has the shortest distance to Sociality at 0.50. Computer from Star Trek has the shortest distance to Information Seeking at 0.63. Search Engine has the shortest distance to Error Recovery at 0.30. Boss has the largest distances across all contexts.}
\caption{The Euclidean distances of each metaphor from the 4 use-contexts and overall. The metaphor with the shortest Euclidean distance to a use-context is bolded.}
\resizebox{\textwidth}{!}{%
\begin{tabular}{l|l|l|l|l|l}
\hline
\textbf{Metaphor} & \textbf{Commands} & \textbf{Sociality} & \textbf{Information Seeking} & \textbf{Error Recovery} & \textbf{Overall} \\ \hline
Admirer & 0.86 & \textbf{0.50} & 1.79 & 1.46 & 0.94 \\ \hline
Assistant & 1.28 & 2.51 & 1.23 & 0.67 & 1.33 \\ \hline
Aunt & 1.85 & 1.27 & 2.07 & 2.37 & 1.74 \\ \hline
Boss & 4.06 & 4.73 & 3.12 & 3.89 & 3.90 \\ \hline
Butler & 2.23 & 3.47 & 1.98 & 1.62 & 2.28 \\ \hline
Child & 2.18 & 1.42 & 3.17 & 2.68 & 2.30 \\ \hline
Coach & 3.05 & 3.41 & 2.31 & 3.10 & 2.87 \\ \hline
Companion & 1.70 & \textbf{0.65} & 2.28 & 2.31 & 1.66 \\ \hline
``Computer'' from Star Trek & 1.25 & 2.47 & \textbf{0.63} & 0.71 & 1.19 \\ \hline
Customer Service Agent & 1.40 & 2.60 & \textbf{0.65} & 0.89 & 1.33 \\ \hline
Encyclopedia & 1.83 & 2.97 & \textbf{0.91} & 1.39 & 1.74 \\ \hline
Family Pet & 2.64 & 1.83 & 3.63 & 3.13 & 2.76 \\ \hline
Flame & 1.70 & \textbf{0.61} & 2.30 & 2.31 & 1.66 \\ \hline
Friend & 2.28 & 1.05 & 2.94 & 2.90 & 2.26 \\ \hline
Genie & \textbf{0.11} & 1.15 & 1.08 & 0.74 & \textbf{0.22} \\ \hline
Librarian & 2.01 & 2.94 & 1.01 & 1.75 & 1.86 \\ \hline
Nurse & 2.25 & 3.07 & 1.29 & 2.07 & 2.10 \\ \hline
Search Engine & 0.41 & 1.65 & 0.68 & \textbf{0.30} & 0.38 \\ \hline
Teacher & 3.11 & 3.80 & 2.18 & 2.96 & 2.95 \\ \hline
Therapist & 2.30 & 3.16 & 1.32 & 2.08 & 2.15 \\ \hline
\end{tabular}%
}
\label{tab:euclidean_distances}
\end{table*}

First, we report the Euclidean distances of each metaphor from the four use-contexts and overall (see Table \ref{tab:euclidean_distances}). Further, to ensure the robustness of our quantitative findings for Study 2, we conducted normality checks (see Table \ref{tab:shapiro}) and reliability (see Table \ref{tab:cronbach}). 
These analyses confirm that all perception measures were internally consistent and that the distributional assumptions required for parametric testing were violated, supporting our use of non-parametric analyses in the main text.

\begin{table*}[htbp]
\begingroup
%\color{blue}
\centering
\Description[Shapiro-Wilk normality test results for paired differences]{A table showing Shapiro-Wilk test statistics and p-values for paired differences between conditions across five measures. All W statistics range from 0.917 to 0.941 and all p-values are below 0.001, indicating significant deviations from normality for all measures.}
\caption{{Shapiro–Wilk test results for normality of paired differences (Metaphor-Fluid – Default) across perception measures.($^{*}p<.05$, $^{**}p<.01$, $^{***}p<.001$)}}
\begin{tabular}{l|c|c}
\hline
\textbf{Measure} & \textbf{$W$} & \textbf{$p$} \\ \hline
Perceived Enjoyment & 0.917 & $<.001^{***}$ \\ \hline
Perceived Intelligence & 0.941 & $<.001^{***}$ \\ \hline
Perceived Trust & 0.931 & $<.001^{***}$ \\ \hline
Perceived Likability & 0.937 & $<.001^{***}$ \\ \hline
Perceived Intention to Adopt & 0.940 & $<.001^{***}$ \\ \hline
\end{tabular}

\vspace{2mm}
\label{tab:shapiro}
\endgroup
\end{table*}
Shapiro–Wilk tests were performed on paired differences between conditions (Metaphor-Fluid – Default). 
All $p$-values were below .05, indicating significant deviations from normality. 
Accordingly, all subsequent analyses used non-parametric Wilcoxon signed-rank tests. 

\begin{table*}[htbp]
\begingroup
\centering
\Description[Cronbach's alpha reliability coefficients for perception measures]{A table showing internal consistency values for Metaphor-Fluid and Default conditions separately and averaged. All alpha values are 0.88 or higher. Perceived Intelligence has the highest average alpha at 0.95. Perceived Trust has the lowest at 0.89. All measures demonstrate excellent internal consistency across both conditions.}
\caption{{Internal consistency (Cronbach’s $\alpha$) for perception measures across both VUI conditions.}}
\begin{tabular}{l|c|c|c}
\hline
\textbf{Measure} & \textbf{$\alpha_{MF}$} & \textbf{$\alpha_{Default}$} & \textbf{$\alpha_{Average}$} \\ \hline
Perceived Enjoyment & .93 & .93 & .93 \\ \hline
Perceived Intelligence & .95 & .94 & .95 \\ \hline
Perceived Trust & .88 & .89 & .89 \\ \hline
Perceived Likability & .91 & .92 & .92 \\ \hline
Perceived Intention to Adopt & .90 & .90 & .90 \\ \hline
\end{tabular}

\vspace{2mm}
\label{tab:cronbach}
\endgroup
\end{table*}

Cronbach’s $\alpha$ coefficients were computed separately for the Metaphor-Fluid and Default VUI conditions. 
All measures exhibited excellent internal consistency ($\alpha\geq.88$). 
Average $\alpha$ values are reported in the main text.

Finally, in Table \ref{tab:wilcoxon_results} we report the results of the Wilcoxon signed-ranked test results comparing Metaphor-Fluid and Default VUIs across all perception measures. 

\begin{table*}[t]
\centering
\Description[Wilcoxon signed-rank test results for five perception measures]{A table showing statistical comparisons between Metaphor-Fluid and Default VUIs. Perceived Enjoyment shows a significant difference with means 5.29 versus 4.79, W equals 542, adjusted p equals 0.0043, and effect size 0.42. Perceived Intention to Adopt shows means 4.40 versus 4.02, adjusted p equals 0.0244, effect size 0.32. Perceived Likability shows means 4.16 versus 3.93, adjusted p equals 0.0477, effect size 0.26. Perceived Trust and Perceived Intelligence show no significant differences.}
\caption{Wilcoxon signed-rank test results comparing Metaphor-Fluid and Default VUIs across perception measures. Mean and standard deviation ($M \pm SD$) for each condition are shown. Statistically significant results are denoted with asterisks (* $p<.05$, ** $p<.01$, *** $p<.001$). \textit{Note.} $p_{adj}$ = Benjamini--Hochberg adjusted $p$-values. $r$ = effect size.}
\label{tab:wilcoxon_results}
\begin{tabular}{l|c|c|c|c|c|c}
\hline
\textbf{Measure} & \textbf{$M_{MF} \pm SD$} & \textbf{$M_{Default} \pm SD$} & \textbf{W} & \textbf{$p$} & \textbf{$p_{adj}$} & \textbf{r} \\ \hline
Perceived Enjoyment & $5.29 \pm 1.22$ & $4.79 \pm 1.47$ & 542.00 & .0009*** & .0043** & .42  \\ \hline
Perceived Intention to Adopt & $4.40 \pm 1.53$ & $4.02 \pm 1.64$ & 656.50 & .0098** & .0244* & .32  \\ \hline
Perceived Likability & $4.16 \pm 0.84$ & $3.93 \pm 0.87$ & 924.50 & .0286* & .0477* & .26  \\ \hline
Perceived Trust & $5.64 \pm 0.93$ & $5.63 \pm 0.84$ & 1393.50 & .5830 & .5830 & .06 \\ \hline
Perceived Intelligence & $4.09 \pm 0.73$ & $4.20 \pm 0.65$ & 798.50 & .2110 & .2638 & .16 \\ \hline
\end{tabular}
\end{table*}

\end{document}